\pdfoutput=1

\documentclass[sigplan,10pt,nonacm,balance=false]{acmart}

\setcopyright{none}
\renewcommand\footnotetextcopyrightpermission[1]{}

\acmConference{}{}{}

\usepackage{graphicx}
\usepackage{subcaption}
\usepackage{tikz}
\usepackage{amsmath}
\usepackage{xspace}
\AtEndEnvironment{figure}{\vspace{-3pt}}
\AtEndEnvironment{figure*}{\vspace{-3pt}}
\AtEndEnvironment{table}{\vspace{-2pt}}
\AtEndEnvironment{table*}{\vspace{-2pt}}

\newcommand{\name}{Stint\xspace}

\newcommand{\suppref}[1]{Appendix~\ref{#1}}

\begin{document}

\title{Making Preemption Productive in Long-Output LLM Serving}

\author{Ziji Shi}
\affiliation{\institution{National University of Singapore}\country{Singapore}}
\email{zijishi@comp.nus.edu.sg}
\author{Xinyi Wan}
\affiliation{\institution{National University of Singapore}\country{Singapore}}
\email{wan.xinyi@u.nus.edu}
\author{Chaoyi Ruan}
\affiliation{\institution{National University of Singapore}\country{Singapore}}
\email{ruancy@comp.nus.edu.sg}
\author{Guangxing Huang}
\affiliation{\institution{Sea AI Lab}\country{Singapore}}
\email{doyoubihgx@gmail.com}
\author{Min Lin}
\affiliation{\institution{Sea AI Lab}\country{Singapore}}
\email{mavenlin@gmail.com}
\author{Jialin Li}
\affiliation{\institution{National University of Singapore}\country{Singapore}}
\email{lijl@comp.nus.edu.sg}

\begin{abstract}
Reasoning LLMs generate long outputs whose key--value (KV) caches occupy
GPU memory for minutes. When memory fills, the server preempts requests,
and a preempted request can repeatedly lose and recover its state while
producing little new output. We call this behavior KV thrashing. In our
replays, existing systems leave more than a fifth of requests stalled for
over a minute. Because preemption cannot be avoided without delaying
admission, we ask how to make it productive.
We present \name, a serving system that preserves a preempted request's
work, bounds how long it waits, and gives it fresh output after it returns.
\name stores fixed-length KV snapshots in a contiguity-first CPU pool with
an exact-extent allocator, and moves them with fused K/V gather/scatter
kernels and double buffering. Age-based readmission brings stalled
requests back, and protection gives them a turn of fresh decoding before
another preemption. Remaining-length estimates from decoding hidden states
guide victim selection.
On olympiad-math replays with Qwen3-32B on H200 GPUs, no request pauses
for more than 50 seconds under \name, and p99 time per output token falls
by 50.7\% relative to recomputation in the same engine. The replay as a
whole finishes 10.7\% sooner (12.0\% higher throughput), because the
requests that baselines starve no longer finish last. On this workload
\name also outperforms LMCache and SGLang with HiCache.
These gains require more host memory and replace rare, unbounded stalls
with more frequent, bounded pauses; they concentrate on long-output
workloads.
\end{abstract}

\maketitle
\pagestyle{plain}

\raggedbottom

\section{Introduction}
\label{sec:intro}

\begin{figure}[t]
  \centering
  \includegraphics[width=\linewidth]{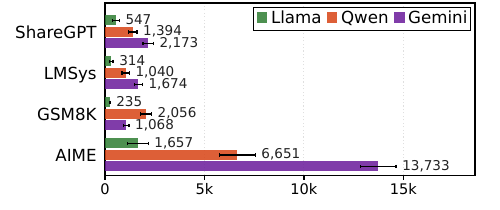}
  \caption{Historical output-length profiles for Llama-3.1-8B,
  Qwen3-14B and Gemini 3 Pro, separate from our serving experiments.
  Numbers label means; error bars show 95\% confidence intervals.}
  \Description{Horizontal grouped bars on a linear axis compare output token lengths on ShareGPT, LMSys, GSM8K, and AIME. Green, orange, and purple bars represent the three models, with exact mean values printed beside each bar. Capped error bars retain the original confidence intervals, recovered from the source figure's vector coordinates.}
  \label{fig:length-dist}
  \vspace{2pt}
  \includegraphics[width=\linewidth]{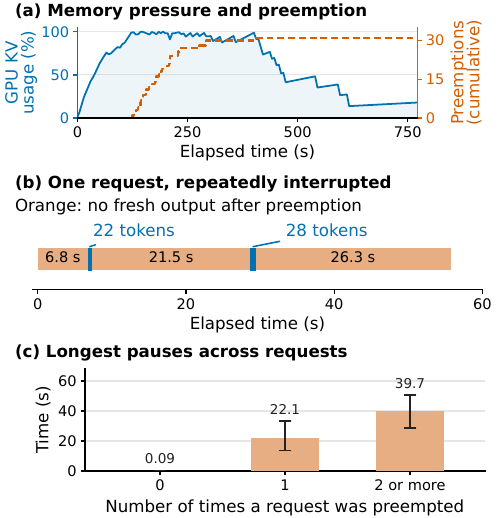}
  \caption{KV thrashing with Qwen3-32B on 200 programming requests from
  LiveCodeBench (LCB)~\cite{jain2024livecodebench}, denoted LCB200.
  GPU KV capacity is reduced and recovery is disabled.
  (a) GPU KV usage and cumulative preemptions over time.
  (b) One request's repeated pauses (orange) and fresh decoding (blue).
  (c) Median longest generation pause by preemption count across three
  replays; whiskers show the middle 50\%.}
  \Description{Three panels connect GPU KV usage, repeated interruption, and output pauses. The first overlays sampled GPU KV usage in blue on the left axis and cumulative preemptions in a dashed orange step line on the right axis. The count reaches 31 during a 772-second replica replay. The second follows three pauses of 6.8, 21.5 and 26.3 seconds in a request on the same replica, separated by 22 and 28 fresh tokens. The last groups requests by the number of times each was preempted: 0, 1, or 2 or more. Their median longest pauses are 0.09, 22.1 and 39.7 seconds, respectively.}
  \label{fig:kv_thrashing}
  \label{fig:preemption-kv-thrashing}
\end{figure}

Reasoning LLMs can solve difficult problems by generating intermediate
steps before reaching an answer~\cite{snell2024scaling,muennighoff2025s1}.
This longer reasoning process also puts pressure on GPU memory. As a
request generates tokens, the serving system stores their keys and values
in a growing KV cache. Longer outputs therefore require both more memory
and more time before that memory can be released.
Figure~\ref{fig:length-dist} illustrates the difference: across the three
models, mean outputs on American Invitational Mathematics Examination
(AIME) problems range from 1{,}657 to 13{,}733 tokens, compared with
314 to 2{,}173 tokens on the two conversational datasets.

When GPU memory runs out, the server must preempt a running request to
make room. It can save the request's KV cache in CPU memory or discard
the cache and recompute it later. In either case, the request must wait
for GPU space and recover its state before generating another token.
Preemption thus interrupts generation even while the server continues
working on other requests. Figure~\ref{fig:kv_thrashing}(a) follows GPU KV
usage during a replay: preemptions accumulate while the KV pool is nearly full.

More importantly, the same request can be preempted repeatedly. While it
waits, other requests keep growing their KV caches and consume the space
released by its eviction. Even after it returns to the GPU, it may soon
be evicted again. We call repeated eviction and recovery with little fresh
output between interruptions \textbf{KV thrashing}.
Figure~\ref{fig:kv_thrashing}(b) shows one request pausing for 21.5 seconds,
producing just 28 fresh tokens, then pausing for another 26.3 seconds.
Swap thrashing has also been observed in preemptive
serving~\cite{wu2026fastserve}; here we study how it disrupts long decoding.
The stakes are high. When we replay 399 olympiad-level math problems with
a median output of 17{,}089 tokens, vLLM with recomputation, LMCache and
SGLang with HiCache each leave 21--23\% of requests stalled for more than a
minute, and about one request in six for more than five minutes. Under
recomputation, a single pause takes up to 72\% of a request's lifetime.

Preemption itself cannot be avoided once long decodes oversubscribe GPU
memory. Admission control can prevent it, but moves the wait to before the
first token: a simple concurrency cap delays first output by several
minutes in our replays (Section~\ref{sec:eval-admission}). We therefore
ask a different question: \emph{how can preemption be made productive?}
A preempted request should keep the work it has already done, return to
the GPU within a bounded time, and produce new output once it returns.
Each condition poses a challenge.

First, the server must choose which request to interrupt. A request with a
large KV cache frees more memory when evicted, but may be close to
completion; its cache size does not reveal how much work remains.
Second, saving and restoring KV must be cheap. CPU memory is larger than
GPU memory but still limited. Reserving large chunks wastes space, while
moving scattered GPU blocks individually adds transfer overhead.
Third, restoration must lead to progress. A saved request can wait
indefinitely while running requests keep growing their caches, and even
after it returns, another preemption can interrupt it before it produces
much output. Fast transfers alone do not solve this scheduling problem.

We present \name, a serving system that makes preemption productive by
connecting the choice of victim, the storage and transfer of its KV, and
its return to useful decoding.

\paragraph{Prediction-guided victim selection.}
\name uses hidden states from ongoing decoding to estimate how much output
each request has left, and prefers to keep requests that appear close to
completion. Sparse online inference refreshes these estimates periodically,
limiting their cost while letting them change as the answer develops.

\paragraph{Contiguity-first CPU KV pool.}
\name stores KV snapshots in contiguous CPU regions. The design follows a
simple observation: \emph{GPU KV grows during decoding, but an offloaded
snapshot has a fixed length until the request resumes.} The allocator can
therefore reserve exactly the snapshot's size, preferring one contiguous
extent, without reserving room for future growth.

\paragraph{Bulk KV transfer.}
Fused K/V kernels gather a request's scattered GPU blocks into staging
buffers and scatter reloaded KV back, so CPU--GPU copies move contiguous
ranges. Double buffering lets one copy proceed while the GPU gathers or
scatters the next group of blocks.

\paragraph{Age-based protected recovery.}
An age-based rule gives a stalled request priority to return to the GPU,
bounding how long it waits. Once selected, the request is protected from
preemption through restoration and a turn of fresh decoding. Reloading KV
or recomputing old context does not count as progress; generating new
tokens does.

We implement \name in vLLM and evaluate it on H200 GPUs with Qwen3-32B.
On the olympiad math replay (IMO-AnswerBench~\cite{luong2025imobench}),
\name raises throughput by 12.0\% and lowers 99th-percentile time per
output token (TPOT) by 50.7\% relative to recomputation in the same engine,
across three matched runs. No request pauses for more than 50 seconds,
versus 11--14 minutes for the worst requests under every baseline.
\name also outperforms LMCache by 14.2\% and SGLang with HiCache by 48.5\%
in throughput, although the latter comparison also spans different
engines. On AIME and LiveCodeBench, p99 TPOT falls by 35--75\% while
throughput changes by $-0.5$\% to $+3.6$\%.

These gains have costs. \name pauses most long requests once, for about
30 seconds but never more than 8.4\% of the request's lifetime, and it
raises first-token and pooled inter-token latency tails. It also needs
more pinned host memory. Benefits concentrate on long-output workloads;
on short-output summarization and BurstGPT replays, \name offers no gain.
In component studies, retaining more snapshots and readmitting stalled
requests raise throughput, while prediction-guided selection lowers it in
exchange for shorter output gaps.

\paragraph{Contributions.}
\begin{itemize}
  \item We characterize KV thrashing during long decoding and show that
  existing systems leave a large fraction of requests stalled for minutes.
  \item We design age-based protected recovery, which bounds how long a
  preempted request waits and guarantees it fresh output after it returns.
  \item We design a contiguity-first CPU KV pool with an exact-extent
  allocator, and bulk KV transfer with fused K/V kernels and double
  buffering, exploiting the fixed length of offloaded snapshots.
  \item We use decoding-time hidden states for prediction-guided victim
  selection, with sparse online inference to limit serving cost.
  \item We evaluate the complete system against three baselines on five
  workloads, including where its benefits end and what they cost.
\end{itemize}
\section{Background}
\label{sec:background}

\subsection{KV Cache Growth During Decoding}

Each new token attends to keys and values from earlier tokens. The serving
system keeps these tensors in a KV cache so it does not have to compute
them again at every step. With full attention, each generated token adds
KV state, and the cache remains live until the request finishes. Long
outputs therefore increase both the size of the cache and how long it
occupies memory.

Table~\ref{tab:kv_size} shows the KV payload for several models, including
Qwen3-32B and Llama-3.1-8B from our main evaluation. A single
10{,}000-token context requires about 2.44\,GiB for Qwen3-32B and
4.81\,GiB for Llama-3.1-405B. Serving many such requests together can
exhaust GPU memory even when each request fits individually.

\begin{table}[t]
\centering
\caption{BF16 KV payload per request, before sharding or replication
across GPUs. Context includes prompt and output tokens. Values exclude
allocation padding and metadata. DeepSeek-R1 uses its compressed MLA cache.}
\begin{tabular}{lrr}
\toprule
\textbf{Model} & \textbf{Per token} & \textbf{10{,}000 tokens} \\
\midrule
DeepSeek-R1~\cite{guo2025deepseek} & 68.625\,KiB & 0.654\,GiB \\
Llama-3.1-405B~\cite{grattafiori_llama_2024} & 504\,KiB & 4.807\,GiB \\
Qwen3-32B~\cite{yang2025qwen3} & 256\,KiB & 2.441\,GiB \\
Llama-3.1-8B~\cite{grattafiori_llama_2024} & 128\,KiB & 1.221\,GiB \\
\bottomrule
\end{tabular}
\label{tab:kv_size}
\end{table}

For Llama and Qwen, the per-token payload is $2L H_{\mathrm{KV}} d b$ bytes: $L$ layers,
$H_{\mathrm{KV}}$ KV heads, head dimension $d$, and $b$ bytes per element,
for both keys and values. These three models have eight KV heads with
dimension 128; the number of layers is 32, 64 and 126 for Llama-3.1-8B,
Qwen3-32B and Llama-3.1-405B, respectively.

DeepSeek-R1 uses multi-head latent attention (MLA). It stores a compressed
vector shared by keys and values, together with a positional key vector.
The table counts this compressed representation. MLA reduces the bytes
per token, but the cache still grows with context length. The physical
memory needed across GPUs also depends on how a serving system shards
or replicates these tensors.

\subsection{KV Cache Management for Serving}

PagedAttention~\cite{kwon2023efficient} divides GPU KV into fixed-size
blocks. A request acquires more blocks as it grows, without reserving a
large contiguous region for its possible future output. This reduces
fragmentation, although the final block can remain partly empty.
Prefix caching~\cite{zheng2024sglang} also allows requests with a common
prefix to reuse KV blocks. Systems such as
LMCache~\cite{cheng2025lmcache,yao_cacheblend_2025} store and transfer KV
across memory tiers and serving workers.

These mechanisms help use available capacity. When capacity is exhausted,
the server still needs to choose a request to preempt and a way to recover
its state later.

\subsection{Current Preemption Mechanisms}

Preemption releases GPU space by removing a request's KV blocks. The
request can recover through either recomputation or offloading.

\paragraph{Recomputation.}
The server discards KV and rebuilds missing state from the saved token
sequence when the request resumes. Any reusable cached prefix reduces
this work. For long contexts, rebuilding still consumes substantial GPU
compute: full causal attention has a quadratic cost component. These
computations restore old state before the request can produce new output.

\paragraph{Offloading.}
The server saves KV in CPU memory and reloads it when the request resumes.
This preserves computed state but uses CPU capacity and host--device
bandwidth. Moving many small GPU blocks can also add dispatch and packing
costs. The choice between these mechanisms depends on how much state must
be rebuilt or moved, as well as the available compute and bandwidth.
\suppref{app:cost-model} derives a compute--transfer cost model and
explains how context length and model geometry affect this tradeoff.

Recovering KV state is only part of resuming a preempted request. The
scheduler must first make enough GPU memory available to reload or
rebuild that state. If running requests continue to occupy the memory,
the preempted request has to wait. Even after it resumes, it may be
preempted again after generating only a few new tokens. The server then
repeats the transfer or recomputation, while the request makes little
progress. The next section examines how this cycle interrupts long
decoding requests.
\section{Challenges with Serving Reasoning Models}
\label{sec:problem}

KV thrashing involves more than the cost of one transfer or one
recomputation. A request can repeatedly lose GPU space, recover its state,
and lose that space again. We first examine the resulting interruptions,
then the decisions that a serving system must make to reduce them.

\subsection{Repeated Preemption Interrupts Decoding}

Repeated preemption matters because a request can spend time recovering
without making much progress. Figure~\ref{fig:kv_thrashing}(b) follows three
consecutive interruptions in one request. It produces only 22 fresh tokens
between its third and fourth evictions, then 28 between its fourth and
fifth. The fourth and fifth evictions interrupt fresh output for 21.5 and
26.3 seconds, respectively. Recovery repeatedly gives this request only a
short opportunity to decode before it must recover again.

Figure~\ref{fig:kv_thrashing}(c) puts these interruptions in context. The
median longest streaming pause is 0.09 seconds for request executions
without preemption, 22.1 seconds for those preempted once, and 39.7 seconds
for those preempted repeatedly. These groups differ in output length and
memory pressure, so the comparison shows an association, not a fixed
latency penalty per preemption. Looking at each request's longest pause
exposes interruptions that an average time per output token can hide.

This illustration uses a separate 200-request LiveCodeBench replay
(LCB200) with recovery disabled and reduced GPU KV capacity, making
repeated preemption easier to observe. It is a motivation experiment, not a comparison of complete
systems at their native capacity. The setup and sampling method are given
in \suppref{app:methods}.

\subsection{Memory Relief and Remaining Work Can Conflict}
\label{sec:challenge-victim}

The first decision is which request to interrupt when GPU memory runs
out. Current KV size measures how much memory a request occupies;
remaining output length estimates how much decoding it has left.
These signals need not rank requests in the same order.

For example, a request with a large cache may be close to completion, while
a request with a smaller cache may have many tokens left to generate.
Preempting the first can free more memory immediately but interrupt a
possible completion. Preempting the second preserves that opportunity but
may free too little space, requiring more evictions. Shared prefix blocks
can further reduce the memory released by an eviction.

The scheduler must therefore balance immediate memory relief against
the opportunity to finish a request and release its cache. It knows how
much memory each request occupies, but how much output remains must be
estimated while the answer is still being generated.

\subsection{Contiguity vs Fragmentation Tradeoff}

After choosing a victim, the server can save its KV in CPU memory to
avoid rebuilding it later. Storing the snapshot in a contiguous region
supports bulk transfers. Each snapshot has a fixed length, but snapshots
from different requests have different sizes and lifetimes. As requests
resume and free their CPU storage, gaps appear between the snapshots that
remain. A new snapshot may not fit in any gap, even when the pool has
enough free space in total.

Large uniform chunks simplify allocation but waste space in partially
filled chunks. Dividing snapshots into many small pieces uses capacity
more flexibly, but requires the transfer engine to handle more regions.
The challenge is to keep snapshots easy to transfer while retaining as
much useful KV as possible in limited CPU memory.

\subsection{Recovery Must Be Efficient and Lead to Progress}
\label{sec:challenge-mechanism}

Saving a snapshot avoids losing computed state, but returning it to the
GPU still takes work. GPU KV is spread across pages, and copying them
individually can require many small transfers, as in vLLM v0.9.2's
block-swapping primitive~\cite{vllmBlockSwap092}. Existing systems reduce
this overhead with larger chunks and pipelines, specialized GPU kernels,
or batched copies~\cite{cheng2025lmcache,sglangHiCacheDesign2026,mooncakeTransferSource2026,yu2026superinfer}.
The CPU layout and transfer method must work together to make restoration
efficient.

Even a fast transfer cannot begin until GPU space is available. While an
offloaded request waits, running requests can consume that space. Once
the snapshot is restored, another eviction may interrupt decoding after
only a few new tokens, as in Figure~\ref{fig:kv_thrashing}(b). Reducing the
cost of each restoration therefore addresses only part of KV thrashing.
The scheduler must also give waiting requests a way to return and enough
time to use their restored state. Giving one request a longer turn,
however, can increase the waiting time of others.
\section{Motivation}

A preempted request leaves behind both completed work and unfinished
output. Our design aims to preserve what has already been computed while
deciding when to spend GPU time on what remains.

\subsection{Use Remaining Work to Choose Victims}

A request close to completion can soon release GPU memory. Its current
cache size, however, does not tell us how much output remains. \name uses
hidden states from ongoing decoding to estimate that remaining work. These
states contain information from the partially generated answer that a
prompt-only estimator cannot use~\cite{dong_emergent_2025}. TRAIL also uses
decoding-time embeddings for length-aware scheduling~\cite{shahout2024dontstop}.

This allows the estimate to change as the answer develops. \name refreshes
it periodically through sparse online inference, while keeping the
predictor's weights fixed (Section~\ref{sec:length-prediction}). The goal
is to use information about remaining work without adding prediction
cost at every token.

\subsection{Preserve KV When Space Is Available}
\label{sec:recompute-offload-model}

\begin{figure*}[t]
  \centering
  \includegraphics[width=\textwidth]{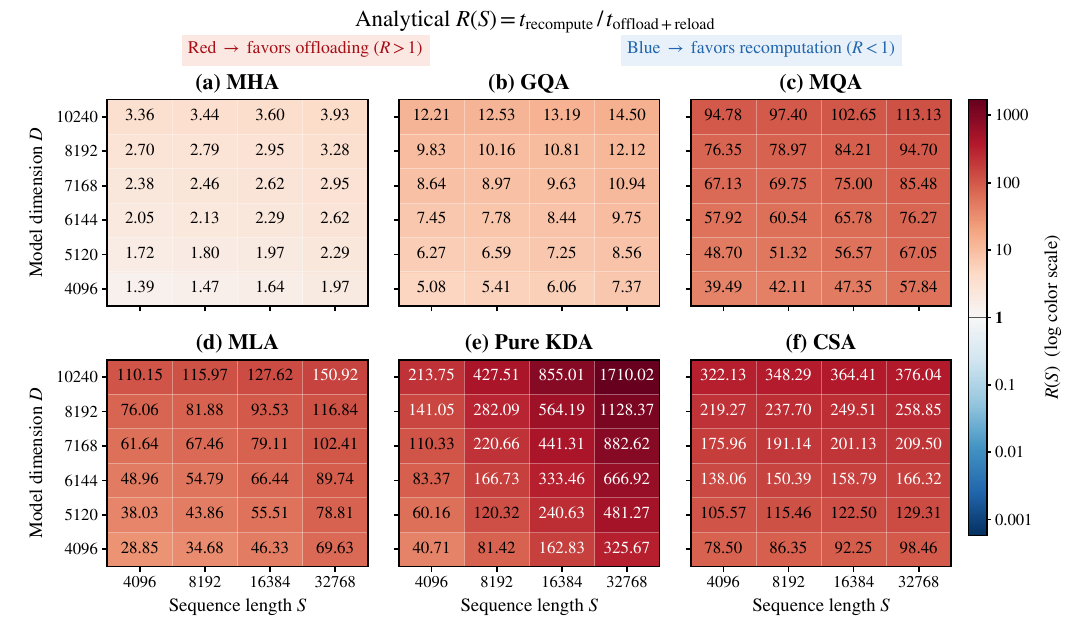}
  \caption{Modeled cost of rebuilding a full context relative to offloading
  and reloading its saved state. Red favors offloading; blue favors
  recomputation. Cell labels show the time ratio, with a shared logarithmic
  color scale centered at equal cost ($R=1$).}
  \label{fig:recompute-offload-comparison}
\end{figure*}

After choosing which request to pause, the server must decide what to
do with the computation it has already completed. Offloading keeps the
resulting KV so that recovery can copy it back; recomputation discards
KV and spends GPU compute to rebuild it. For conventional full attention,
KV bytes grow linearly with
context length, while dense
attention adds a quadratic component to recomputation. Long contexts can
therefore favor offloading, though CPU capacity, bandwidth, prefix reuse
and interference with other requests also affect the cost.

Figure~\ref{fig:recompute-offload-comparison} compares these two costs across
six attention mechanisms. Each cell divides the time to rebuild a full
context by the time to offload and reload its saved state. A value of two,
for example, means rebuilding takes twice as long as the transfer round
trip in this model.

All plotted cases favor transferring saved state at the assumed rates.
Compact state can be cheap to move even when rebuilding it is costly:
MLA and CSA store compressed history, while pure KDA saves a fixed-size
state. For pure KDA, a longer context adds rebuild work without increasing
the snapshot size. Smaller state can also reduce memory pressure, so these
ratios alone do not show how often preemption occurs.

To compute these ratios, we give all panels the same feed-forward layer
at each model width, with an intermediate width four times the hidden
width. We assume the GPU can perform 12,500 floating-point operations in
the time the CPU--GPU link moves one byte. MLA, KDA and CSA use attention
dimensions drawn from DeepSeek-V3~\cite{deepseek2024v3},
Kimi Linear~\cite{kimiteam2025linear} and DeepSeek-V4~\cite{deepseek2026v4},
respectively. We keep their head sizes and compression dimensions fixed
while varying the hidden width. These are comparisons of individual
blocks, not measured serving speedups for the full models.
\suppref{app:cost-model} gives the dimensions and derivation.

Preserving state also gives us a useful allocation opportunity. GPU KV
grows during decoding, but an offloaded snapshot has stopped growing.
Its known size lets the CPU pool reserve a few contiguous regions without
leaving room for future tokens. Those regions can then support bulk
transfers from and back to the paged GPU cache.

The cost model motivates retaining KV; it is not an online decision rule.
\name offloads when CPU snapshot allocation succeeds and recomputes
otherwise (Section~\ref{sec:victim-policy}).

\subsection{Make Restoration Lead to New Output}

Retaining KV can make restoration cheaper, but its value depends on what
happens next. If a request waits indefinitely for GPU space, its saved
state cannot help it finish. If it returns only to be evicted again after
a few tokens, the system repeatedly pays to restore the same accumulated
work.

This motivates coupling restoration with an opportunity to generate new
output. \name uses stall age to give a waiting request priority to return,
then protects it for a turn of new tokens. The turn starts with state
restoration, but only newly generated tokens count toward completing it.
This connects the CPU pool and transfer engine to the scheduler: they
preserve and restore the state, while recovery scheduling gives the
request time to use it (Section~\ref{sec:ageing}).
\section{Design}
\label{sec:design}

\name follows a request from preemption back to decoding: choose a victim,
save its KV if possible, then restore it and give it time to produce new
output. Prediction guides the first decision; stall age and protection
guide the return.

The common goal is useful progress. A fast KV reload is wasted if the
request is evicted again before it can use that state. Conversely, a
protected request still has to recover its state before decoding. The
scheduler, CPU pool and transfer engine therefore work along the same
recovery path.

\subsection{Design Goals and System Overview}
\label{sec:design-goals}

\begin{figure*}[t]
  \centering
\begingroup
\definecolor{archink}{RGB}{35,49,61}
\definecolor{archblue}{RGB}{86,180,233}
\definecolor{archorange}{RGB}{230,159,0}
\definecolor{archgreen}{RGB}{0,158,115}
\begin{tikzpicture}[
  x=1in,y=1in,
  font=\fontsize{10.5}{12}\selectfont\sffamily,
  text=archink,
  line width=0.75pt,
  box/.style={draw=archink!80,rounded corners=2pt,align=center,
    inner sep=2pt,minimum width=1.88in,minimum height=0.54in},
  note/.style={align=center,inner sep=1pt,fill=white},
  flow/.style={->,>=stealth,draw=archink,line width=0.9pt}
]
\path[use as bounding box] (0,-0.16) rectangle (6.90,1.38);

\node[box,minimum width=1.60in,minimum height=0.88in,fill=black!3]
  (runner) at (0.90,0.47)
  {\textbf{Model runner}\\[3pt]Prefill and decode};
\node[box,fill=archblue!14] (predictor) at (3.32,0.80)
  {\hyperref[sec:length-prediction]{\textbf{Length predictor} (\S\ref*{sec:length-prediction})}\\[2pt]Sparse online inference};
\node[box,fill=archorange!14] (scheduler) at (5.88,0.80)
  {\hyperref[sec:scheduling]{\textbf{Scheduler} (\S\ref*{sec:scheduling})}\\[2pt]
   SRJF victim policy\\
   Age-based protected recovery};
\node[box,fill=archgreen!12] (pool) at (5.88,0.14)
  {\hyperref[sec:cpu-kv-pool]{\textbf{KV manager} (\S\ref*{sec:cpu-kv-pool})}\\[2pt]Contiguity-first CPU KV pool\\Exact-extent allocator};
\node[box,fill=archblue!8] (transfer) at (3.32,0.14)
  {\hyperref[sec:async-transfer]{\textbf{Transfer engine} (\S\ref*{sec:async-transfer})}\\[2pt]Fused K/V gather/scatter\\Double-buffered bulk copies};

\draw[flow] (scheduler.north) -- (5.88,1.28)
  -- (0.90,1.28) -- (runner.north);
\node[note] at (3.32,1.28) {Scheduled requests};

\draw[flow] (1.70,0.80) -- (predictor.west);
\node[note] at (2.04,1.02) {Hidden\\states};
\draw[flow] (predictor.east) -- (scheduler.west);
\node[note] at (4.60,1.02) {Length\\estimates};
\draw[flow] (scheduler.south) -- (pool.north);
\draw[flow] (pool.west) -- (transfer.east);
\node[note] at (4.60,0.39) {Block\\maps};
\draw[flow] (transfer.west) -- (1.70,0.14);
\node[note] at (2.04,0.39) {KV ready\\events};
\end{tikzpicture}
\endgroup
   \caption{One replica combines prediction-guided victim selection, a
  contiguity-first CPU KV pool, bulk KV transfer, and age-based protected
  recovery. Arrows connect prediction, state preservation and return to decoding.}
  \Description{Five boxes show the model runner, predictor, scheduler, KV manager and transfer engine. Hidden states flow from the runner through prediction to the scheduler, which sends scheduled requests back. The scheduler requests preemption or restoration from the KV manager. Block maps reach the transfer engine, whose layer-ready events order attention in the model runner.}
  \label{fig:arch}
\end{figure*}
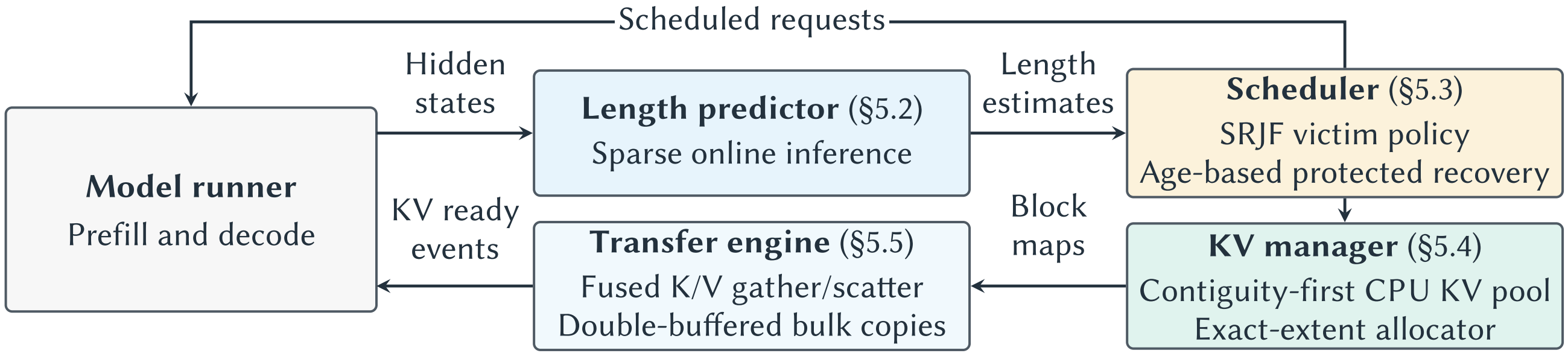

Figure~\ref{fig:arch} shows how the components interact within one replica.

\paragraph{Prediction-guided victim selection.}
On GPU allocation failure, \name uses hidden-state estimates of remaining
output length to choose a victim, extending vLLM's first-come-first-served
scheduler (Section~\ref{sec:scheduling}).

\paragraph{Contiguity-first CPU KV pool.}
The KV manager stores fixed-length snapshots in CPU \emph{extents}, or
contiguous ranges of block positions. It prefers one extent and falls back
to several, then to recomputation if allocation fails
(Section~\ref{sec:cpu-kv-pool}).

\paragraph{Bulk KV transfer.}
Fused K/V kernels gather scattered GPU blocks for bulk copying into CPU
extents and scatter them back on reload. Double buffering overlaps these
kernels with copying (Section~\ref{sec:async-transfer}).

\paragraph{Age-based protected recovery.}
The scheduler selects an overdue stalled request, makes GPU space for it,
and protects it through KV restoration and a turn of fresh decoding
(Section~\ref{sec:ageing}).

\subsection{Remaining-Length Prediction During Decoding}
\label{sec:length-prediction}

To choose which request to preempt, the scheduler needs to know which
requests are close to finishing. KV cache size cannot answer this question:
two requests with similar caches may have very different amounts of output
left. \name therefore estimates remaining length while the answer is
being generated.

\paragraph{Estimating how much output remains.}
We use hidden states, the internal representations the model already
computes during token generation. Prior work suggests that these
representations encode output structure~\cite{dong_emergent_2025}.
A small predictor combines information from selected layers to estimate
what fraction of the answer has been generated. This requires no
additional forward pass through the LLM.

For example, if a request has generated 1,000 tokens and the predictor
estimates that it is halfway finished, about 1,000 tokens remain. More
generally, after generating $d$ tokens, an estimated completed fraction
$\hat{p}$ gives a total length of $d/\hat{p}$. Subtracting the tokens
already generated gives the estimated remaining length $\hat{r}_i$.
We smooth successive estimates to reduce fluctuations. The predictor's
weights stay fixed during serving; only its estimates change as the
answer develops.

\paragraph{Keeping prediction inexpensive.}
Hidden states are already available during generation, but collecting them
and running a predictor at every token still adds overhead. \name
therefore reuses predictions across decoding steps. It caches the predicted
total length and subtracts newly generated tokens to update the remaining
length. This \emph{sparse online inference} normally refreshes predictions
every $K$ decoding steps, with $K=64$ in our evaluated configuration.
A request receives an earlier update when it has no estimate or is
approaching its predicted end.

Reducing predictor calls alone would still leave the cost of collecting
its inputs. \name avoids this work by deciding which requests need an
update before the LLM's forward pass. During that pass, it saves only the
required token representations from selected layers for those requests.
When no update is needed, it skips both feature collection and predictor
execution. Thus, most decoding steps reuse an estimate without preparing
another prediction.

When several requests need updates, \name processes them together in one
batch on the serving GPU. The small predictor first compresses the selected
hidden states into smaller vectors, then estimates progress. Keeping these
features on the GPU avoids transferring large hidden-state tensors to the
CPU; only the resulting scalar predictions need to be returned.

The interval $K$ balances prediction cost against freshness. Longer
intervals reduce regular prediction work, while early updates let requests
revise estimates they are about to outgrow.
Section~\ref{sec:eval-predictor-cost} measures the serving overhead, and
\suppref{app:prediction-interval} examines this tradeoff across different
intervals.

\paragraph{Supporting different models.}
Qwen and Llama use their own trained predictors, but both provide the
scheduler with the same information: a predicted total output length.
This lets us change the predictor without changing the scheduling policy.

These estimates help the scheduler choose which request to interrupt.
Completing the recovery cycle also requires deciding when that request
should return and giving it time to produce new output. The next section
explains how \name connects these decisions.

\subsection{Victim Selection and Recovery}
\label{sec:scheduling}

When GPU memory fills, the scheduler must decide which request to pause
and how that request will resume. \name uses predicted remaining length
to favor keeping requests that are close to completion. It also tracks
the progress of preempted requests, giving those that remain stalled
priority to return and a protected turn to generate new output.

\subsubsection{Prediction-Guided Victim Selection}
\label{sec:victim-policy}

When memory runs out, \name favors keeping requests that appear close to
completion. It preempts the eligible request with the longest predicted
remaining output. For the eligible set $\mathcal{E}$, this rule is
\begin{equation}
  \text{victim} = \arg\max_{i \in \mathcal{E}} \hat{r}_i.
  \label{eq:eviction-score}
\end{equation}
This is the shortest-remaining-job-first (SRJF) victim policy: evicting
the longest remaining job favors keeping near-completion requests on the GPU.
The eligible set contains other running requests, excluding those already
selected for the current step and the request protected by recovery.
Ties prefer earlier arrivals, then request ID. If any eligible candidate
lacks a usable prediction, the whole decision falls back to
\emph{physical-largest}: choose the request that releases the most GPU
blocks. If no alternative is eligible, the ordinary request needing
memory can itself be preempted.

\paragraph{Remaining work versus memory relief.}
A request with much output left can have a small current cache; evicting
it may free little memory. A nearly finished request can hold a large
cache that is expensive to move. SRJF favors possible completions, while
physical-largest favors immediate memory relief.
Section~\ref{sec:eval-victim} compares them with storage, transfers and
recovery held fixed.

\paragraph{Preserve KV when CPU space is available.}
After selecting a victim, the scheduler allocates CPU storage for its
snapshot. Success leads to offloading; failure due to capacity, extent or
request limits leads to discarding KV and queuing recomputation
(Section~\ref{sec:recompute-offload-model}). This reacts to GPU allocation
failure; it does not predict future offload times or use a per-victim
compute--transfer threshold.

\subsubsection{Age-Based Protected Recovery}
\label{sec:ageing}

Choosing a victim frees memory for other requests, but leaves the
interrupted request waiting. Those other requests continue growing their
KV caches, so the waiting request may struggle to regain GPU space.
Even after it returns, another eviction can interrupt it before it
generates much output. It then pays for recovery again after making
little progress: the KV thrashing described in Section~\ref{sec:problem}.

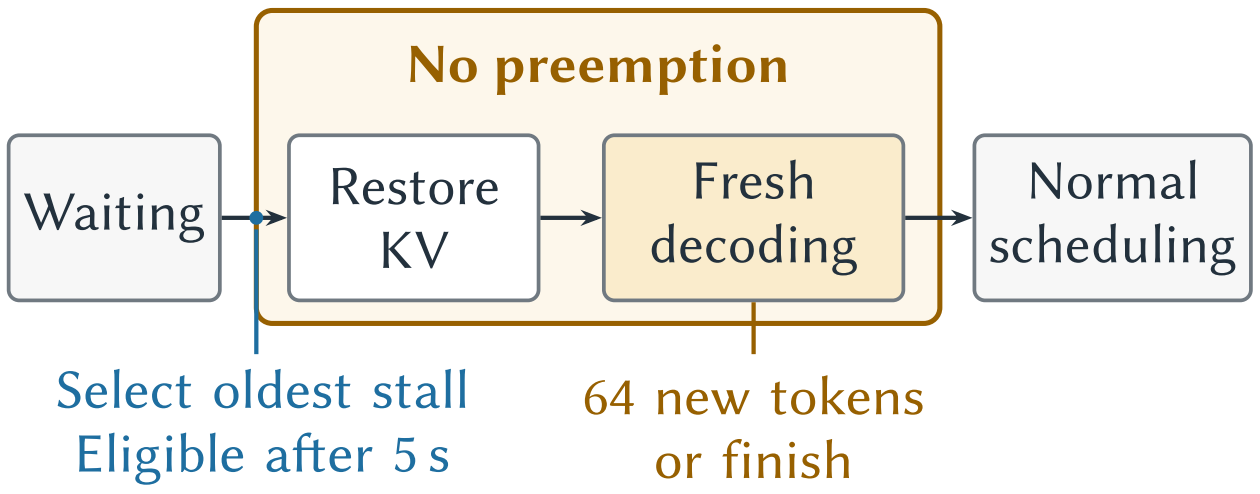
\begin{figure}[t]
  \centering
\begingroup
\usetikzlibrary{arrows.meta}
\definecolor{rescueink}{RGB}{35,49,61}
\definecolor{rescueblue}{RGB}{30,110,160}
\definecolor{rescueorange}{RGB}{230,159,0}
\definecolor{rescuedark}{RGB}{151,96,0}
\begin{tikzpicture}[
  x=1in,y=1in,
  font=\fontsize{10.5}{12}\selectfont\sffamily,
  text=rescueink,
  line width=0.7pt,
  stage/.style={draw=rescueink!65,rounded corners=2pt,
    align=center,inner sep=2pt,minimum height=.43in},
  flow/.style={-{Stealth[length=4pt,width=3pt]},line width=.8pt,draw=rescueink},
  note/.style={align=center,inner sep=0pt}
]
\path[use as bounding box] (0,-.13) rectangle (3.30,1.20);

\draw[draw=rescuedark,fill=rescueorange!8,rounded corners=3pt,line width=.95pt]
  (.685,.345) rectangle (2.465,1.16);
\node[note,text=rescuedark,font=\fontsize{10.5}{12}\selectfont\sffamily\bfseries]
  at (1.575,1.005) {No preemption};

\node[stage,fill=black!3,minimum width=.55in] (wait) at (.315,.62)
  {Waiting};
\node[stage,fill=white,minimum width=.65in] (restore) at (1.095,.62)
  {Restore\\KV};
\node[stage,fill=rescueorange!20,minimum width=.78in] (decode) at (1.98,.62)
  {Fresh\\decoding};
\node[stage,fill=black!3,minimum width=.72in] (normal) at (2.915,.62)
  {Normal\\scheduling};
\draw[flow] (wait.east) -- (restore.west);
\draw[flow] (restore.east) -- (decode.west);
\draw[flow] (decode.east) -- (normal.west);

\draw[draw=rescueblue,line width=.8pt] (.685,.59) -- (.685,.265);
\fill[rescueblue] (.685,.62) circle[radius=1.3pt];
\node[note,text=rescueblue,text width=1.38in] at (.70,.075)
  {Select oldest stall\\Eligible after 5\,s};

\draw[draw=rescuedark,line width=.8pt] (decode.south) -- (1.98,.265);
\node[note,text=rescuedark,text width=1.15in] at (1.98,.075)
  {64 new tokens\\or finish};
\end{tikzpicture}
\endgroup
   \caption{Age-based protected recovery: readmit the oldest eligible stalled
  request and protect KV restoration and 64 fresh tokens, ending early on
  completion. Unfinished requests return to normal scheduling.}
  \Description{A horizontal sequence follows one request from waiting to KV
  restoration, fresh decoding and normal scheduling. The oldest stalled
  request becomes eligible after five seconds. An orange outline starts
  at selection, before restoration, and covers both restoration and fresh
  decoding. Preemption is disabled through 64 new tokens or completion.
  Completed requests exit; unfinished requests return to normal scheduling.}
  \label{fig:starvation-recovery}
\end{figure}

\name connects the return to the GPU with a turn of new output.
\emph{Age-based readmission} gives a stalled request priority to return.
\emph{Protection} then prevents preemption while it restores KV and
generates new tokens. The first addresses prolonged waiting; the second
lets the request use the state it has recovered.
Figure~\ref{fig:starvation-recovery} follows one request through this process.

\paragraph{1. Give the oldest stalled request a turn.}
The first preemption starts a stall timer. Reloading KV or recomputing
old context leaves it running; only a new output token clears it.
Resetting the timer at restoration would make a repeatedly interrupted
request appear newly active even if it had produced no new output.
Keeping the timer running lets the scheduler recognize that it is still
waiting to make progress.

When no request is protected, the scheduler selects the oldest stall that
has reached $T_{\mathrm{age}}=5$ seconds in the evaluated configuration.
Each replica's scheduler protects one request at a time. Another protected request
may still be finishing its turn, so five seconds is an eligibility
threshold, not a bound on waiting. Offloaded requests can also reload
normally before this threshold when GPU space and scheduling budget permit.

\paragraph{2. Make room and restore its state.}
Protection starts as soon as the request is selected, before restoration.
The scheduler reserves a running-request slot and, if necessary, preempts
other requests to free GPU blocks. For this step it uses physical-largest
victim selection, even when ordinary eviction uses SRJF. The immediate
goal is to obtain enough space for the selected request to return.

The request then reloads its saved CPU snapshot. If its KV was discarded,
it reuses any cached prefix and recomputes the missing state. Both paths
recover previous work. Protection remains active throughout this work so
that the request can proceed to decoding.

\paragraph{3. Protect a turn of new output.}
Once its state is ready, the request is scheduled first and cannot be
preempted until it generates $Q$ new tokens or finishes. We use $Q=64$
in the evaluated configuration. Reloaded KV and recomputed tokens do not
count toward this turn: they restore existing context. The first new
token clears the stall timer, but protection
continues until the request has generated all $Q$ new tokens since selection.

Other requests can decode alongside it when resources permit. If more
space is needed as its KV grows, the scheduler can preempt those other
requests. After the protected turn, the request keeps its KV and continues
under ordinary scheduling; it is not automatically evicted. A later
preemption starts a new stall timer.

The length of the protected turn controls how much new output a request
can produce between recovery cycles. A short turn can be followed quickly
by another eviction. A longer turn gives the request more progress for
the cost of restoring its state, but can make other requests wait longer.
Section~\ref{sec:eval-recovery} evaluates this tradeoff;
\suppref{app:recovery-implementation} gives additional scheduling details.

Recovery therefore needs both efficient state restoration and time to use
the restored state. The following sections explain how the CPU pool and
transfer engine support the restoration step.

\subsection{Contiguity-First CPU KV Pool}
\label{sec:cpu-kv-pool}

Recovery can reuse a saved snapshot only while the CPU pool has room to
retain it. The pool must therefore support efficient transfers without
wasting space that could hold another request's state.

GPU KV grows with each generated token, so fixed-size blocks let requests
acquire space as needed~\cite{kwon2023efficient}. An offloaded CPU snapshot
does not grow: decoding has stopped. \name can therefore allocate its
known size in contiguous CPU ranges, without reserving room for future
tokens. After restoration, growth resumes on the GPU; a later offload saves
a new snapshot at its current size.

\begin{figure}[t]
  \centering
  \includegraphics[width=\columnwidth]{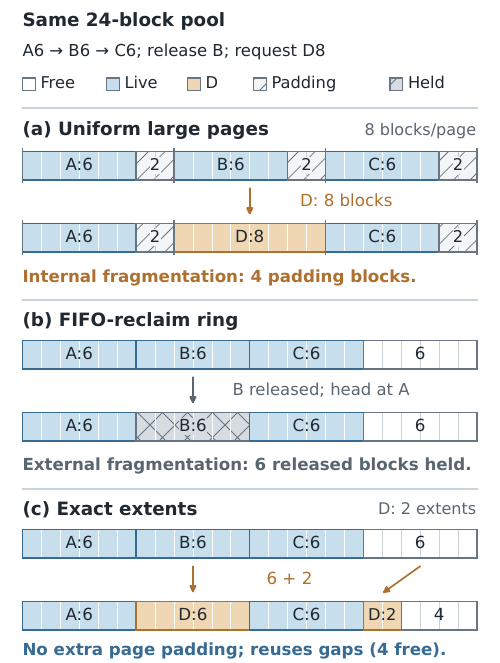}
  \caption{Allocation in the same 24-block pool: A, B and C each need
  six blocks; release B, then request eight blocks for D.
  (a)~Uniform eight-block chunks admit D but retain four padding blocks
  (internal fragmentation).
  (b)~The FIFO-reclaim ring holds B's released gap behind A
  (external fragmentation under this policy), so D waits.
  (c)~Exact extents admit D using $6+2$ blocks and leave four free.
  All sizes are illustrative KV-block counts.}
  \Description{Three vertically stacked panels show allocation before and after releasing snapshot B and requesting space for D. All use a 24-block pool and six-block snapshots A, B and C. Uniform eight-block chunks admit the eight-block D but reserve two unused blocks in each of A and C. A ring that reclaims in allocation order cannot reuse B while A remains live, leaving only six usable tail blocks, so D waits. Exact extents reuse six blocks from B and two from the tail, admitting D in two extents and leaving four blocks free.}
  \label{fig:exact-extent-allocator}
\end{figure}

Figure~\ref{fig:exact-extent-allocator} shows why a fixed snapshot size
does not by itself solve allocation. Uniform large pages leave unused
space inside partially filled chunks: \emph{internal fragmentation}.
A ring that reclaims space in allocation order cannot reuse a released
region while an older snapshot blocks reclamation. This creates
\emph{external fragmentation} under that FIFO policy. The ring is an
illustrative comparison, not an evaluated baseline or a claim about all
ring designs.

\paragraph{Exact-extent allocator.}
\name first tries to place the whole snapshot in one contiguous CPU
region, called an \emph{extent}. If no single region is large enough,
it combines a few free regions. This keeps a snapshot contiguous when
possible while still using gaps left by earlier requests.

The allocator searches for the smallest free extent that fits the
snapshot and returns the unused portion to the pool. This best-fit rule
preserves larger free regions for other snapshots. If none fits, it
takes space from the largest free extents, using at most eight per
snapshot in the optimized configuration. Bounding the number of extents
limits how many separate copies the transfer engine must issue.

Each allocation reserves the snapshot's current block count, including
its final partially filled KV block. It adds no padding to a fixed-size
request slot. If the pool cannot provide a complete allocation within
its capacity, extent and request-count limits, the request falls back
to recomputation.

\paragraph{Capacity and reuse.}
After a snapshot is reloaded, its CPU regions can hold other requests.
\name waits until the model step completes before releasing them, then
merges neighboring free regions. It does not move live snapshots to
compact the pool, so fragmentation can still prevent some allocations.

Each GPU worker has its own pool of pinned CPU memory. The configured
budget counts usable KV storage, which differs from the memory reserved
by the underlying allocator. In our implementation, an 800\,GiB usable
pool requires 1,024\,GiB of backing memory. This pool-level overhead is
separate from the per-snapshot padding that exact extents avoid.
\suppref{app:kv-implementation} describes the data layout and memory
accounting; Section~\ref{sec:eval-natural} reports measured host-memory use.

\subsection{Bulk KV Transfer}
\label{sec:async-transfer}

\begin{figure}[t]
  \centering
\begin{tikzpicture}[
  x=1in,y=1in,
  font=\fontsize{10.5}{12}\selectfont\sffamily,
  line width=0.6pt,
  box/.style={draw,minimum height=0.27in,inner sep=2pt,align=center},
  note/.style={inner sep=0pt,align=center},
  title/.style={note,anchor=west,font=\fontsize{10.5}{12}\selectfont\sffamily\bfseries},
  arrow/.style={->,>=stealth,line width=0.7pt},
  reuse/.style={arrow,dashed}
]
\path[use as bounding box] (0,-0.14) rectangle (3.25,4.37);

\node[title] at (0,4.22) {(a) HiCache: direct sparse reload};
\node[note] at (0.40,3.96) {GPU pages};
\node[note] at (2.85,3.96) {CPU pages};
\foreach \y/\id in {3.63/2,3.36/1,3.09/3} {
  \draw[fill=blue!9] (0.20,\y-0.10) rectangle +(0.40,0.20);
  \node[note] at (0.40,\y) {\id};
}
\foreach \y/\id in {3.63/1,3.36/2,3.09/3} {
  \draw[fill=orange!15] (2.65,\y-0.10) rectangle +(0.40,0.20);
  \node[note] at (2.85,\y) {\id};
}
\draw[arrow] (2.61,3.63) -- (0.64,3.36);
\draw[arrow] (2.61,3.36) -- (0.64,3.63);
\draw[arrow] (2.61,3.09) -- (0.64,3.09);
\node[note,fill=white,inner sep=3pt] at (1.625,3.40)
  {UVA copy kernel\\indexed loads/stores};

\node[title] at (0,2.77) {(b) \name: bulk copy, then scatter};
\foreach \x/\id in {0.02/1,0.19/{},0.36/2,0.53/{},0.70/3} {
  \draw[fill=blue!9] (\x,1.98) rectangle +(0.17,0.29);
  \node[note] at (\x+0.085,2.125) {\id};
}
\foreach \x/\id in {1.30/1,1.48/2,1.66/3} {
  \draw[fill=green!10] (\x,1.98) rectangle +(0.18,0.29);
  \node[note] at (\x+0.09,2.125) {\id};
}
\foreach \x/\id in {2.60/1,2.78/2,2.96/3} {
  \draw[fill=orange!15] (\x,1.98) rectangle +(0.18,0.29);
  \node[note] at (\x+0.09,2.125) {\id};
}
\draw[arrow] (1.25,2.125) -- (0.92,2.125);
\node[note] at (1.06,2.49) {Fused K/V\\scatter kernel};
\draw[arrow] (2.55,2.125) -- (1.89,2.125);
\node[note] at (2.24,2.49) {cudaMemcpyAsync\\(H2D DMA)};
\node[note] at (0.45,1.73) {GPU pages\\growing KV};
\node[note] at (1.57,1.73) {GPU staging\\contiguous};
\node[note] at (2.84,1.73) {CPU extent\\fixed snapshot};

\node[title] at (0,1.15) {(c) Overlap with two staging pairs};
\node[note,anchor=east] at (0.58,0.73) {DMA};
\node[note,anchor=east] at (0.58,0.29) {Scatter};
\draw[arrow,black!45] (0.66,0.73) -- (3.18,0.73);
\draw[arrow,black!45] (0.66,0.29) -- (3.18,0.29);
\node[box,fill=white,minimum width=0.54in] (ca) at (0.96,0.73) {copy A};
\node[box,fill=black!12,minimum width=0.54in] (cb) at (1.56,0.73) {copy B};
\node[box,fill=white,minimum width=0.54in] (ca2) at (2.16,0.73) {copy A};
\node[box,fill=white,minimum width=0.54in] (sa) at (1.56,0.29) {use A};
\node[box,fill=black!12,minimum width=0.54in] (sb) at (2.16,0.29) {use B};
\node[box,fill=white,minimum width=0.54in] (sa2) at (2.76,0.29) {use A};
\draw[arrow] (ca.south east) -- (sa.north west);
\draw[arrow] (cb.south east) -- (sb.north west);
\draw[arrow] (ca2.south east) -- (sa2.north west);
\draw[reuse] (sa.north east) -- (ca2.south west);
\node[note,anchor=south east] at (3.18,0.90) {time};
\end{tikzpicture}
   \caption{KV reload (one layer; K or V shown): (a)~HiCache's indexed
  unified virtual addressing (UVA) kernel; (b)~\name's bulk DMA followed
  by scatter; (c)~double buffering overlaps copying and scattering.
  Arrows denote readiness and safe reuse, not durations. HiCache's
  separate staged-DMA offload path is not shown.}
  \Description{Three panels compare transfer methods and explain overlap. HiCache reloads indexed CPU pages directly with a UVA copy kernel. \name reloads a contiguous CPU extent with cudaMemcpyAsync H2D DMA, followed by a fused K/V scatter kernel into paged GPU KV. Two staging pairs alternate so copying B can overlap scattering A. Solid dependency arrows wait for data readiness; a dashed arrow waits until scattering has finished before a buffer is reused.}
  \label{fig:offload-comparison}
\end{figure}
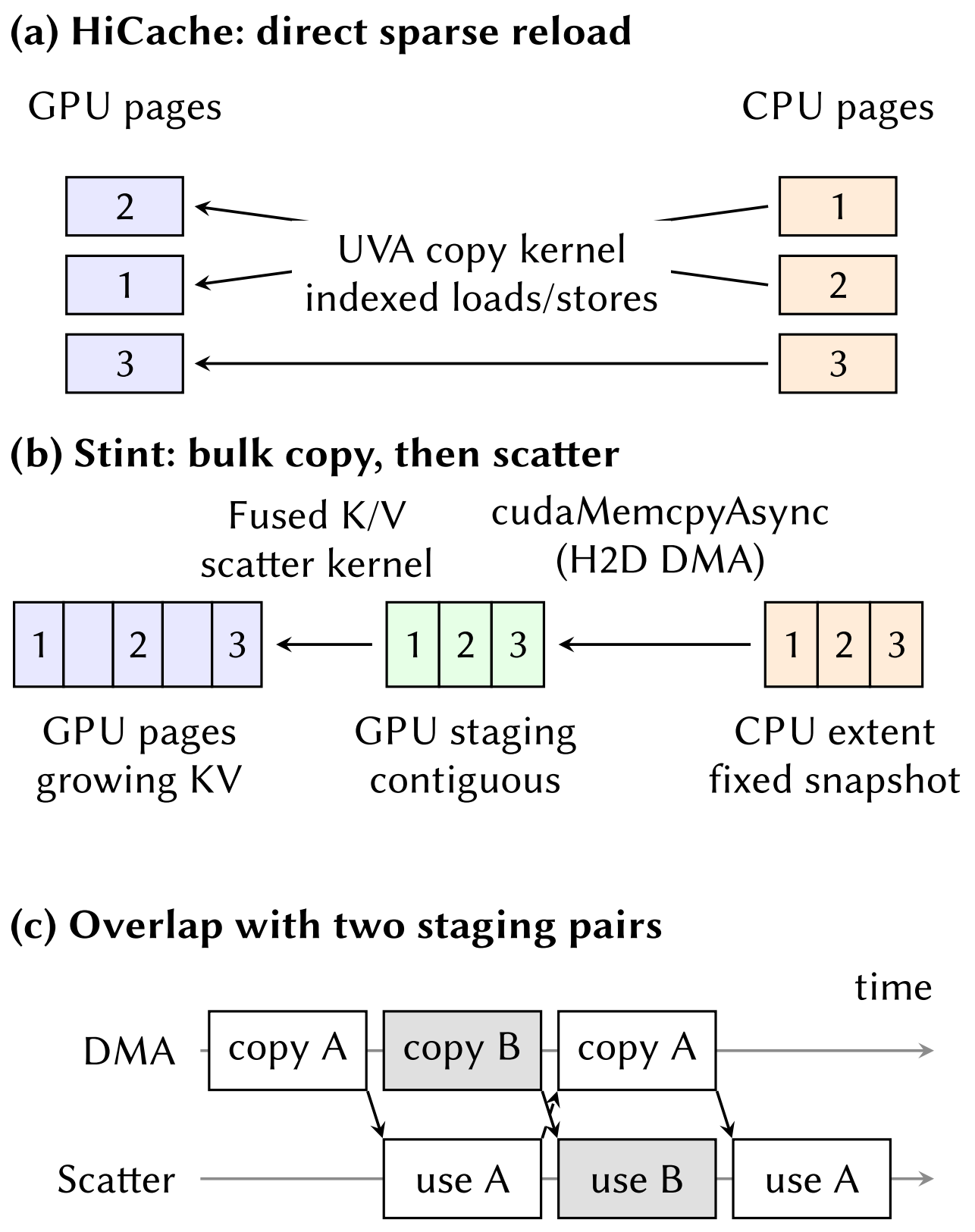

The CPU pool gives each snapshot a few contiguous regions, but its GPU
blocks remain scattered. Copying each block separately would turn a large
snapshot into many small transfers. \name instead gathers the GPU blocks
into a temporary GPU buffer and copies them to the CPU in bulk. Reload
reverses these steps: copy into the temporary buffer, then scatter the
data into the request's GPU blocks (Figure~\ref{fig:offload-comparison}).
These \emph{staging buffers} keep scattered accesses within GPU memory
while transfers over the CPU--GPU link use contiguous ranges.

\paragraph{Fused K/V gather/scatter kernels.}
Staging adds GPU work to gather and scatter the data. To reduce launch
overhead, \name handles K and V together in one kernel invocation.
Transfers proceed layer by layer, with the engine grouping CPU regions
into chunks that fit the staging buffers. One kernel gathers both K and V
into separate contiguous buffers; \texttt{cudaMemcpyAsync} then copies
each range to pinned CPU memory using direct memory access (DMA).
Reload uses the corresponding fused scatter kernel. K and V share the
kernel launch but retain separate DMA copies. Free gaps between extents
are never copied.

\paragraph{Double buffering.}
With one staging pair, gathering the next chunk must wait until the
current copy finishes. Reload has the same problem: the next copy must
wait until scattering releases the buffers. These operations serialize
even though copying and gather/scatter use different GPU resources.

\name alternates between two staging pairs, each containing a K buffer
and a V buffer. During offload, the GPU gathers the next chunk into B
while DMA copies A to the CPU. During reload, DMA fills B while the GPU
scatters A into the request's KV blocks. The two stages can therefore
work on different chunks at the same time.

\paragraph{Protect data until transfers finish.}
A staging buffer must not be overwritten while a copy or scatter still
reads it. Likewise, attention must wait until incoming KV is ready and
outgoing KV is safe to reuse. CUDA events enforce these dependencies.
Later layers' transfers can continue while the model proceeds, but the
dependencies can still delay attention. Double buffering overlaps copying
with gather/scatter; it does not hide every transfer behind model
computation. \suppref{app:kv-implementation} gives buffer sizes and
execution details.

\subsection{Prediction Errors and Limits}
\label{sec:misprediction}

Prediction errors change who keeps GPU memory. An overestimate can cause
a nearly finished request to be evicted; restoring its accumulated KV
may still be expensive. An underestimate can keep a long-running request
resident while other requests give up memory. Sparse refresh supplies
new information but may not correct either error in time.

Physical-largest fallback handles missing or invalid estimates, while
protected recovery gives a selected stalled request time to produce fresh
output. Valid but inaccurate estimates can still hurt victim selection.
We therefore evaluate the cost and effect of prediction separately from
the rest of the system, measuring throughput and latency alongside
prediction accuracy.
\section{Evaluation}
\label{sec:eval}

We evaluate whether \name makes preemption productive in long-output
serving. We ask four questions. Does \name complete long outputs faster,
with a shorter tail of slow requests (Section~\ref{sec:eval-e2e})? Does it
bound how long preempted requests stall (Section~\ref{sec:eval-pauses})?
Where do its benefits end, and what do they cost
(Section~\ref{sec:eval-natural})? Which components contribute
(Sections~\ref{sec:eval-allocator}--\ref{sec:eval-admission}, summarized in
Section~\ref{sec:eval-attribution})?

\subsection{Evaluation Setup}
\label{sec:eval-method}

\paragraph{Models and workloads.}
We use three long-output workloads. IMO-AnswerBench contains
olympiad-level problems with short verifiable
answers~\cite{luong2025imobench}; we replay all 399 valid questions from
its public release~\cite{superhumanRepository}. Its median output is
17,089 tokens, and 213 of 399 outputs exceed 16,384 tokens. AIME1000
replays 1,000 requests drawn from 355 distinct AIME tasks, and LCB880
replays 880 LiveCodeBench programming tasks~\cite{jain2024livecodebench}.
The main experiments serve BF16 Qwen3-32B on eight H200 GPUs as four
two-GPU replicas (DP4/TP2). Finite Poisson arrivals enter at 4 requests/s
across the deployment, with round-robin routing.

\paragraph{Fixed-output replay.}
Each system receives identical prompts and the output length that the same
model produced in a natural-completion run with greedy decoding. Requests
use temperature 0 and ignore EOS until that length, so every system performs
the same output work; answer quality is not measured. Greedy trajectories
still diverge as batch composition changes: in matched runs, only 4 of 1,197
IMO-AnswerBench responses have identical text under recompute and \name.
Length predictions therefore see hidden states from a trajectory that
resembles, but need not match, the one that defined the request's length.
We test natural completion separately on 175 LiveCodeBench v6 tasks at
1 request/s on one TP2 replica. Further studies cover Qwen/arXiv
summarization, Qwen/BurstGPT, Llama-3.1-8B-Instruct at DP8/TP1, and
Cachewise structural replay; settings are in \suppref{app:methods}.

\paragraph{Predictor training.}
Weights remain fixed during serving. The IMO-AnswerBench predictor is
trained on 100 randomly selected questions, and all 399 enter the serving
comparison. Restricting the analysis to the 299 questions outside the
training set gives the same direction: relative to recompute, p99 TPOT
falls by 53.56\% and p99 completion time by 4.16\%
(\suppref{app:superhuman-exposure}). AIME1000 and LCB880 also draw on
development tasks used for predictor validation or training.

\paragraph{Systems.}
Recompute uses \name's maintained vLLM 0.9.2 build with default FCFS
scheduling, no predictor and no CPU KV pool, so it differs from \name only
in the mechanisms this paper adds. \name uses SRJF victims, K64 inference,
exact-extent storage, fused K/V transfers, double buffering and age-5/Q64
recovery. We compare them on all three main workloads.
IMO-AnswerBench also compares LMCache~\cite{cheng2025lmcache} with upstream
vLLM, and SGLang with HiCache. Both can restore a prefix and rebuild missing
state; \name preserves the complete decode snapshot. The HiCache comparison
also includes differences between the SGLang and vLLM engines. LMCache's
runs include the resumed-request metadata repair described in
\suppref{app:baseline-corrections}; \suppref{app:methods} gives baseline settings.

\paragraph{Resources.}
Recompute and \name have equal GPU KV capacity. LMCache and HiCache
retain their native capacities, so the four-system results compare complete
stacks. CPU KV payload budgets are similar, but \name requires more pinned
backing (Section~\ref{sec:eval-natural}).

\paragraph{Metrics and repetitions.}
Throughput counts output tokens from first scheduled arrival through
final completion, including drain. TTFT starts at HTTP dispatch; TPOT is
each request's average output-token interval after its first token. ITL
pools client gaps between token-bearing chunks across all requests. Each
request's longest gap exposes interruptions hidden by its average.
Completion time covers the whole request; arrival-to-first-output also
includes admission waiting. Each main comparison uses three matched runs;
percentage changes average paired ratios. Figures show every run, with the
arithmetic mean and sample SD. Component studies state their own
repetition counts.

\subsection{Long-Output Serving Performance}
\label{sec:eval-e2e}

\begin{figure*}[t]
  \centering
  \includegraphics[width=\linewidth]{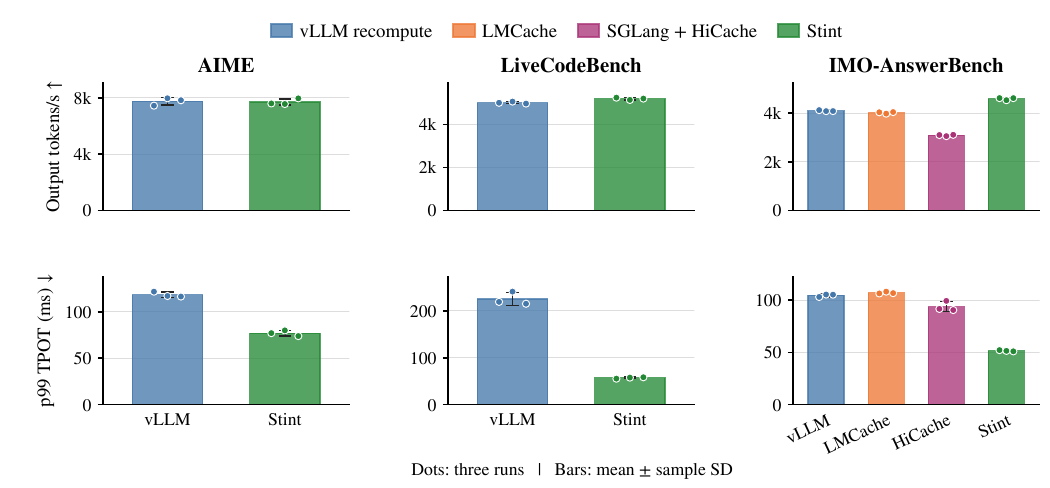}
  \caption{Long-output fixed replays on Qwen3-32B, H200 DP4/TP2, at
  4 requests/s. Higher throughput and lower p99 TPOT are better.
  Dots are three runs; bars and whiskers show mean and sample SD.
  AIME and LiveCodeBench compare recompute with \name; IMO-AnswerBench adds
  LMCache and HiCache. All systems complete the same prescribed output per workload.}
  \Description{Six linear bar panels show throughput and p99 time per output token for AIME, LiveCodeBench and IMO-AnswerBench. \name has lower p99 TPOT in all three workloads. Throughput is close to recompute on AIME, modestly higher on LiveCodeBench, and higher than all three baselines on IMO-AnswerBench. Every run is plotted.}
  \label{fig:qwen3-e2e}
\end{figure*}

We first measure how quickly each system completes a fixed amount of
output work and how slowly its most affected requests generate tokens.
Figure~\ref{fig:qwen3-e2e} shows throughput and p99 TPOT on all three
workloads; Table~\ref{tab:superhuman-metrics} reports the remaining
IMO-AnswerBench metrics, including those that \name makes worse.

\begin{table}[t]
\centering
\caption{IMO-AnswerBench serving metrics, mean of three matched runs
(Qwen3-32B, H200 DP4/TP2, 4 requests/s). Lower is better except throughput.
HiCache does not report preemption events. Host memory is the sampled
node/cgroup peak, not process RSS.}
\label{tab:superhuman-metrics}
\setlength{\tabcolsep}{2.5pt}
\begin{tabular}{@{}lrrrr@{}}
\toprule
Metric & Recomp. & LMCache & HiCache & \name \\
\midrule
Throughput (tok/s) & 4,100 & 4,023 & 3,092 & \textbf{4,592} \\
Median TPOT (ms) & \textbf{38.2} & 39.5 & 49.7 & 40.9 \\
p99 TPOT (ms) & 104.5 & 107.1 & 93.8 & \textbf{51.5} \\
Pooled p99 ITL (ms) & \textbf{47.9} & 111.7 & 57.4 & 78.0 \\
Mean TTFT (ms) & \textbf{49.4} & 74.7 & 147.4 & 56.7 \\
p99 TTFT (ms) & \textbf{67.9} & 535.5 & 567.9 & 246.7 \\
p99 completion (s) & 1,588 & 1,631 & 2,101 & \textbf{1,522} \\
Preemptions per run & 510 & 509 & -- & 1,376 \\
Host memory (GiB) & \textbf{328} & 1,131 & 1,134 & 1,357 \\
\bottomrule
\end{tabular}
\end{table}

\paragraph{Throughput and tail TPOT.}
On IMO-AnswerBench, \name raises throughput by 11.99\%, 14.15\% and
48.53\% relative to recompute, LMCache and HiCache, and lowers p99 TPOT by
50.69\%, 51.86\% and 44.96\%. Both gains hold in every matched run against
every baseline. On AIME and LCB, p99 TPOT falls by 34.99\% and 74.55\%
relative to recompute, while throughput changes by $-0.45\%$ and
$+3.64\%$; AIME's difference changes sign across runs. Tail TPOT improves
on all three workloads, while throughput benefits most from IMO-AnswerBench's
longer outputs. These finite replays measure a fixed amount of work,
not sustainable arrival capacity.

\paragraph{What gets worse.}
\name preempts 2.7 times as often as recompute, but retains KV in every
preemption: it records zero recomputation events. These more frequent
interruptions cost typical latency. Median TPOT rises from 38.2 to 40.9\,ms,
pooled p99 ITL rises from 47.9 to 78.0\,ms, and mean TTFT rises by 14.81\%,
with run-level p99 TTFT of 181--281\,ms versus 65--70\,ms
(Table~\ref{tab:superhuman-metrics}). Section~\ref{sec:eval-pauses} shows
how these frequent pauses differ from recompute's rare ones.

\begin{figure*}[t]
  \centering
  \includegraphics[width=\linewidth]{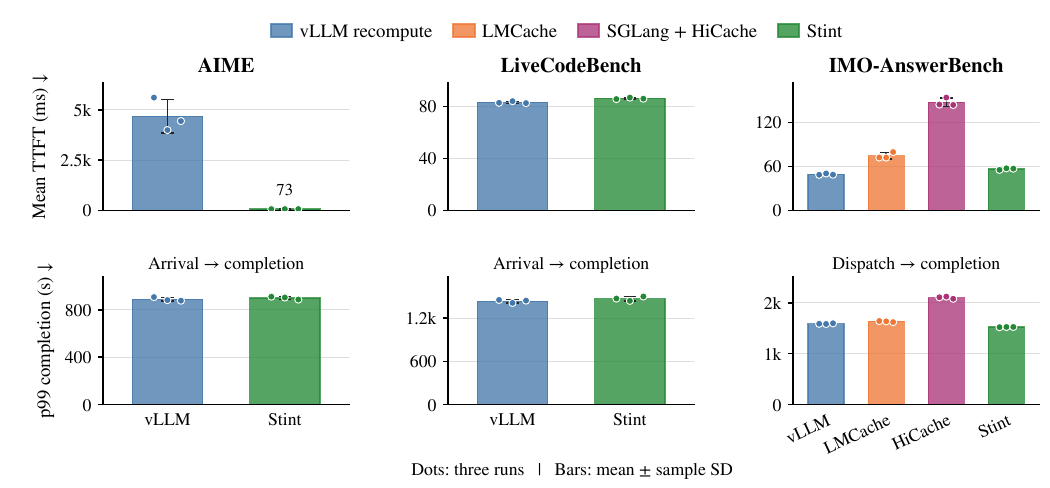}
  \caption{Latency tradeoffs in the same runs as Figure~\ref{fig:qwen3-e2e};
  lower is better. TTFT starts at HTTP dispatch. Completion time starts
  at scheduled arrival for AIME and LiveCodeBench, and at HTTP dispatch for
  IMO-AnswerBench; comparisons are within each workload. Dots show three
  runs, with mean and sample SD.}
  \Description{Six bar panels show mean TTFT and p99 completion time for AIME, LiveCodeBench and IMO-AnswerBench. \name's TPOT gains do not imply that every request starts or finishes sooner. Completion tails worsen on AIME and LiveCodeBench; IMO-AnswerBench completion tails improve while TTFT is worse than recompute.}
  \label{fig:latency-tradeoffs}
\end{figure*}

\paragraph{Starting and finishing a request.}
A request can wait long to begin, or finish later even if its average token
spacing improves (Figure~\ref{fig:latency-tradeoffs}). \name lowers
IMO-AnswerBench's p99 completion time by 4.16\%. The gain is concentrated
in the slowest requests. Pairing each request with its recompute run,
74\% of the 1,197 requests finish later under \name, with a median
slowdown of 5\%; p90 completion time rises by 4\%, while the maximum
falls by 8\%. \name trades a small delay for most requests against the
multi-minute delays of the starved ones. On AIME, mean TTFT falls
from 4.69\,s to 73\,ms; on LCB it stays near 85\,ms. AIME and LCB completion-time
p99 rises by 1.30\% and 2.43\%: lower TPOT does not ensure an earlier finish.

\subsection{Are Stalls Bounded?}
\label{sec:eval-pauses}

\begin{figure*}[t]
  \centering
  \includegraphics[width=\linewidth]{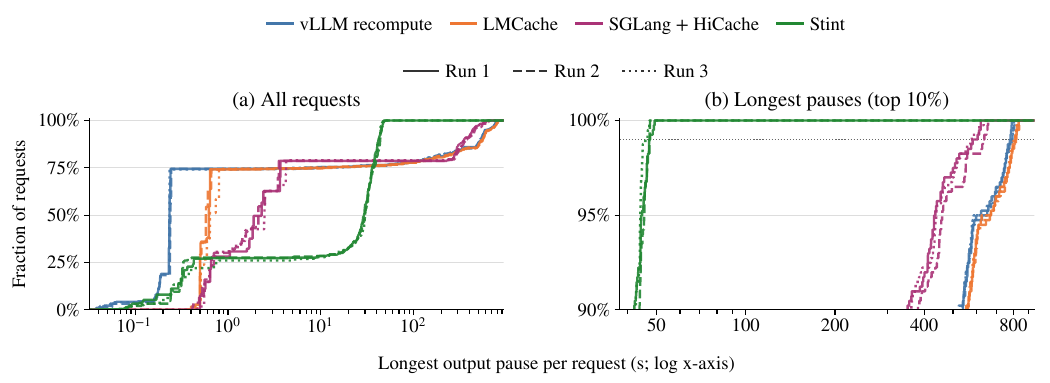}
  \caption{Longest output pause per request in the IMO-AnswerBench replays.
  Each request contributes one maximum inter-chunk gap, excluding time
  before its first output. Curves retain the three runs separately.
  \name removes the multi-minute stalls while making the typical request's
  longest pause larger.}
  \Description{Empirical distributions and a tail view show each request's maximum output gap. \name removes the very large pauses seen in the baselines, but its median request maximum is larger. The metric includes scheduling and transport effects and is not request completion time.}
  \label{fig:request-pauses}
\end{figure*}

KV thrashing starves requests: a request preempted under memory pressure
can wait while others keep growing their caches. A low average time per
token can hide such a stall. We therefore examine each request's longest
output pause, the largest gap between token-bearing chunks after its first
output (Figure~\ref{fig:request-pauses} and Table~\ref{tab:pause-bounds}).

\begin{table}[t]
\centering
\caption{Longest output pause per IMO-AnswerBench request, pooled over three
runs (1,197 requests per system). Share of lifetime divides the pause by the
request's dispatch-to-completion time.}
\label{tab:pause-bounds}
\setlength{\tabcolsep}{2.5pt}
\begin{tabular}{@{}lrrrr@{}}
\toprule
 & Recomp. & LMCache & HiCache & \name \\
\midrule
Median (s) & \textbf{0.24} & 0.62 & 2.31 & 29.9 \\
p99 (s) & 788 & 813 & 613 & \textbf{47.3} \\
Maximum (s) & 808 & 840 & 658 & \textbf{49.3} \\
Requests $>60$\,s & 23.2\% & 23.3\% & 21.2\% & \textbf{0\%} \\
Requests $>300$\,s & 15.9\% & 16.2\% & 15.5\% & \textbf{0\%} \\
Max.\ share of lifetime & 72.4\% & -- & -- & \textbf{8.4\%} \\
\bottomrule
\end{tabular}
\end{table}

\paragraph{Baselines leave requests stalled for minutes.}
Every baseline leaves 21--23\% of requests with a pause longer than one
minute and about 16\% with one longer than five minutes; the longest
single pause reaches 11--14 minutes. Under recompute, the 90th-percentile
request spends 45\% of its lifetime in one pause, and the worst spends 72\%.
These requests have no path back to the GPU while other requests hold the
memory they need.

\paragraph{\name bounds every stall.}
No \name request pauses for more than 49.3\,s, and no pause takes more than
8.4\% of the request's lifetime (median 3.1\%). Age-based readmission
provides this bound. The price is that \name pauses more requests: 72\% of
requests have a pause above 10\,s, and the median longest pause is about
30\,s. These pauses fall on long requests, as prediction-guided selection
intends. Dividing requests into output-length terciles, a pause above 10\,s
affects 30\% of the shortest third, 87\% of the middle third and 99\% of the
longest third under \name, versus 21--31\% of each third under recompute.

\name thus replaces rare, unbounded stalls with frequent, bounded ones.
Which is preferable depends on the client. Batch, agentic and evaluation
workloads that consume complete answers benefit; an interactive client
that streams tokens to a reader sees a pause of about 30 seconds in most
long responses. The readmission threshold and protected-turn length tune
this tradeoff (Section~\ref{sec:eval-recovery}). Pooled ITL tells the same
story from a different angle: it counts every gap, so \name's more frequent
pauses raise its p99 even though its worst pauses are far shorter.

\paragraph{Why can throughput improve?}
Superficially, more preemptions should cost throughput. \name records
1,344--1,392 swaps and zero recomputations per run, versus 506--514
recomputation preemptions with recompute, yet completes the same output
work 12\% faster. Saving KV preserves work that would otherwise need
rebuilding, while readmission brings stalled requests back into decode
batches instead of leaving them idle. Section~\ref{sec:eval-attribution}
summarizes the component evidence for each mechanism.

\subsection{Natural Completion and Workload Limits}
\label{sec:eval-natural}

\paragraph{Natural completion.}
The tail benefit also appears when requests stop naturally. On 175
LiveCodeBench v6 tasks with a 32,768-token output ceiling, \name with SRJF
delivers 3.46\% higher throughput and 68.07\% lower p99 TPOT than recompute
across three matched runs on one H200 TP2 replica at 1 request/s. Mean
arrival-to-completion time falls by 4.16\%; median TPOT and p99 TTFT rise
by 13.99\% and 17.90\%. Recompute's generated programs pass 304 of 525
evaluations, \name's physical-largest policy passes 308, and SRJF passes
311; outcomes change in both directions across runs, so this shows
neither a quality gain nor a quality loss. Within \name, SRJF lowers mean
completion time by 6.78\% relative to physical-largest while reducing
throughput by 7.37\%. \suppref{app:natural-quality} reports per-run results.

\begin{figure*}[t]
  \centering
  \includegraphics[width=\linewidth]{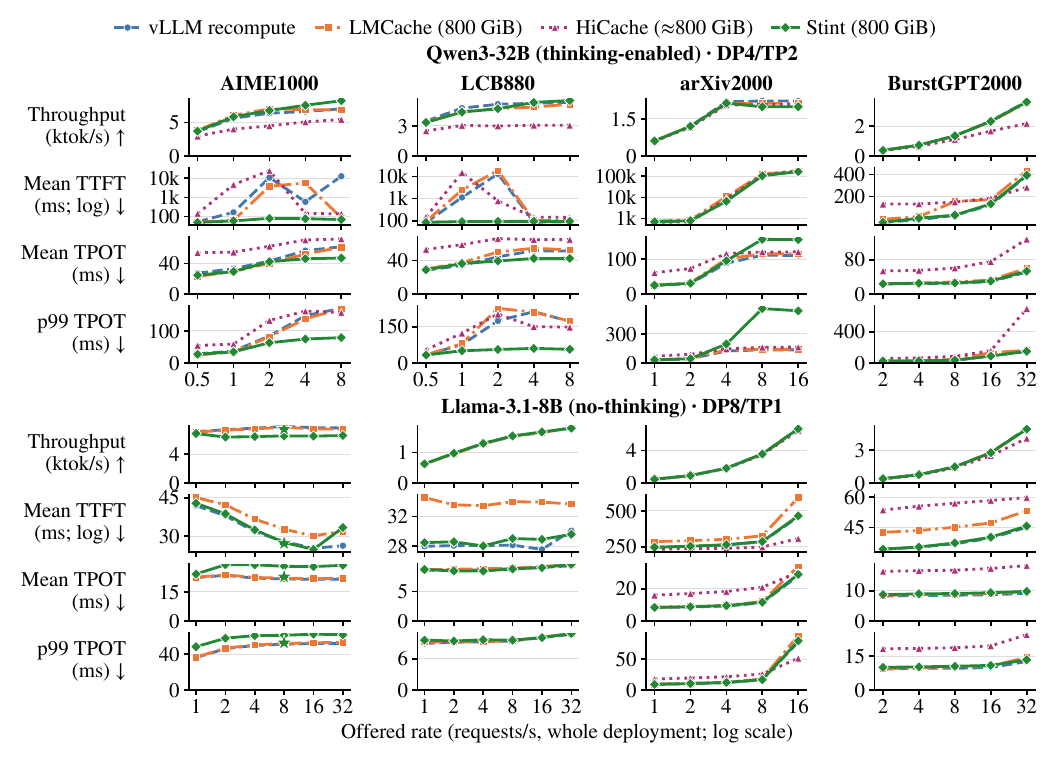}
  \caption{Earlier end-to-end fixed-output load sweeps, one run per curve
  point. \name curves use Q64; the Llama/AIME star at 8 requests/s uses
  Q1024 (mean $\pm$ population SD, two runs). arXiv and BurstGPT use
  2,000-request sweeps. TTFT uses a logarithmic scale. Llama/HiCache on
  AIME and LCB is unmeasured. LMCache curves predate the resume-path repair.}
  \Description{Thirty-two load-curve panels compare the system with maintained-build vLLM recompute, LMCache with an 800-GiB host pool, and SGLang HiCache with a 797-GiB host pool. Columns are AIME1000, LCB880, arXiv2000, and BurstGPT2000. Each model has rows for throughput, mean TTFT on a logarithmic scale, mean TPOT and p99 TPOT. The continuous Llama AIME curve includes the original Q64 result at 8 requests/s; a separate star shows the Q1024 result. Missing Llama HiCache AIME and LCB curves are unmeasured. Gains and regressions are retained.}
  \label{fig:load-sweep}
\end{figure*}

\paragraph{Load sweeps.}
Figure~\ref{fig:load-sweep} sweeps the offered rate across four workloads
and two models, one run per point. These sweeps predate the repeated
comparisons above, and their LMCache curves predate the resume-path repair
in \suppref{app:baseline-corrections}; one repaired rerun at 8 requests/s
raised LMCache throughput by 7.80\% on AIME1000 and 1.71\% on LCB880.
We therefore use them to locate where \name helps, not to quantify its
margin over LMCache.

On Qwen reasoning and coding, \name's p99 TPOT advantage appears as soon as
preemption begins, at 1 request/s on LCB880 and 2 requests/s on AIME1000,
and grows with load: at 4 requests/s its
p99 TPOT is 74.81\,ms on AIME1000 and 60.31\,ms on LCB880, versus 138.09
and 210.10\,ms for LMCache. The gains depend on workload. \name offers no
consistent gain on arXiv summarization, whose outputs were capped at 1,024
tokens during collection and leave less decode work over which scheduling
can offset its cost. In larger high-rate Qwen/arXiv cohorts, \name has
19--21\% lower throughput than recompute. High-rate BurstGPT has no
preemptions in either system and 2.5--3.5\% lower throughput with \name:
its added work has no recovery cost to offset.

On Llama/AIME, \name has lower throughput and higher TPOT than LMCache.
Llama has no explicit thinking mode, and many saved AIME outputs repeat
text until the generation limit, which may weaken the hidden-state cues for
predicting completion; this comparison does not isolate that effect.
Long requests also pass repeatedly through short protected turns.
Increasing the protection quantum to 1,024 tokens improves throughput with
the same predictor (Section~\ref{sec:eval-recovery}).

\paragraph{Other workloads.}
Repeated confirmations agree with the sweeps. The Qwen/arXiv, Qwen/BurstGPT
and Llama/AIME confirmations do not improve throughput over recompute. On
Cachewise's shared-prefix structural replay, \name outperforms the measured
LMCache and HiCache configurations in throughput and p99 TPOT, but HiCache
has lower TTFT. \suppref{app:tuning} and \suppref{app:cachewise} give
details.

\paragraph{Host-memory cost.}
Preserving snapshots requires substantial host memory. On IMO-AnswerBench,
\name's sampled host-memory peak is 1,357\,GiB on average, versus 328\,GiB
for recompute and 1,131--1,134\,GiB for LMCache and HiCache. Most of the
difference from the other caches comes from the pinned allocator rounding
each 1.5625\,GiB per-layer buffer to 2\,GiB, which turns an 800\,GiB payload
into 1,024\,GiB of backing (\suppref{app:methods}). This overhead belongs to
the current implementation rather than to exact extents, and allocating
each buffer at its exact size would remove it.

\subsection{Contiguity-First CPU KV Pool}
\label{sec:eval-allocator}

We test whether reducing wasted CPU space lets the system retain more
snapshots and avoid recomputation. We compare exact extents with uniform
8K-token chunks and a mix of 8K-, 16K- and 32K-token chunks. All three
designs store snapshots in contiguous regions, so this comparison tests
how well they use the available capacity.

This study uses AIME1000 and the first
800 tasks of the LiveCodeBench replay LCB880, denoted LCB800.
Each design serves Qwen3-32B on H200 TP4 with an 800\,GiB pool at
4 requests/s, with three runs per configuration
(Tables~\ref{tab:allocator-serving} and~\ref{tab:allocator-lcb}).

Fallbacks count preemptions that use recomputation. Padding divides the
unused blocks within snapshot allocations by all their reserved blocks,
accumulating both counts across allocations. It excludes the 16-token
block rounding shared by all designs.

\begin{table}[t]
\centering
\caption{AIME1000 allocator comparison at 800\,GiB: mean $\pm$ population SD across three runs.}
\label{tab:allocator-serving}
\setlength{\tabcolsep}{3pt}
\begin{tabular}{@{}lrrr@{}}
\toprule
Allocator & Tok/s & Fallbacks & Padding (\%) \\
\midrule
Uniform 8K & $4603\pm123$ & $521\pm12$ & $68.3\pm1.6$ \\
Mixed sizes & $4615\pm35$ & $1093\pm82$ & $77.8\pm0.2$ \\
Exact extents & \textbf{4792\,$\pm$\,61} & \textbf{0\,$\pm$\,0} & \textbf{0\,$\pm$\,0} \\
\bottomrule
\end{tabular}
\end{table}

\begin{table}[t]
\centering
\caption{LCB800 allocator comparison at 800\,GiB: mean $\pm$ population SD across three runs.}
\label{tab:allocator-lcb}
\setlength{\tabcolsep}{3pt}
\begin{tabular}{@{}lrrr@{}}
\toprule
Allocator & Tok/s & Fallbacks & Padding (\%) \\
\midrule
Uniform 8K & $3301\pm43$ & $0\pm0$ & $48.0\pm1.4$ \\
Mixed sizes & $3188\pm61$ & $433\pm13$ & $64.9\pm1.2$ \\
Exact extents & \textbf{3293\,$\pm$\,29} & \textbf{0\,$\pm$\,0} & \textbf{0\,$\pm$\,0} \\
\bottomrule
\end{tabular}
\end{table}

Exact extents eliminate recorded recomputation fallbacks on AIME1000 and
raise throughput by 4.09\% over uniform chunks, with one CPU region per
offload. On LCB800, exact and uniform designs avoid fallbacks and throughput
is nearly unchanged ($-0.25\%$). Reclaiming padding helps when it retains
additional snapshots; once both pools hold them, compact storage adds little
throughput.
\suppref{app:storage-transfer} reports smaller-pool placement limits.

\subsection{Bulk KV Transfer}
\label{sec:eval-dense-kv-transfer}

Retaining a snapshot is useful only if the system can move it efficiently.
We compare the time to transfer identical BF16 payloads on H200, including
gathering, copying, scattering and completion. Each method has three
fresh-process repetitions, with no model computation during the transfer.

Moving the bytes is only part of this work: the system must also prepare
the mappings between source and destination blocks.
Table~\ref{tab:dense-kv-transfer} therefore reports two timings.
``MemCopy'' starts after the adapter's mappings and indices are prepared;
``Prepare+MemCopy'' includes that preparation. Both include HiCache's
internal descriptor construction.

\begin{table}[t]
  \centering
  \caption{Local KV transfer latency (ms), averaged over three repetitions.
  Each operation moves 160\,MiB of K/V (one layer, 2560 scattered GPU
  pages). The two cohorts run on different hosts, each with its own
  \name control; compare within a cohort. Mooncake adapters are local;
  SparDA uses 16 CTAs. Standard deviations are in
  \suppref{app:storage-transfer}.}
  \label{tab:dense-kv-transfer}
  \setlength{\tabcolsep}{2.5pt}
  \begin{tabular}{@{}lrrrr@{}}
    \toprule
    & \multicolumn{2}{c}{MemCopy} & \multicolumn{2}{c}{Prepare+MemCopy} \\
    \cmidrule(lr){2-3}\cmidrule(lr){4-5}
    Method & D2H & H2D & D2H & H2D \\
    \midrule
    \multicolumn{5}{@{}l}{\textit{HiCache cohort (host A)}} \\
    \name & 3.680 & \textbf{3.231} & \textbf{4.243} & \textbf{3.961} \\
    HiCache layer-first & \textbf{3.274} & 3.361 & 8.929 & 9.075 \\
    HiCache page-first & 15.793 & 3.360 & 21.948 & 9.086 \\
    \midrule
    \multicolumn{5}{@{}l}{\textit{Local-adapter cohort (host B)}} \\
    \name & \textbf{3.642} & \textbf{3.231} & \textbf{4.210} & \textbf{3.957} \\
    Mooncake batch & 4.753 & 4.919 & 27.064 & 27.068 \\
    Mooncake fallback & 16.358 & 17.308 & 38.177 & 39.164 \\
    SparDA-derived & 16.344 & 4.546 & 52.268 & 6.322 \\
    \bottomrule
  \end{tabular}
\end{table}

Once mappings are prepared, \name and HiCache layer-first are close and
neither wins both directions: HiCache offloads 11\% faster (3.274 versus
3.680\,ms), while \name reloads 4\% faster. Both move 46--52\,GB/s, so
either path nearly saturates the host link. Contiguous
extents help mainly by simplifying preparation, since a snapshot needs a
few CPU ranges per layer rather than one index per page; with preparation
included, \name takes less than half of HiCache's time in both directions.
\name is faster than the Mooncake and SparDA-derived adapters in every
column of their cohort. Across the full HiCache sweep, \name wins
72 of 96 large fragmented configurations and 126 of 324 overall, while
HiCache often wins at smaller sizes. The Mooncake/SparDA cohort also depends
on size and tuning; \suppref{app:methods} and \suppref{app:storage-transfer}
give the full comparisons.

\paragraph{Fused K/V kernels and double buffering.}
The transfer design reduces two sources of overhead: separate kernel
launches for gathering or scattering K and V, and waiting to reuse staging
buffers. To examine both, we compare separate and fused kernels with one
or two staging pairs, each holding K and V buffers. All four paths use
bulk copying. Fusion also batches CPU regions, while double buffering
overlaps gather/scatter with DMA.
The diagnostic moves 2\,GiB across 64 layers
between 1,024 scattered GPU blocks and eight CPU regions, gathering on
offload and scattering on reload.

\begin{table}[t]
\centering
\caption{K/V kernel and buffering ablation: transfer latency (ms),
mean $\pm$ population SD of ten operations, not serving runs.
D2H offloads KV; H2D reloads it.}
\label{tab:transfer-ablation}
\setlength{\tabcolsep}{3pt}
\begin{tabular}{@{}lcrr@{}}
\toprule
K/V kernels & Staging pairs & D2H & H2D \\
\midrule
Separate & 1 & $60.438\pm0.121$ & $53.065\pm0.050$ \\
Fused & 1 & $48.434\pm0.136$ & $44.170\pm0.128$ \\
Separate & 2 & $47.792\pm0.079$ & $43.140\pm1.091$ \\
Fused & 2 & $46.425\pm0.022$ & $42.503\pm0.122$ \\
\bottomrule
\end{tabular}
\end{table}

Using fusion and two staging pairs gives the lowest offload and reload
times, about 23.2\% and 19.9\% below separate K/V kernels with one pair
(Table~\ref{tab:transfer-ablation}). Earlier profiling records 64
gather/scatter launches with fusion versus 1,024 without it, while payload
DMA calls remain at 1,024. The gains from fusion and double buffering
are not additive.

Faster local transfers need not translate into higher serving throughput:
requests also wait for GPU space and compete for compute. We therefore
test the transfer changes during serving as well. The ablation uses
300-request AIME and 200-request LiveCodeBench
replays, denoted AIME300 and LCB200. On Qwen3-32B, p99 ITL falls by
4.02\% and 8.14\%, respectively, in two paired runs per arm
(H200 DP2/TP1, 4 requests/s). Throughput changes are small ($-0.96\%$
and $+0.33\%$) and vary in sign across seeds; maximum pauses increase
in three of four pairs. Faster transfers help recovery, while scheduling
determines when requests can use the restored state.

\subsection{Remaining-Length Prediction}
\label{sec:eval-prediction}

Victim selection needs an ordering by remaining work. We compare length
predictions from \name with DSMN
(single-layer hidden states), LTR~\cite{fu_efficient_2024} (OPT-350M),
S3 (DistilBERT length buckets) and PiA (target-model LoRA); the latter
three use prompts. All use Qwen3-32B generations and validation-selected
checkpoints, with no online training. The plotted AIME300/LCB200
populations combine validation and test requests used during development.

\begin{figure*}[t]
  \centering
  \includegraphics[width=\linewidth]{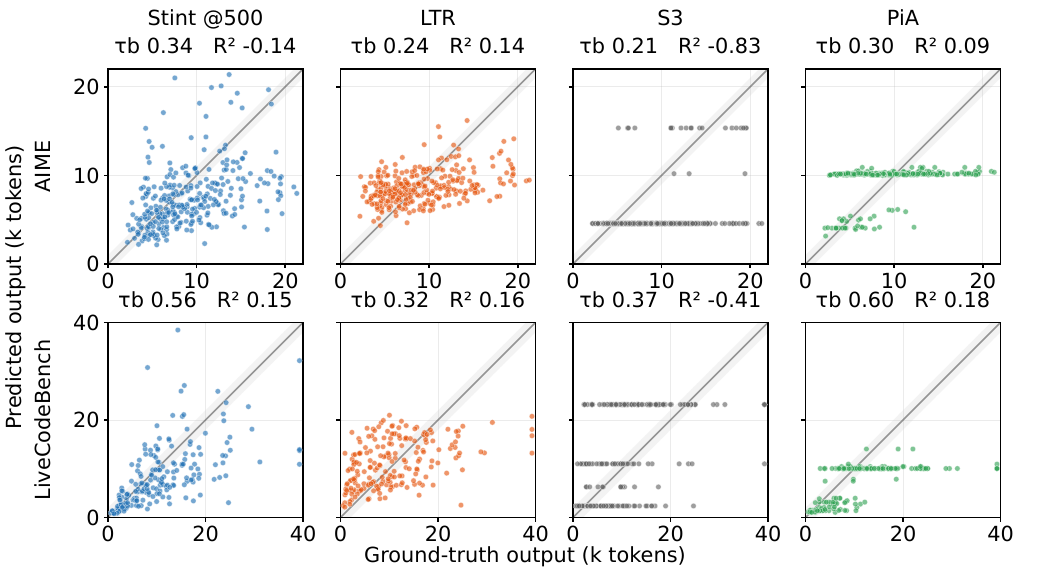}
  \caption{Output-length predictions on common step-500 survivors:
  300 AIME requests (top), and 199 of 200 LiveCodeBench
  requests (bottom).
  \name predicts at step 500; LTR, S3 and PiA use prompts.
  Kendall's $\tau_b$ measures ordering; $R^2$ measures numerical fit.
  Higher is better for both; 1 is perfect. DSMN results are in
  \suppref{app:predictor-details}.}
  \Description{Eight square scatter plots compare predicted and actual output lengths for AIME and LiveCodeBench across \name, LTR, S3, and PiA. Each panel reports Kendall tau-b for ranking quality and R-squared for numerical fit. Higher is better for both.}
  \label{fig:pred-scatter}
\end{figure*}

\paragraph{Accuracy.}
\name ranks AIME lengths best ($\tau_b=0.34$), although LTR has lower
absolute error. On LCB, \name ($\tau_b=0.56$, MAE 4,744 tokens)
is close to the leading PiA predictor ($\tau_b=0.60$, MAE 4,660).
Hidden states provide useful ordering information, with prompt-time
prediction remaining competitive on code.
\suppref{app:predictor-details} gives the complete regression metrics.

\paragraph{Serving cost.}
\label{sec:eval-predictor-cost}
Sparse prediction has a small aggregate cost: enabling K64 reduces
throughput by 0.58\% on AIME1000 and 0.26\% on LCB880 across three
Qwen runs at DP4/TP2 and 4 requests/s. We keep physical-largest victims,
storage, transfers and
recovery fixed, computing predictions without using them. This includes
scheduling effects as well as predictor execution.

A separate three-run K1/K64/K128 study on Qwen/LCB880 shows that K64
makes 97.13\% fewer batched prediction calls than K1. Serving changes are
small and mixed, and call counts do not measure time saved. The interval trades inference
cost against decision freshness;
\suppref{app:prediction-interval} reports the complete sensitivity results.

\subsection{Prediction-Guided Victim Selection}
\label{sec:eval-victim}

Producing length estimates has a cost; using them changes which requests
keep GPU memory. We separate these effects by enabling prediction in both
configurations and changing only victim selection: SRJF uses predicted
remaining length, while physical-largest uses the memory an eviction
would release.

In the same three-run controls, SRJF lowers p99 ITL by 13.86\% on AIME
and 14.57\% on LiveCodeBench, but reduces throughput and increases p99
TPOT. Here, the tail of individual output gaps improves at the cost of
overall output rate and the tail of per-request average token spacing.
\suppref{app:victim-comparison} gives the full results.

Prediction-guided victim selection therefore offers a scheduling
tradeoff within \name; these controls do not explain the full system's
throughput gains by prediction alone. Both policies retain the same CPU
storage, transfer and recovery mechanisms. We examine recovery next.

\subsection{Age-Based Protected Recovery}
\label{sec:eval-recovery}

\paragraph{Readmission versus protection.}
Recovery must bring a request back and give it time to produce output.
We compare disabled recovery, age-only readmission and age-5/Q64 protection
on 50 LCB requests with Qwen3-32B, H200 TP2 and reduced GPU KV capacity.
Each mode uses SRJF/K64 and identical transfers, with one run per mode.

\begin{table}[t]
\centering
\caption{Qwen/LCB recovery controls, one run per mode. Wait is p99
eviction-to-next-fresh-output time. Gap $>10$\,s is the fraction of
requests whose longest client chunk gap exceeds 10\,s.}
\label{tab:recovery-modes}
\setlength{\tabcolsep}{3pt}
\begin{tabular}{lrrrr}
\toprule
Mode & Tok/s & Wait (s) & Offloads & Gap $>10$\,s \\
\midrule
Disabled & 475.43 & 45.11 & 71 & 22\% \\
Age-only & 484.45 & 5.20 & 494 & 0\% \\
Protected(Q=64) & \textbf{494.10} & 13.75 & 286 & 68\% \\
\bottomrule
\end{tabular}
\end{table}

Protection raises throughput and reduces offload cycles relative to
age-only readmission (Table~\ref{tab:recovery-modes}). Age-only shortens
waits but raises offloads from 71 to 494; protection reduces them to 286
and increases throughput by 1.99\% relative to age-only. Age-only still
has the lowest wait p99, and protection leaves more requests with gaps
above 10\,s: one request's protected progress makes others wait. All modes
record zero repeated evictions without fresh output, so this sample does
not establish that protection is needed to prevent zero-progress evictions.

Scheduling dominates the observed recovery wait: time before reload is
scheduled accounts for 98.12--99.70\% of summed eviction-to-next-token
intervals across the three modes. Protection also reduces repeated
evictions from 458 to 251, although short decode turns remain. This
supports improving readmission alongside transfer speed. The overlapping
request intervals are not a wall-clock breakdown of RAM residence or DMA,
and this small study does not quantify protection's role in the full DP4 gains.

\paragraph{Length of the protected turn.}
Longer protected turns can sharply reduce repeated rescues. On
Llama/AIME1000 at DP8/TP1 and 8 requests/s, increasing Q64 to Q1024
lowers mean repeat rescues from 4,622.5 to 37.5 per run. Across two runs
per arm, throughput rises 16.68\% and p99 TPOT falls 16.30\%.
The mean run-maximum streaming gap rises from 143.43 to 400.71\,s, and
Q1024 still trails same-node recompute in throughput and p99 TPOT.
This tuned development replay shows both the value of protected progress
and its waiting cost.
\suppref{app:protection-quantum} retains the full counters and controls.

\paragraph{How recovery affects throughput.}
Readmission can also keep decode batches fuller. In two Qwen/SRJF
comparisons with matched seeds, Q64 raises throughput by 3.93\% and
6.78\%, while mean decode batches grow from 11.91 to 12.50 and from
12.58 to 13.77 requests. Both pairs have
more offloads and no recomputation events, so fewer transfers cannot
explain the gains. Larger batches are consistent with the improvement;
GPU utilization was not measured. The second pair has higher
mean TPOT and p99 ITL, and other victim policies have mixed throughput
effects; \suppref{app:recovery-batching} gives the full controls.

\subsection{Admission Control and Continuing Arrivals}
\label{sec:eval-admission}

\name also avoids the long admission wait of a simple concurrency cap.
FIFO64 keeps the recompute engine fixed and admits at most 64
dispatched-but-unfinished requests per replica, or 256 across DP4; the rest wait in an external
FIFO queue. This is one policy, not a tuned upper bound on admission control.

Across three AIME and LCB runs, \name has higher throughput and much
earlier first output than FIFO64. Mean arrival-to-first-output time is
0.074\,s and 0.087\,s with \name, versus 260.63\,s and 225.19\,s with
FIFO64. The cap achieves lower p99 TPOT after admission, while
dispatch-based TTFT omits its queue. Both endpoints are needed to locate waiting;
\suppref{app:admission} gives the full comparison.

The tail TPOT benefit persists with continuing arrivals. We replay LCB
at 0.25, 0.5 and 1 request/s for 900 seconds per phase,
then drain every request. Across three runs, \name lowers p99 TPOT
by 47.10\% relative to recompute, with $-0.28\%$ throughput and
$+6.60\%$ p99 completion time. Requests span phase boundaries, precluding
independent steady-state capacity estimates. \suppref{app:continuing-arrivals}
reports phase and drain results: the tradeoff between progress and waiting
extends beyond a single arrival burst.

\subsection{Where Do the Gains Come From?}
\label{sec:eval-attribution}

Table~\ref{tab:attribution} collects the controlled comparisons above.
Each changes one mechanism inside \name while holding the others fixed,
but on smaller or different workloads than the main comparison. We have not
ablated the full system on IMO-AnswerBench, so no row explains its 12\%
throughput gain by itself.

\begin{table*}[t]
\centering
\caption{Controlled comparisons within \name. Each row changes one mechanism
and holds the others fixed. Throughput changes are relative to the second
configuration; paired values follow the workload order. Runs are per arm.}
\label{tab:attribution}
\setlength{\tabcolsep}{4pt}
\begin{tabular}{@{}llcrl@{}}
\toprule
Change & Workloads & Runs & Throughput (\%) & Other effect \\
\midrule
Exact extents vs.\ uniform 8K & AIME1000, LCB800 & 3 & $+4.1$, $-0.3$ & AIME fallbacks $521\to0$ \\
Fused K/V, two pairs vs.\ legacy & AIME300, LCB200 & 2 & $-1.0$, $+0.3$ & p99 ITL $-4\%$, $-8\%$ \\
Prediction enabled but unused & AIME1000, LCB880 & 3 & $-0.6$, $-0.3$ & -- \\
SRJF vs.\ physical-largest victims & AIME1000, LCB880 & 3 & $-7.9$, $-7.4$ & p99 ITL $-14\%$, $-15\%$ \\
Protected recovery (Q64) vs.\ none & LCB50, 2 pairs & 1 & $+3.9$, $+6.8$ & mean decode batch $+5\%$, $+9\%$ \\
Protected vs.\ age-only readmission & LCB50 & 1 & $+2.0$ & offloads $494\to286$ \\
Protected turn Q1024 vs.\ Q64 & Llama AIME1000 & 2 & $+16.7$ & max.\ gap $\times2.8$ \\
\bottomrule
\end{tabular}
\end{table*}

Three patterns emerge. First, retaining more snapshots and readmitting
stalled requests raise throughput where they apply: exact extents help
when they avoid recomputation fallbacks, and protected recovery keeps
decode batches fuller. Second, faster transfers mainly shorten output
gaps; they barely change throughput because requests spend most of their
recovery time waiting to be scheduled rather than moving data
(Section~\ref{sec:eval-recovery}). Third, prediction-guided selection
lowers throughput by 7--8\% in exchange for 14--15\% shorter p99 ITL.
It acts as a latency policy rather than a source of throughput: a
deployment that values throughput can choose physical-largest victims
while keeping the same storage, transfer and recovery mechanisms, as the
natural-completion study also suggests (Section~\ref{sec:eval-natural}).
An ablation that adds mechanisms one at a time on IMO-AnswerBench would
attribute the full-system gain directly; we leave it to future work.
\section{Related Work}
\label{sec:related}

We compare how prior systems schedule requests, predict remaining work
and manage KV during recovery.

\subsection{LLM Serving Systems}

Orca~\cite{yu_orca_nodate} schedules at each model iteration.
Sarathi-Serve~\cite{agrawal_taming_nodate} and
DeepSpeed-FastGen~\cite{holmes_deepspeed-fastgen_nodate} combine prompt
chunks with decoding to improve batching. Splitwise~\cite{patel_splitwise_2024}
and DistServe~\cite{zhong2024distserve} place prefill and decode on separate
GPU pools; Llumnix~\cite{sun2024llumnix} reschedules requests across
instances. \name focuses on repeated eviction and recovery within a worker.

FastServe~\cite{wu2026fastserve} uses multilevel feedback queues and
service quanta, promotes starved requests, and proactively swaps KV based
on estimated next execution time.
SuperInfer~\cite{yu2026superinfer} rotates requests according to
SLO-derived urgency and current memory and transfer budgets.
TokenFlow~\cite{chen2026tokenflow} uses generated-output buffers and client
consumption rates to coordinate admission, preemption and resumption.
InferCept~\cite{abhyankar2024infercept} pauses requests during external
calls and chooses among preserving, discarding and swapping their KV to
minimize wasted memory and recomputation. FastSwitch~\cite{shen2024fastswitch}
reduces the overhead of preemption-induced context switching in
fairness-aware serving, including transfers of fragmented KV blocks.
CacheOPT~\cite{shen2025cacheopt} orders preemption victims by latency SLO
and estimated remaining time, and chooses between swapping and recomputation.
These systems decide how to preempt or how to switch contexts cheaply.
\name adds a guarantee about what happens afterward: stall age bounds how
long a preempted request waits, age survives reload or recomputation, and
only fresh output counts toward the protected quantum. Our controls
evaluate these rules within \name; we do not benchmark these complete
systems.

\subsection{KV Cache Management Systems}

\paragraph{KV Cache Compression.}
KIVI~\cite{liu2023kivi} reduces KV precision, while
H$_2$O~\cite{zhang2023h2o} and Scissorhands~\cite{liu2023scissorhands}
retain selected tokens. These methods reduce stored state and can change
attention results. \name preserves KV precision and restores the full
available state for dense attention, rebuilding discarded state if needed.

\paragraph{Inference on Sparse KV Cache.}
StreamingLLM~\cite{xiao2024streamingllm} retains attention sinks and a
sliding window. Quest~\cite{tang2024quest} selects pages by query similarity;
InfiniGen~\cite{lee2024infinigen} predicts important tokens and prefetches
their KV. These methods skip or discard entries, so outputs can differ
from dense attention. SparDA~\cite{fu2026sparda} predicts the next layer's
KV block choices and prefetches them during the current layer's execution.
Our SparDA-derived adapter measures equal-payload transport only, excluding
its sparse attention, prediction and overlap with model computation.

\subsection{Output Length Prediction}

Length estimates for scheduling can come from
prompts~\cite{fu_efficient_2024,qiu_efficient_2024} or intermediate model states.
TRAIL, from \emph{Don't Stop Me Now}~\cite{shahout2024dontstop},
feeds target-model embeddings to a lightweight remaining-length classifier
and uses its predictions to order requests in a preemptive scheduling
policy. \name uses a multilayer regression head with sparse online
inference, and applies its estimates only to choose which running request
to evict under memory pressure. Age-based recovery then readmits victims
regardless of prediction, which bounds the cost of a wrong estimate. Section~\ref{sec:eval-prediction} compares predictors on the
same recorded populations.

\subsection{KV Cache Offloading}

FlexGen~\cite{sheng2023flexgen} plans GPU--CPU--disk placement for batch
inference. LoongServe~\cite{wu_loongserve_2024} uses elastic sequence
parallelism; Infinite-LLM~\cite{lin_infinite-llm_2024} distributes KV storage.
Mooncake~\cite{qin2025mooncake} separates prefill and decode, using shared
CPU/SSD KV storage and cache-aware request placement.

LMCache~\cite{cheng2025lmcache} gathers scattered GPU pages into chunks
and pipelines prefix-aware KV storage and loading with execution. It also
supports delayed saving of decode KV, disabled in our measured configuration
(Section~\ref{sec:eval-method}). \name also packs and pipelines transfers,
but allocates CPU space around each snapshot's known length.

SuperInfer's DuplexKV~\cite{yu2026superinfer} makes a KV block's layers
contiguous and batches transfers on GH200. \name retains paged GPU KV and
uses bounded exact extents with matching offsets in each layer's CPU
buffers. Two staging pairs overlap gather/scatter with copying. Our local transfer
measurements do not compare the complete SuperInfer system.

Pie~\cite{xu2024pie} prefetches upcoming layers' CPU KV during computation.
TokenFlow~\cite{chen2026tokenflow} chooses reload versus recomputation by
estimated cost and overlaps loading with eviction. \name offloads on GPU
allocation failure when CPU snapshot allocation succeeds, otherwise falling
back to recomputation. Protected recovery then gives restored requests time
to generate fresh output.

\paragraph{Earlier workshop version.}
An earlier three-page workshop paper~\cite{shi2025workshop}
described cascading preemption, hidden-state length prediction, an analytic
recomputation/offload tradeoff, and asynchronous layerwise transfers with
packing. This paper adds the contiguity-first CPU pool and exact-extent
allocator, age-based protected recovery, and sparse predictor execution.
It also adds repeated system comparisons, component controls and studies
of natural completion, output pauses and workload limits.

\section{Discussion and Limitations}
\label{sec:discussion}

\paragraph{Who benefits.}
\name helps when outputs are long enough that requests are preempted while
much decoding remains. It then shortens the tail of slow requests and
bounds every stall, at the cost of more frequent pauses of about 30 seconds
for long requests (Section~\ref{sec:eval-pauses}). Batch, agentic and
evaluation workloads that consume complete answers gain the most. An
interactive client that streams tokens to a reader may prefer shorter,
more frequent rescues, which a lower readmission threshold or a different
protected-turn length can provide. Short-output workloads gain nothing:
without preemption, prediction and scheduling add work but save no recovery.

\paragraph{Scope of the evaluation.}
We evaluate dense models (Qwen3-32B and Llama-3.1-8B) with up to two GPUs
per replica, or four in the allocator study, on vLLM 0.9.2. Reasoning
deployments increasingly use mixture-of-experts models with compressed or
latent attention. Our cost model suggests that preserving their smaller
per-token state remains favorable (Section~\ref{sec:recompute-offload-model}), but we
have not measured them. Disaggregated serving places decoding on dedicated
instances where the same memory pressure arises; combining \name with
prefill--decode disaggregation is future work. The main comparisons use
finite arrival bursts at one rate; the earlier load sweeps
(Figure~\ref{fig:load-sweep}) and continuing-arrival study
(Section~\ref{sec:eval-admission}) cover other rates with fewer runs.

\paragraph{Fixed-output replay and predictor exposure.}
Fixed-output replay gives every system identical work, but it does not
measure answer quality, and replayed greedy trajectories diverge from the
generation that defined each length (Section~\ref{sec:eval-method}). The
predictors were trained or validated on tasks that also appear in the
serving workloads. The 299 IMO-AnswerBench questions outside the training
set show the same direction of change, and the natural-completion study
shows the tail benefit with real termination, but neither is a fully
held-out generalization test.

\paragraph{Attribution.}
Component studies change one mechanism at a time on smaller or different
workloads (Section~\ref{sec:eval-attribution}). Several recovery studies
use one or two runs and reduced GPU capacity. An ablation on
IMO-AnswerBench that adds mechanisms one at a time would attribute the
full-system gain directly.

\paragraph{Baselines.}
Baselines run their native configurations. The HiCache comparison also
spans the SGLang and vLLM engines, and LMCache's measured configuration
includes a resume-path repair (\suppref{app:baseline-corrections}).
We compare against FastServe, TokenFlow, SuperInfer, InferCept and
CacheOPT only in related work.

\paragraph{Host memory.}
Retaining snapshots needs host memory in proportion to the preempted KV.
In the current implementation, allocator rounding adds 224\,GiB to the
800\,GiB payload, which roughly matches \name's excess host memory over
LMCache and HiCache. Each worker also manages its own pool; sharing
host capacity across replicas could help when pressure is uneven, at the
cost of coordination.
 
\section{Conclusion}
\label{sec:conclusion}

Long decoding makes preemption unavoidable, and existing systems let
preempted requests stall for minutes. \name makes preemption productive:
it preserves a preempted request's work in a contiguity-first CPU pool,
moves it back with bulk transfers, bounds how long the request waits, and
gives it fresh output after it returns. On olympiad-math replays, this
completes the same work faster with a much shorter tail of slow requests,
and no request stalls for more than a minute. The price is host memory and
more frequent, bounded pauses. Preserving state and guaranteeing time to
use it must be designed together.

\bibliographystyle{ACM-Reference-Format}

\clearpage
\appendix
\noindent\textbf{Guide to the supplementary material.}
The first sections expand the main serving results and explain the roles
of recovery, storage, transfers and prediction. The cost model and exact
settings follow. Earlier extensions and historical experiments are kept
separate from the current system's evidence.

\medskip
\noindent
\begin{tabular}{@{}p{0.76\columnwidth}p{0.18\columnwidth}@{}}
\textbf{Evidence or detail} & \textbf{Section} \\
Natural completion and workload scope & \ref{app:workload-details} \\
Admission waiting and continuing arrivals & \ref{sec:supp-controls} \\
Recovery, protected turns and batching & \ref{app:recovery-details} \\
CPU allocation and KV transfers & \ref{app:storage-transfer} \\
Prediction accuracy, policy and interval & \ref{app:predictor-details} \\
Recompute--offload cost model & \ref{sec:appendix-cost-model} \\
Evaluation settings and implementation & \ref{app:methods} \\
Earlier online-training extension & \ref{supp:online-training} \\
Historical serving experiments & \ref{supp:historical} \\
Earlier predictor variants & \ref{supp:earlier-predictors} \\
\end{tabular}
\medskip

Workloads include American Invitational Mathematics Examination (AIME)
math problems, LiveCodeBench (LCB) programming tasks, and IMO-AnswerBench
olympiad problems~\cite{luong2025imobench}.
AIME1000 and LCB880 denote the main 1,000- and 880-request replays.
Numeric suffixes give replay request counts, not repetition counts;
each smaller study states its own population.
\section{Additional Serving Results}
\label{app:workload-details}

The natural-completion study extends the tail-TPOT result beyond fixed
output replay. We then examine predictor training exposure and additional
workloads to explain where the serving benefits apply. These comparisons
retain their task outcomes, latency tradeoffs and resource costs.

\subsection{Natural Completion and Coding Quality}
\label{app:natural-quality}

Fixed-length replay holds output work constant, but real programs can
finish early or reach the generation limit. We therefore serve 175 new
LiveCodeBench v6 tasks (LCB175) with natural EOS, then run their program
tests. Recompute,
physical-largest \name (P0), and SRJF \name (S) each run three matched
arrival seeds. The setting is Qwen3-32B, one H200 TP2 replica, 1 request/s,
a 32{,}768-new-token ceiling, and 200\,GiB CPU KV for P0/S. Both \name
arms use age-5/Q64 recovery; only S runs the frozen $K=64$ predictor.

\begin{figure*}[t]
\centering
\includegraphics[width=\linewidth]{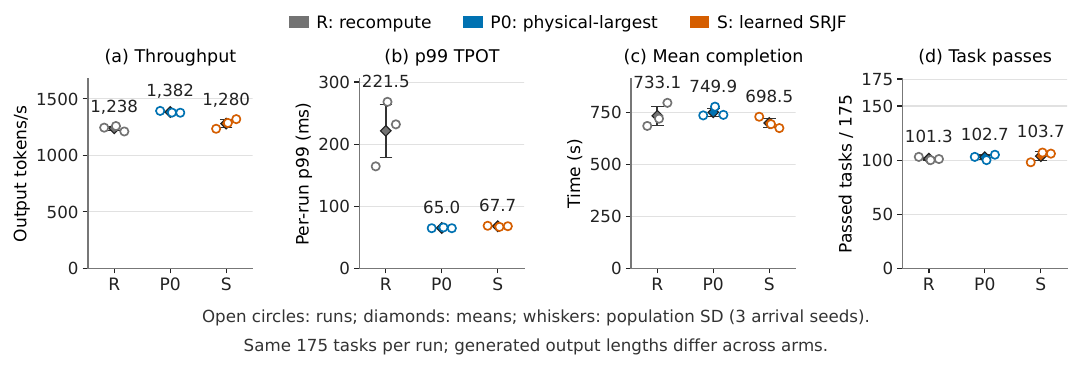}
\caption{Natural-completion LCB175: serving performance and coding
outcomes across three arrival seeds. Circles show individual runs;
diamonds and whiskers show means and population standard deviations.
The same 175 tasks repeat, but generated lengths differ across arms.}
\Description{Four panels compare recompute, physical-largest \name and
SRJF \name on output throughput, p99 TPOT, mean arrival-to-completion
time and the number of programs that pass their tests.}
\label{fig:app-natural-lcb}
\end{figure*}

Against recompute, S reduces p99 TPOT by 68.07\%. Its mean completion
time from arrival falls 4.16\%, and throughput rises 3.46\%.
Median TPOT instead increases 13.99\%, and p99 TTFT increases 17.90\%.
P0 has higher throughput than S in every block. Compared with P0, S has
7.37\% lower throughput and 4.14\% higher p99 TPOT, but 6.78\% lower
mean completion time. Prediction-guided ordering therefore offers a
tradeoff in this study; it does not improve every endpoint over the
same recovery system with physical-largest victims.

\begin{table}[t]
\centering
\caption{Programs passing the offline tests, out of 175 tasks per run.
All evaluations have a valid outcome. Repeated seeds do not create new
independent tasks.}
\label{tab:app-natural-quality}
\begin{tabular}{lrrr}
\toprule
Arrival seed & Recompute & P0 & SRJF \\
\midrule
30 & 103 & 103 & 98 \\
31 & 100 & 100 & 107 \\
32 & 101 & 105 & 106 \\
\bottomrule
\end{tabular}
\end{table}

Pass counts vary in both directions (Table~\ref{tab:app-natural-quality}).
Of 67 recompute/SRJF pass--fail changes, 60 occur when neither response
hits the output limit. The results therefore establish neither better
nor equivalent task quality. They are also not an identical-work
comparison: natural output amounts differ across arms.

As an additional descriptive metric, we divide passing responses by
full serving time, including time spent on unsuccessful responses.
S improves this quantity by 6.62\% over recompute on average, but the
three changes are $-8.35\%$, $+7.96\%$ and $+20.26\%$. It is 5.12\%
lower than P0 on average. This excludes offline scoring time and is not
whole-agent workflow throughput.

Both \name arms record no preemption before a protected turn ends,
but repeated rescues remain common: 955 for P0 and 913 for S across the
three runs. Mean selection-time stall ages are 38.94 and 44.16\,s;
mean rescue-acquisition-to-next-token times are 0.1640 and 0.1653\,s.
These are separate histograms, not a joined breakdown of RAM waiting
and transfer time. The 5-second threshold makes a request eligible for
rescue; it does not bound its pause to five seconds.

\subsection{IMO-AnswerBench Training Exposure}
\label{app:superhuman-exposure}

We use all 399 usable IMO-AnswerBench questions to create enough
concurrent long decodes for the serving comparison. A fixed random draw
(seed 20261003) selects 100 questions for predictor training. Within
those 100, an 80/20 split selects the training epoch using validation
MAE; a fresh head then refits all 100 for that many epochs. Qwen selects
22 epochs. The head remains frozen during the four-system serving runs.
Llama and Hunyuan heads were also trained, but this study reports
serving results only for Qwen.

The shared output-length freeze comes from one natural-completion run.
It retains all 30 requests that reach the generation limit. Median
output length is 17{,}089 tokens; 213 of 399 requests exceed 16{,}384.
Subsequent system comparisons replay the same 7{,}279{,}811 output
tokens, so these are performance measurements rather than a new
answer-quality comparison.

Table~\ref{tab:app-exposure} reports the two groups separately within
the same three recompute/\name pairs. Both groups share scheduling and memory pressure. Of the 299,
28 were used in an earlier calibration, so training-unseen does not
mean fully held out.

\begin{table*}[t]
\centering
\caption{IMO-AnswerBench subgroup changes with \name relative to recompute.
Percentages average three within-block ratios; negative latency changes
are better. Both groups share each full 399-request replay.}
\label{tab:app-exposure}
\begin{tabular}{lrrrrr}
\toprule
Group & Mean TTFT & Mean TPOT & p99 TPOT & \shortstack{Mean dispatch-\\to-completion} & \shortstack{p99 dispatch-\\to-completion} \\
\midrule
Training100 & $+7.42\%$ & $+1.34\%$ & $-36.47\%$ & $+3.85\%$ & $-3.52\%$ \\
Training-unseen299 & $+17.37\%$ & $-8.45\%$ & $-53.56\%$ & $-4.10\%$ & $-4.16\%$ \\
\bottomrule
\end{tabular}
\end{table*}

Both groups have lower p99 TPOT, while their mean completion-time
changes differ. Their TTFT and median TPOT worsen on average.
The grouping does not isolate predictor benefit: offloading and recovery
also differ between systems. Nor does it measure predictor accuracy or
held-out generalization. We report throughput only for the full replay,
because the two groups share the same execution resources and duration.

\subsection{Selected Configurations on Other Workloads}
\label{app:tuning}

We use the best-supported settings from the separate tuning study, then
compare each with unchanged recompute in three additional runs. Each
comparison reuses its frozen tasks, output lengths, arrival schedule and
routing. These experiments test the selected configuration; they do not
isolate every optimization or re-tune the baseline.

\paragraph{Qwen/arXiv: fewer preemptions do not ensure better tails.}
This study uses 2{,}000 requests at 16 requests/s on H200 DP4/TP2.
The selected setting waits for enough space to admit the full prompt,
reserves headroom of 64 blocks, and sets the long-prefill threshold to
zero. Qwen thinking is disabled when preparing these prompts.
Compared with recompute, throughput changes by $-0.57\%$, mean TTFT
falls 25.07\%, and mean arrival-to-completion falls 16.96\%.
However, p99 TTFT rises 5.73\%, p99 TPOT rises 6.51\%, and p99 completion
time rises 10.53\%. Preemptions fall by 98.60\%, yet throughput does not
improve. This shows why preemption count alone cannot explain performance.

\paragraph{Llama/AIME: longer protected turns reduce extreme pauses.}
This comparison uses 1{,}000 requests at 16 requests/s on H200 DP8/TP1,
with 54{,}257 GPU KV blocks per worker. The selected \name setting uses
$K=64$, age 5\,s, a 1{,}024-token protected turn, early rotation disabled,
and 800\,GiB CPU KV. Throughput changes by $+0.10\%$, while the p99 of
per-request longest output gaps falls 54.63\%. Mean TPOT, p99 TPOT and
p99 completion time instead increase 3.89\%, 2.74\% and 2.62\%.
The small throughput difference varies in sign across runs; it does not
establish either a general gain or equivalence.

\paragraph{Qwen/BurstGPT: limited benefit without long output pauses.}
This comparison uses 2{,}000 synthetic length-matched requests at
32 requests/s on H200 DP4/TP2. Prediction turns on above 80\% KV
pressure and off below 60\%, with 64-step hysteresis.
The selected setting remains 2.04\% below recompute throughput, with
5.20\% higher mean TPOT and 0.72\% higher p99 TPOT. The longest output
gaps stay below one second in both systems. This workload leaves little
scope for reducing the severe interruptions that motivate \name.

\begin{figure*}[t]
\centering
\includegraphics[width=0.94\linewidth]{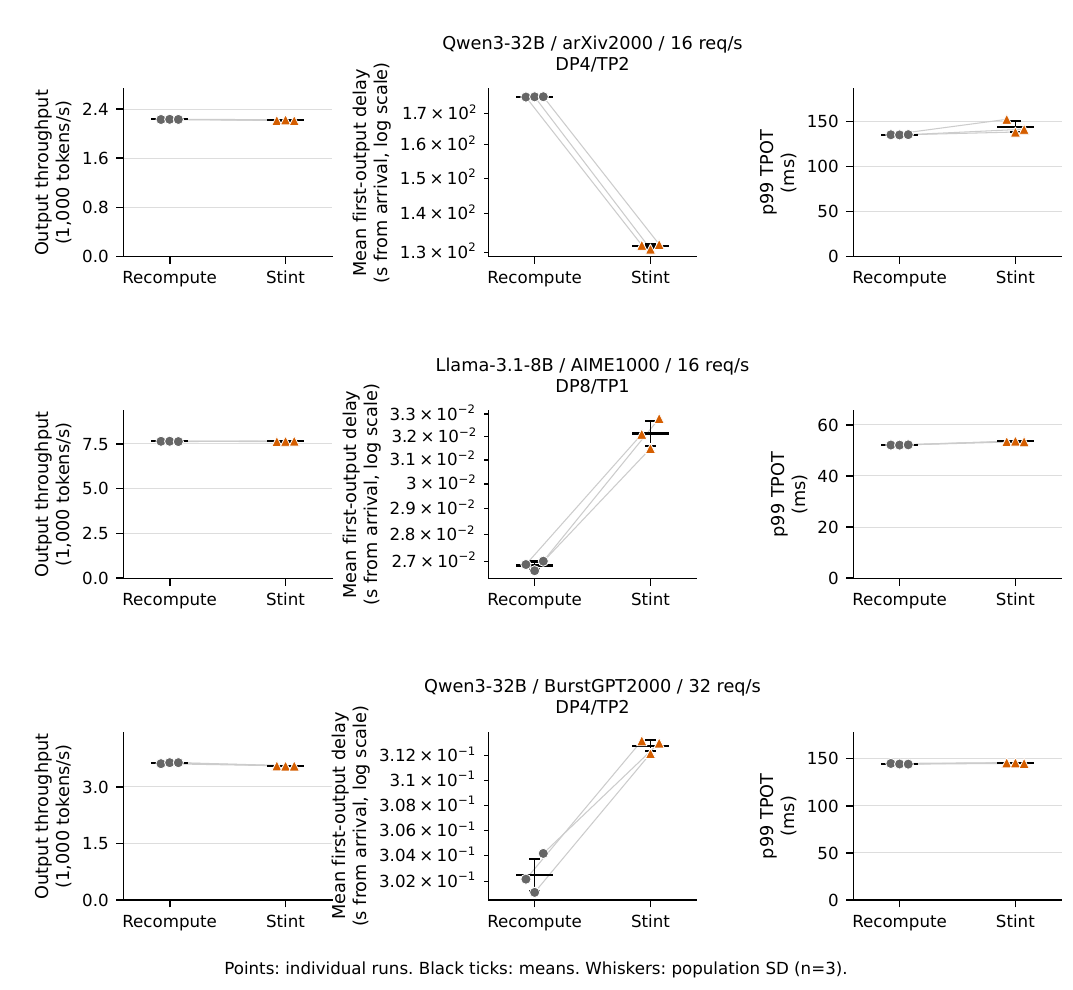}
\caption{Three-run confirmation of the selected configurations.
Each row retains its own workload and arrival rate. Points are runs;
black ticks and whiskers show means and population standard deviations.
First-output delay uses a logarithmic axis; the other axes are linear.}
\Description{Three rows compare recompute and selected \name settings
on Qwen/arXiv, Llama/AIME and Qwen/BurstGPT. Columns show throughput,
mean arrival-to-first-output time and p99 TPOT.}
\label{fig:app-tuning}
\end{figure*}

\subsection{Cachewise: Shared Prefixes and Short Outputs}
\label{app:cachewise}

Cachewise coding-agent traces contain redacted request text. We therefore
replay synthetic tokens that preserve the recorded lengths, known shared
prefixes and parent--child dependencies. Each run contains 1{,}418 calls
in 65 sessions and 94 root branches. Roots arrive at 4 branches/s;
a child becomes eligible only after its own parent succeeds, with no
added think time. Thus, 4 roots/s is not 4 total requests/s, and child
release times depend on the serving system's progress.

The four systems run three matched blocks on Qwen3-32B/H200 DP4/TP2.
They complete the same 151{,}247 output tokens per run, averaging only
106.66 tokens per call. \name uses physical-largest victims with
prediction and feature capture disabled, retaining age-5/Q64 recovery
and 800\,GiB CPU KV. The other cache systems use the settings in
Appendix~\ref{app:methods}.

\begin{figure*}[t]
\centering
\includegraphics[width=\linewidth]{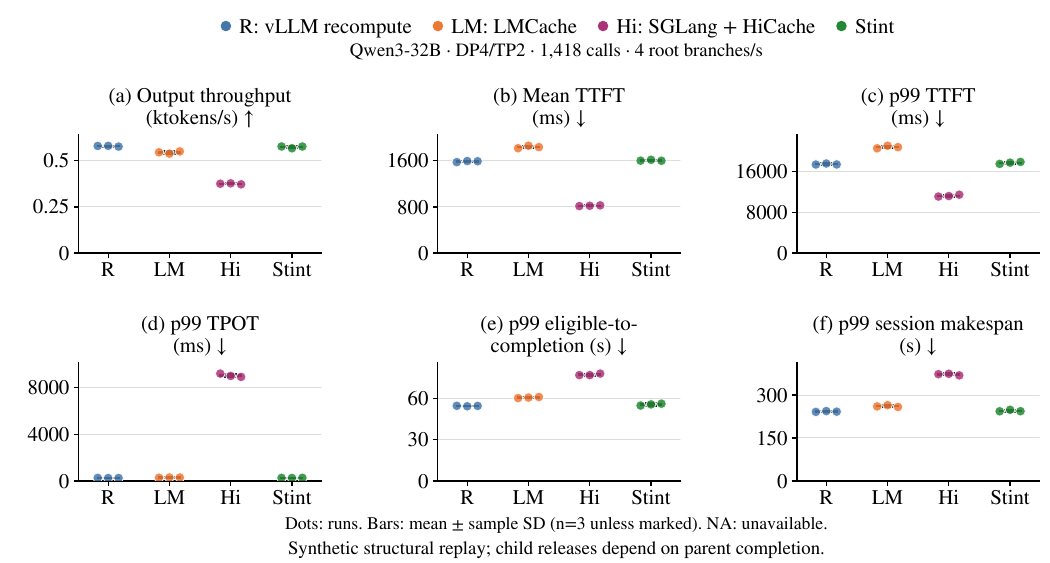}
\caption{Cachewise structural replay, three matched blocks per system.
Bars show means with sample standard deviation; points retain individual
runs. Eligible-to-completion includes waiting after a parent permits a
call, while session makespan also includes the dependency chain.}
\Description{Six panels compare four systems on output throughput, mean
and p99 TTFT, p99 TPOT, p99 eligible-to-completion time and p99 session
makespan. HiCache has lower TTFT but much higher p99 TPOT.}
\label{fig:app-cachewise}
\end{figure*}

Against recompute, \name changes throughput by $-0.89\%$, mean TTFT
by $+1.14\%$, and p99 TPOT by $+2.59\%$. It shows no advantage in
these metrics.
Against LMCache, throughput rises 5.33\% and p99 TPOT falls 9.87\%.
Against HiCache, throughput rises 52.97\% and p99 TPOT falls 97.13\%,
but mean TTFT rises 95.73\%. HiCache's faster first output remains an
important counterpoint to its longer later delays in this replay.
Mean sampled memory-guard peaks are 330.12, 1{,}133.69, 1{,}135.16 and
1{,}358.39\,GiB for recompute, LMCache, HiCache and \name, respectively.
These are node/cgroup peaks, not isolated pool measurements.

All nine recompute/LMCache/\name runs record zero preemptions; HiCache's
requested counters are unavailable. Zero preemptions do not imply that
no CPU lookup or reload occurs. The study provides a low-pressure,
shared-prefix comparison, not evidence of the thrashing mechanism,
predictor generalization or original coding-task quality. Each run has
1{,}294 defined TPOT samples; the 124 single-output-token calls are
excluded rather than assigned zero TPOT.
\section{Admission Waiting and Continued Arrivals}
\label{sec:supp-controls}

Shorter pauses after a request starts do not necessarily mean a shorter
response. Admission control can improve streaming by keeping requests
outside the engine. These experiments account for that waiting and then
ask whether the tradeoff persists while new requests continue to arrive.

\subsection{A FIFO Admission Control}
\label{app:admission}

We compare recompute, FIFO64 and \name on AIME1000 and LCB880.
Each arm has three matched runs on Qwen3-32B, H200 DP4/TP2, at
4 requests/s. FIFO64 changes only client admission: each replica allows
64 dispatched but unfinished requests. The recompute runtime stays fixed.
All arms complete the same inputs and prescribed output lengths.

\begin{table*}[t]
\centering
\caption{FIFO admission reduces post-admission TPOT but adds waiting before
output begins. Values average three runs. Arrival-to-first-output includes
the admission queue; the memory column is the sampled node/cgroup guard,
not process RSS or CPU KV capacity.}
\label{tab:supp-admission-context}
\setlength{\tabcolsep}{5pt}
\begin{tabular}{llrrrr}
\toprule
Workload & System & Output tok/s & \shortstack{Mean arrival-to-\\first-output (s)} & p99 TPOT (ms) & \shortstack{Memory guard\\peak (GiB)} \\
\midrule
AIME1000 & Recompute & 7,741.27 & 4.687 & 118.26 & 417.18 \\
 & FIFO64 recompute & 6,493.93 & 260.629 & 32.16 & 416.35 \\
 & \name & 7,701.16 & 0.074 & 76.86 & 1,446.90 \\
\midrule
LCB880 & Recompute & 5,007.50 & 0.084 & 225.14 & 451.33 \\
 & FIFO64 recompute & 4,703.59 & 225.188 & 39.99 & 451.01 \\
 & \name & 5,189.30 & 0.087 & 57.16 & 1,481.02 \\
\bottomrule
\end{tabular}
\end{table*}

FIFO64 has lower p99 TPOT than \name, but its throughput is lower and
its mean arrival-to-first-output time is several minutes
(Table~\ref{tab:supp-admission-context}). Measuring TTFT only after
dispatch would hide most of this delay. \name starts output sooner by
keeping more requests in flight, at the cost of substantially more host
memory and less regular streaming than FIFO64. This compares one FIFO
limit; it does not establish that protected recovery beats every
admission policy.

\subsection{Continued Arrivals}
\label{app:continuing-arrivals}

The finite main replay eventually stops admitting work. We therefore
repeat the three-system comparison while requests arrive for 2{,}700
seconds: 900 seconds each at 0.25, 0.5 and 1 request/s, followed by a drain.
The model, DP4/TP2 topology and storage/recovery settings remain the same.
Three matched seeds complete 1{,}591, 1{,}577 and 1{,}583 requests per arm,
with no censored requests.

\begin{figure*}[t]
\centering
\includegraphics[width=\linewidth]{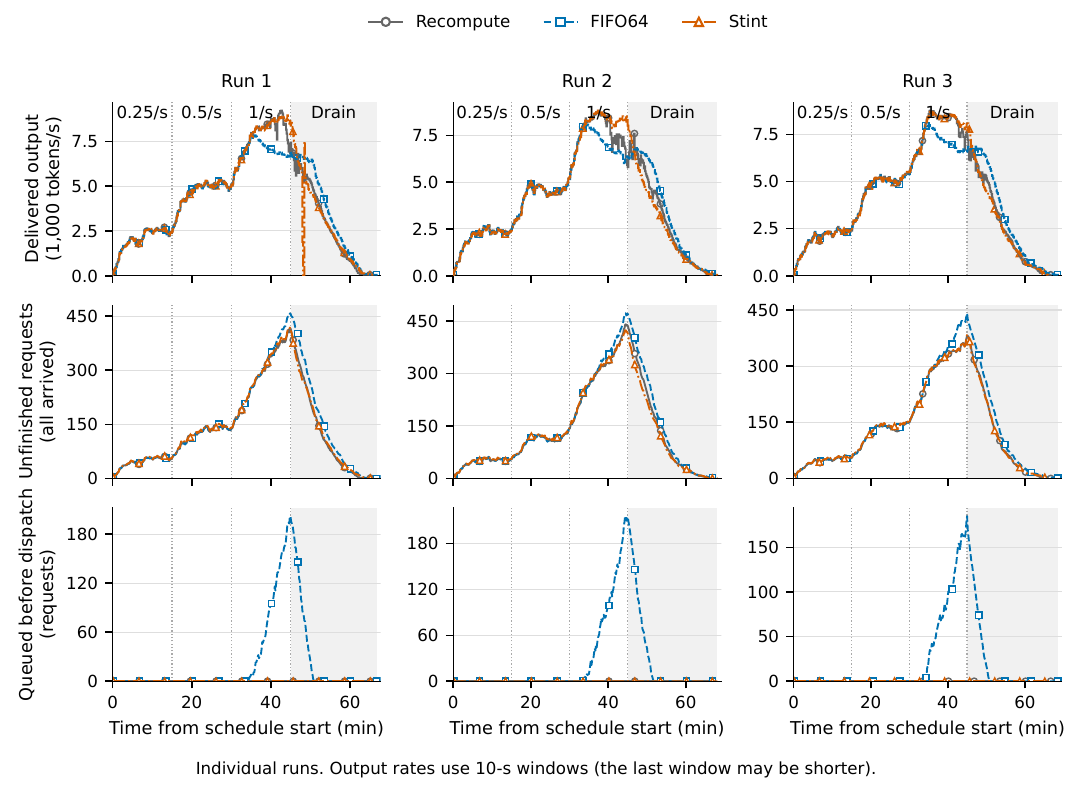}
\caption{Three repeated runs with continued arrivals followed by a drain.
Each column shows one matched block: delivered output, unfinished
requests, and the client admission queue. Output rates use 10-second
windows. Earlier phases can leave requests active in later phases.}
\Description{Three columns follow arrival seeds 30, 31 and 32. Rows show delivered
output per second, unfinished requests and requests queued before
dispatch for recompute, FIFO64 recompute and \name.}
\label{fig:app-continuing-arrivals}
\end{figure*}

Relative to recompute, \name reduces p99 TPOT by 47.10\% and the
longest observed output gap by 80.55\%. Whole-run throughput changes by
$-0.28\%$ on average, with mixed changes across the three pairs.
The tradeoff remains. Completion p99, measured from arrival, increases
6.60\%; pooled p99 ITL increases 4.83\%; and median TPOT increases
1.67\%. Smaller extreme pauses do not mean every response finishes sooner.

\name delivers 1.07\% more output per second during the arrival period,
but its drain is longer in two of the three runs. At the end of arrivals,
\name/recompute have 413/416, 416/437 and 385/382 unfinished requests.
FIFO64 further reduces streaming pauses while adding 55.11--65.60\,s
of mean admission waiting. That waiting stays in the arrival-based metrics.
These are finite, changing-load replays: overlapping cohorts prevent us
from treating each phase as an independent steady state or a capacity limit.
\section{Additional Recovery Evidence}
\label{app:recovery-details}

Age-based readmission and protection address different parts of a
request's return to decoding. The main paper compares their effects on
waiting and repeated offload. Here we show the pause distributions,
longer protected turns, and batching observations behind that discussion.
These studies also show how progress for one request can increase waiting
for others. Their configurations and repetition counts remain separate.

\subsection{Historical Recovery Across Pause Thresholds}
\label{app:recovery-thresholds}

\begin{figure*}[t]
  \centering
  \includegraphics[width=\linewidth]{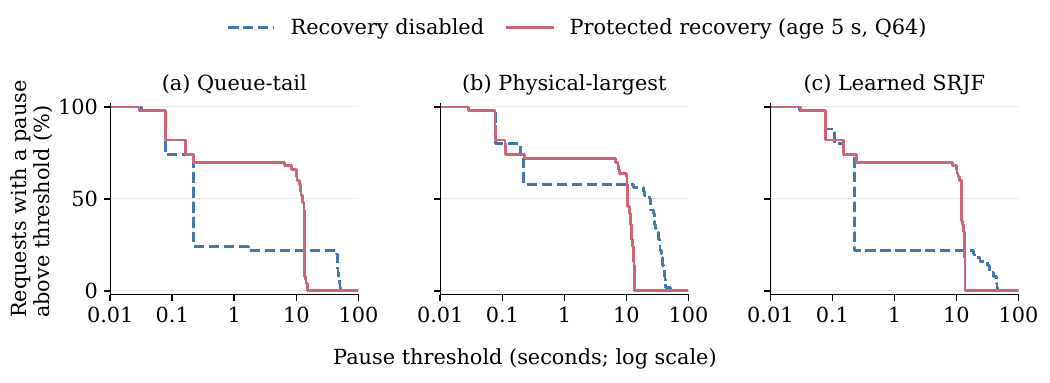}
  \caption{Longest client-chunk gaps in the six LCB recovery controls.
  Each curve is one complete run of 50 requests; the x-axis is logarithmic.
  Protected recovery uses age 5\,s and a 64-fresh-token quantum.
  A lower curve means fewer requests exceed that pause threshold.
  In these runs, protected recovery reduces the longest pauses while
  increasing the fraction of requests with moderate pauses.}
  \Description{Three panels show queue-tail, physical-largest and learned
  SRJF victims. Dashed blue curves disable recovery and solid red curves
  enable protected recovery. In all three panels, the protected curve is
  higher at a ten-second threshold but reaches zero before thirty seconds.}
  \label{fig:supp-recovery-thresholds}
\end{figure*}

For each request, we take its longest observed gap between token-bearing
client chunks. This excludes waiting before the first output.
Figure~\ref{fig:supp-recovery-thresholds} plots the fraction of
requests whose longest gap exceeds each threshold. It uses every observed
maximum, rather than selecting a single latency target.

This earlier reduced-capacity study is separate from the repeated full-deployment experiments. All six runs use Qwen3-32B BF16, one H200 TP2 replica, 11{,}251 GPU KV
blocks per worker, and 100\,GiB CPU KV per worker. They replay the same
50 requests and 544{,}661 output tokens, using replica 0's offsets from
the finite Poisson 16-request/s DP4 schedule with seed 10. K64 prediction
and the transfer path stay fixed. Each mode has one observation; the
curves describe those requests, not uncertainty across repeated runs.

With SRJF, protected recovery raises the fraction above 10\,s from 22\% to 68\%,
while reducing the fraction above 30\,s from 16\% to zero. Queue-tail
and physical-largest show the same direction at these thresholds.
Thus, reducing the longest pauses can come with more frequent moderate
interruptions. These distributions do not establish a waiting-time bound.
The separate age-only comparison in the main paper shows that readmission
without protection can have still shorter waits, at the cost of more
offloads in that observation.

\subsection{Earlier Protection-Quantum Controls}
\label{app:protection-quantum}
\paragraph{Give a recovered request enough time to progress.}
This earlier H200 study isolates the protected-turn setting; the later
three-run selected-configuration confirmation is reported in
Appendix~\ref{app:tuning}. A short protected turn can still be followed
by another recovery cycle.
To test this, we change the fresh-output quantum from 64 to 1{,}024
tokens on Llama-3.1-8B/AIME1000. This full-deployment study uses eight
H200 GPUs at DP8/TP1, finite Poisson arrivals at 8 requests/s, and an
800\,GiB aggregate CPU KV pool. Each \name arm and a same-node
recompute control have two complete runs.

The longer turn reduces mean repeat rescues from 4{,}622.5 to 37.5 per
run. Swap preemptions fall from 5{,}348.5 to 889.5, and recomputation
preemptions from 1{,}143.5 to 8.5. Throughput rises from
$6{,}412\pm8$ to $7{,}482\pm23$ tokens/s ($+16.68\%$), while p99 TPOT
falls from $63.00\pm0.25$ to $52.73\pm0.01$\,ms ($-16.30\%$).
Here, changing the protected turn produces fewer recovery cycles
together with higher throughput. The counters do not measure how much
GPU time those avoided cycles save.

The cost appears in the longest pauses. The mean run-maximum streaming
gap rises from 143.43 to 400.71\,s. Q1024 also remains 2.81\% below
same-node recompute throughput and 2.37\% above its p99 TPOT, although
its maximum-gap summary is shorter than recompute's 910.45\,s.
These controls isolate the protected-turn setting; the later confirmation
of the selected configuration appears in Appendix~\ref{app:tuning}.

The predictor, K64 cadence, transfer path, GPU capacity and 5\,s age
threshold stay fixed. Runs use the same seed-10 replay, with arm order
reversed between repetition blocks. Summaries report arithmetic means
and population SD; p99 summaries average run-level p99 values. Tuning
and confirmation use the same population, whose 413 generation-capped
profiles supply 96.44\% of output tokens. This result supports balancing
repeated recovery against other requests' waiting; it does not establish
one quantum for all workloads or a waiting-time bound under continued
arrivals.

\subsection{Recovery, Batching and Victim Policies}
\label{app:recovery-batching}
\paragraph{Throughput can also change through batching.}
In the seed-10 Qwen/SRJF recovery comparison, Q64 increases throughput by 3.93\% over
disabled recovery while increasing the recorded mean decode batch
from 11.91 to 12.50 requests. A second within-seed pair shows the same
direction: throughput rises by 6.78\%, the mean batch grows from 12.58
to 13.77, and eviction-wait p99 falls from 77.32 to 28.59\,s.
Both pairs record more offloads and no recomputation events.
Fewer transfers or recorded recomputations therefore do not explain
these gains. Larger decode batches are consistent with the improvement,
but these measurements do not isolate GPU utilization.

The second pair also has latency costs: offloads rise from 141 to 454,
requests with gaps above 10\,s increase from 46\% to 86\%, mean TPOT
rises 15.10\%, and pooled p99 ITL rises 46.47\%. Its p99 TPOT falls
16.58\%. These metrics describe different aspects of the request
experience; a lower eviction-wait p99 does not mean every latency
measure improves.

Recovery also shortens waits with physical-largest victims. The
eviction-wait p99 falls from 42.66 to 13.51\,s at seed 10, and from
82.68 to 24.50\,s at seed 11.
Throughput changes by $+0.85\%$ and $-0.51\%$, and gaps above 10\,s
affect more requests in both pairs. Seeds route different subsets to
replica 0, and the initial screen and follow-up differ in concurrent
lane activity. We compare treatments within each setting, without
pooling them as identical-workload repetitions. The unflushed terminal
batch sample is excluded from batch summaries.
With queue-tail eviction,
recovery reduces the mean batch from 13.86 to 12.46 and throughput by 9.90\%.
The initial screen uses single seed-10 pairs at reduced GPU capacity (11{,}251 blocks per
worker), with 100\,GiB CPU KV per worker, using one replica's offsets from
the original finite Poisson 16-request/s DP4 schedule. The scheduler
summaries are consistent with a batching explanation; they do not measure
GPU utilization or establish its share of the end-to-end gains.
\section{CPU KV Pool and Bulk KV Transfer}
\label{sec:supp-extra-eval}
\label{app:component-details}
\label{app:storage-transfer}

The CPU pool and transfer engine support the same recovery path: retain
a request's state, then move it back when the scheduler makes room. We
first examine storage under greater CPU pressure, then the broader
transfer comparisons. Each study retains its own configuration and
repetition count.

\subsection{Allocation Under Greater CPU Pressure}
\paragraph{A smaller CPU pool.}
A separate AIME1000 study reduces the pool to 200\,GiB on H200 TP4 at
4 requests/s. Across three runs, exact extents reduce recomputation
fallbacks from $1377\pm25$ to $260\pm29$ relative to uniform 4K chunks,
while throughput rises from $4297\pm52$ to $4524\pm46$ tokens/s.
Exact extents use 1.04 regions per offload on average, although some
allocations still fail because free space is fragmented.

This pressure study also limits each snapshot to eight regions. Uniform
4K chunks therefore cannot represent snapshots above 32K tokens within
that limit. The arms run concurrently on different fixed NUMA/GPU
placements, so the throughput change cannot be assigned solely to the
allocator. Fewer fallback events show that more state was retained;
they do not measure how much GPU recomputation time was saved.

\subsection{Additional Transfer Comparisons}

Table~\ref{tab:supp-dense-kv-transfer} reports the two transfer
cohorts of the main paper's table with their standard deviations, each with its own matched \name control. Both use CUDA 12.8,
but they run on different H200 hosts and use different GPU/NUMA
placements. Comparisons stay within a cohort and direction.

\begin{table*}[t]
  \centering
  \caption{Local KV transfer latency in milliseconds (mean $\pm$ population
  SD, three repetitions). Each operation moves one layer of 2{,}560 pages
  (160\,MiB of K/V), with scattered GPU pages and up to eight CPU regions.
  The cohorts use separate hosts and separately measured \name controls.}
  \label{tab:supp-dense-kv-transfer}
  \setlength{\tabcolsep}{5pt}
  \begin{tabular}{lrrrr}
    \toprule
    & \multicolumn{2}{c}{Prepared} & \multicolumn{2}{c}{With preparation} \\
    \cmidrule(lr){2-3}\cmidrule(lr){4-5}
    Method & D2H & H2D & D2H & H2D \\
    \midrule
    \multicolumn{5}{l}{\textit{HiCache cohort: H200 host A, CUDA 12.8}} \\
    \name & $3.680\pm0.001$ & $3.231\pm0.001$ & $4.243\pm0.009$ & $3.961\pm0.007$ \\
    HiCache layer-first & $3.274\pm0.002$ & $3.361\pm0.002$ & $8.929\pm0.068$ & $9.075\pm0.086$ \\
    HiCache page-first & $15.793\pm0.009$ & $3.360\pm0.001$ & $21.948\pm0.012$ & $9.086\pm0.054$ \\
    \midrule
    \multicolumn{5}{l}{\textit{Local-adapter cohort: H200 host B, CUDA 12.8}} \\
    \name & $3.642\pm0.003$ & $3.231\pm0.002$ & $4.210\pm0.009$ & $3.957\pm0.003$ \\
    Mooncake local batch & $4.753\pm0.010$ & $4.919\pm0.015$ & $27.064\pm0.776$ & $27.068\pm0.770$ \\
    Mooncake local fallback & $16.358\pm0.027$ & $17.308\pm0.075$ & $38.177\pm0.440$ & $39.164\pm0.481$ \\
    SparDA-derived (16 CTAs) & $16.344\pm0.070$ & $4.546\pm0.003$ & $52.268\pm0.335$ & $6.322\pm0.015$ \\
    \bottomrule
  \end{tabular}
\end{table*}

\paragraph{Results.}
These measurements show that bulk copying is only part of
transfer cost. At 160\,MiB, HiCache layer-first has lower D2H latency
after mappings are prepared, but \name is faster when adapter preparation
is included. For the 64-layer case (10\,GiB), prepared offload takes
202.236\,ms in \name and 204.433\,ms in HiCache layer-first. Reload takes
196.935\,ms and 209.548\,ms, respectively. Across the HiCache matrix,
\name has the lowest mean latency
in 72 of 96 large fragmented configurations, but only 126 of 324 overall;
these counts weight configurations equally. HiCache often wins at smaller
sizes.

In the separate local-adapter cohort, the 64-layer prepared offload
means are 201.550\,ms for \name and 218.266\,ms for Mooncake local batch.
Reload takes 196.806 and 222.676\,ms, respectively.
Mean sample bandwidth is 53.27/54.56\,GB/s for \name and
50.11/49.32\,GB/s for the local batch adapter (D2H/H2D).
These values use that cohort's own control. SparDA tuning also matters:
at 256 pages and eight CPU regions, its 32-CTA variant beats \name on
prepared H2D (0.4164 versus 0.4274\,ms).
These results compare complete transfer paths. They do not isolate the
effects of packing and double buffering, measure overlap with model
computation, or establish an end-to-end serving speedup.
\section{Additional Predictor Results}
\label{app:predictor-details}

The scheduler uses predicted remaining length to order requests. We
report the complete accuracy comparison, then the serving effect of
using that ordering, and finally the cost--freshness tradeoff of sparse
inference. The accuracy comparison uses 300 AIME requests (AIME300)
and a 200-request LiveCodeBench population (LCB200); 199 of the latter
reach the plotted decoding step.

\subsection{Complete Prediction Metrics}
\paragraph{Regression accuracy.}
Table~\ref{tab:supp-pred-regression} compares all methods on the same
step-500 survivors: 300 AIME requests and 199 LCB requests.
No method dominates both workloads. On AIME300, \name gives the
best ordering ($\tau_b=0.34$), while LTR has the lowest absolute error.  On
LCB200, PiA gives the best ordering and lowest absolute error
($\tau_b=0.60$, MAE 4,660 tokens), closely followed by \name
($\tau_b=0.56$, MAE 4,744). The main paper's predictor scatter plot
also exposes the quantized outputs of S3 and the
narrow prediction bands of the prompt-only regressors.

\begin{table*}[t]
\centering
\caption{Output-length prediction on the common step-500 populations.  \name
and DSMN predict at step 500; the other methods predict at prefill time.  These
development populations combine validation and test requests.}
\label{tab:supp-pred-regression}
\begin{tabular}{llrrrrr}
\toprule
\textbf{Model} & \textbf{Workload} & \textbf{$n$} & \textbf{MAE} & \textbf{RMSE} & \textbf{$R^2$} & \textbf{$\tau_b$} \\
\midrule
\name & AIME300 & 300 & 3,471 & 4,604 & -0.14 & \textbf{0.34} \\
DSMN  & AIME300 & 300 & 3,942 & 5,580 & -0.68 & 0.24 \\
LTR   & AIME300 & 300 & \textbf{3,179} & \textbf{3,993} & \textbf{0.14} & 0.24 \\
S3    & AIME300 & 300 & 4,459 & 5,833 & -0.83 & 0.21 \\
PiA   & AIME300 & 300 & 3,377 & 4,114 & 0.09 & 0.30 \\
\midrule
\name & LCB200 & 199 & 4,744 & 7,286 & 0.15 & 0.56 \\
DSMN  & LCB200 & 199 & 5,914 & 8,869 & -0.26 & 0.46 \\
LTR   & LCB200 & 199 & 5,664 & 7,224 & 0.16 & 0.32 \\
S3    & LCB200 & 199 & 7,333 & 9,365 & -0.41 & 0.37 \\
PiA   & LCB200 & 199 & \textbf{4,660} & \textbf{7,146} & \textbf{0.18} & \textbf{0.60} \\
\bottomrule
\end{tabular}%
\end{table*}

\paragraph{Ranking quality.}
The victim rule needs to rank requests by remaining length. We therefore
report Kendall's $\tau_b$ alongside absolute error: a numerically imperfect
estimate can still put two requests in the correct order. On these common
step-500 populations, \name has the best ordering on AIME, while PiA leads
on LCB. Decoding-time hidden states provide useful information, but
prompt-time methods remain competitive on code generation. These
accuracy measurements alone do not establish a serving benefit.

\paragraph{Victim selection.}
\label{app:victim-comparison}
We compare SRJF with physical-largest victims in three matched runs
on each workload, AIME1000 and LCB880. Both arms serve Qwen3-32B with
prediction enabled.
SRJF lowers p99 ITL by 13.86\% on AIME and 14.57\% on LCB, while
throughput falls by 7.88\% and 7.41\% and p99 TPOT rises by 13.68\%
and 0.62\%, respectively. Thus shorter tail chunk gaps come at a
throughput cost in these controls. Favoring nearly finished requests may
free less memory immediately, but the comparison does not isolate this
mechanism. The full-system gains cannot be attributed to learned victim
selection alone.

\subsection{Sensitivity to the Prediction Interval}
\label{sec:supp-k-sweep}
\label{app:prediction-interval}

Sparse online inference reduces how often the predictor runs, but also
changes how fresh its estimates are. We vary $K$ on Qwen3-32B/LCB880,
using H200 DP4/TP2 at 4 requests/s. Each of three arrival seeds (30--32)
has fresh $K=1$, 64 and 128 runs, for nine observations. All runs complete
880 requests and 9{,}074{,}823 output tokens. They keep 45{,}105 GPU KV
blocks per worker, 800\,GiB CPU KV, age-5/Q64 recovery, fused K/V kernels
and double buffering.

\begin{figure*}[t]
\centering
\includegraphics[width=0.84\linewidth]{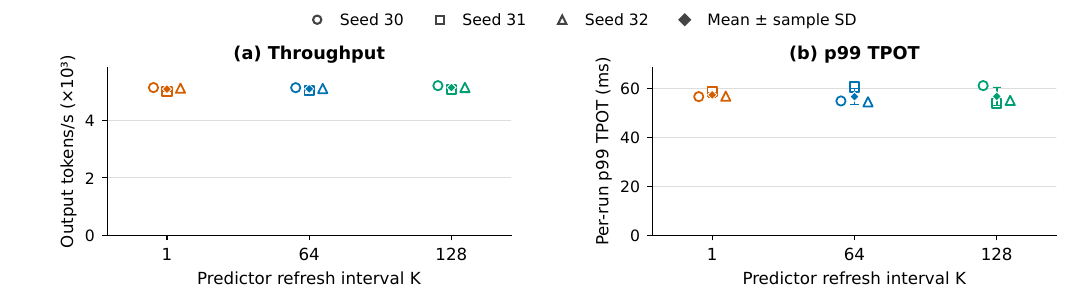}
\caption{Prediction-interval sensitivity with the current H200 transfer
path. Points show the three arrival seeds; diamonds and error bars show
means and sample standard deviations. Throughput and p99 TPOT change
much less than predictor activity.}
\Description{Two panels compare K1, K64 and K128 on output throughput and
p99 TPOT. All three run values remain visible around each mean.}
\label{fig:app-k-sensitivity}
\end{figure*}

Relative to $K=64$, $K=1$ changes throughput by $-0.12\%$, mean TPOT
by $+1.83\%$, and p99 TPOT by $+1.45\%$, averaging matched ratios.
$K=128$ changes them by $+0.90\%$, $-0.05\%$, and $+0.53\%$.
Its throughput is higher in all three pairs, but its p99 TPOT changes
range from $-10.88\%$ to $+11.44\%$. These small, mixed serving effects
do not establish one best interval for every workload.

Predictor activity changes much more. $K=64$ records 97.13\% fewer
batched invocations and 98.41\% fewer selected rows than $K=1$.
$K=128$ selects 46.32\% fewer rows than $K=64$, but makes 32.65\% more
calls because exceptional refreshes bypass the regular interval.
The counters are cumulative snapshots taken after the run, summed from
the output TP ranks; no corresponding before snapshots were recorded.
They support a reduction in recorded predictor activity, not a measured
reduction in predictor execution time. Because $K$ also changes the
estimates used for scheduling, this is a whole-system sensitivity study,
not an isolated head-latency benchmark.
\section{Recompute vs.\ Offload Cost Model}
\label{sec:appendix-cost-model}
\label{app:cost-model}

A preempted request can recover by rebuilding missing KV or by reloading
saved KV. This simple model compares rebuilding a full context with one
offload--reload round trip. It explains the compute--transfer tradeoff;
the serving system does not use its crossover to choose a recovery action.

\paragraph{Use the model's actual dimensions.}
Let $S$ be context length, $L$ the number of layers, $D$ the hidden width,
and $I$ the intermediate width of the SwiGLU MLP. Let $H_q$ and $H_{\rm KV}$
be the numbers of query and KV heads, with head dimension $d$.
The combined query width is $A=H_qd$ and the combined KV width is
$C=H_{\rm KV}d$. A KV element occupies $P$ bytes ($P=2$ for BF16).

These widths need not equal the hidden width. Qwen3-32B, for example,
has $D=5{,}120$ but $A=8{,}192$, and its MLP uses $I=25{,}600$.
Table~\ref{tab:cost-model-dimensions} gives the dimensions used here.

\begin{table*}[t]
\centering
\caption{Dense-model dimensions and BF16 KV payload before TP sharding.
The payload counts both K and V across all layers.}
\label{tab:cost-model-dimensions}
\begin{tabular}{lrrrrrr}
\toprule
Model & $L$ & $D$ & $I$ & $A$ & $C$ & KiB/token \\
\midrule
Qwen3-32B & 64 & 5,120 & 25,600 & 8,192 & 1,024 & 256 \\
Llama-3.1-8B & 32 & 4,096 & 14,336 & 4,096 & 1,024 & 128 \\
\bottomrule
\end{tabular}
\end{table*}

\paragraph{Bytes moved.}
The full KV cache occupies $2SLCP$ bytes: one key and one value per token
in each layer. Offloading and later reloading that cache therefore moves
\begin{equation}
  V_{\rm roundtrip}=4SLCP.
\end{equation}
This count applies to the full available state. Partial snapshots and
reused prefixes require accounting for the state actually transferred.

\paragraph{Work to rebuild KV.}
Count a multiply--add as two FLOPs. The three SwiGLU projections cost
$6SDI$ FLOPs per layer. Query and output projections cost $4SDA$;
key and value projections cost $4SDC$. The leading term for the two
causal-attention matrix products is $2S^2A$. Thus,
\begin{equation}
  F_{\rm layer}(S) \simeq 6SDI+4SD(A+C)+2S^2A.
\end{equation}
The attention term counts useful lower-triangular work. Actual kernels
may perform additional work because of tiling and padding. Normalization,
rotary embeddings, activation functions and the output vocabulary head
are omitted.

\paragraph{Comparing modeled times.}
Let $B_{\rm comp}$ be effective compute throughput in FLOPs/s and
$B_{\rm mem}$ effective host--device bandwidth in bytes/s. With one
bandwidth for both transfer directions,
\begin{equation}
  t_{\rm recomp}\simeq\frac{LF_{\rm layer}(S)}{B_{\rm comp}},
  \qquad
  t_{\rm transfer}\simeq\frac{4SLCP}{B_{\rm mem}}.
\end{equation}
If the two directions have different effective bandwidths,
$t_{\rm transfer}=2SLCP(1/B_{\rm D2H}+1/B_{\rm H2D})$ instead.
For the common-bandwidth approximation, their ratio is
\begin{equation}
  R(S)=\frac{t_{\rm recomp}}{t_{\rm transfer}}
  \simeq\frac{B_{\rm mem}}{B_{\rm comp}}
     \frac{6DI+4D(A+C)+2SA}{4CP}.
\end{equation}
Values above one favor the transfer round trip within this model.
Writing $\rho=B_{\rm comp}/B_{\rm mem}$, the modeled crossover is
\begin{equation}
  S^\star=\frac{4CP\rho-6DI-4D(A+C)}{2A}.
\end{equation}
When $S^\star\leq0$, this approximation favors transfer for positive
context lengths. That conclusion depends on the assumed effective rates.
Peak hardware specifications alone do not establish those rates.

\paragraph{Relation to the simpler formula.}
If $A=D$, $I=4D$, and $r=H_{\rm KV}/H_q$, then $C=rD$ and the ratio
reduces to
\begin{equation}
  R(S)\simeq\frac{B_{\rm mem}}{B_{\rm comp}}
      \frac{D(7+r)+S/2}{Pr}.
\end{equation}
These assumptions explain the earlier expression. They are not the
actual Qwen3-32B dimensions. MHA, GQA and MQA differ in $H_{\rm KV}$;
the general expression above handles each without assuming $A=D$.

\subsection{Compressed and Recurrent Attention}
\label{app:attention-cost-extensions}

The six-panel comparison in Figure~\ref{fig:recompute-offload-comparison}

uses hypothetical blocks with a common SwiGLU width $I=4D$ and
$\rho=12{,}500$ FLOPs/byte. MHA, GQA and MQA use 32 query heads,
head width $D/32$, and 32, 8 and 1 KV heads, respectively.
For MLA, KDA and CSA, we instead keep the source models' attention
dimensions fixed while varying $D$. Thus the panels compare specified
block designs, not equally capable models or complete hybrid/MoE models.

Let $M(S)$ denote one layer's saved-state bytes and $F(S,D)$ its
contribution to rebuilding the full context. Each panel uses
$R=F/(2\rho M)$: rebuilding from scratch versus one complete
offload--reload round trip. Cache vectors use BF16; recurrent and
compressor state use FP32 where required by the reference implementations.

\paragraph{MLA: save the latent state.}
DeepSeek-V3's MLA~\cite{deepseek2024v3} stores a shared 512-element
KV latent and a 64-element RoPE key per token. It does not need separate
expanded keys and values in the saved snapshot. We use 128 query heads,
128-dimensional non-RoPE keys and values, and query rank 1,536.
Rebuilding uses expanded prefill: temporary per-head keys and values
support attention, while the retained cache remains compressed. Including
the projections and leading causal-attention work gives
\begin{align}
 M_{\rm MLA}(S)&=1{,}152S,\\
 F_{\rm MLA}(S,D)&\simeq
 S(24D^2+36{,}992D+109{,}051{,}904)\notag\\
 &\quad+40{,}960S^2.
\end{align}
The compact decode path has a different attention cost; we do not use
single-token decode work to estimate a full prefill.

\paragraph{Pure KDA: save the recurrent state.}
Kimi Delta Attention~\cite{kimiteam2025linear} keeps a recurrent matrix
and short-convolution history instead of a growing token cache.
We use 32 heads of width 128, convolution width four, gate rank 128,
and the published chunked-prefill approximation with chunk size 64.
Saving the FP32 matrices and three BF16 convolution buffers costs
\begin{align}
 M_{\rm KDA}&=4\cdot32\cdot128^2
             +2\cdot3\cdot4{,}096\cdot4\notag\\
             &=2{,}195{,}456,\\
 F_{\rm KDA}(S,D)&\simeq
 S(24D^2+33{,}344D+6{,}258{,}688).
\end{align}
The work includes projections, gates, convolutions and the published
chunked core, rather than replaying the single-token recurrent kernel.
This is a pure KDA block; Kimi Linear also contains MLA layers whose
cache grows with context length.

\paragraph{CSA: save all compressed history.}
DeepSeek-V4-Flash's CSA~\cite{deepseek2026v4} compresses every four
tokens into one shared key--value entry, then selects at most 512 entries
for each query. The snapshot must keep all compressed entries and indexer
keys, because the next query may select a different subset. It also keeps
a 128-token sliding window and the overlapping compressor state.
We use 64 main heads of width 512, 64 indexer heads of width 128,
query rank 1,024, and eight output groups of width 1,024. With the
reference implementation's full FP32 compressor buffers,
\begin{align}
 M_{\rm CSA}(S)&=1{,}280\lfloor S/4\rfloor
       +1{,}024\min(S,128)\notag\\
       &\quad+81{,}920.
\end{align}
The final term saves both main and indexer compressor buffers, including
slots that a more compact snapshot could omit. Define
$n_t=\lfloor t/4\rfloor$ and
$a_t=\min(t,128)+\min(512,n_t)$. The corresponding work is
\begin{align}
 F_{\rm CSA}(S,D)&\simeq
 S(24D^2+24{,}704D+150{,}994{,}944)\notag\\
 &\quad+131{,}072\sum_{t=1}^{S}a_t
       +16{,}384\sum_{t=1}^{S}n_t.
\end{align}
The last term scans the compressed history for indexer scores; top-$k$
does not remove this work. These counts use the reference BF16 caches,
not the production model's low-precision storage or arithmetic speedups.

\paragraph{Scope.}
These counts describe unsharded blocks. With tensor parallelism, each
worker needs its own state, work and bandwidth accounting. Communication
and KV replication prevent a simple division of all costs by the TP degree.
Full hybrid and MoE models also need their actual layer mix and expert
computation.

The model omits cached-prefix savings, waiting for GPU space, launch and
packing costs, overlap, and interference with other requests. It also
omits softmax, normalization, nonlinearities, CSA selection/pooling work,
and small state metadata. A saved prefix or checkpoint reduces rebuild
work. Compute rates can also differ across kernels.
The plotted ratio therefore does not predict serving throughput or total
interruption time. Those outcomes require measured system comparisons.
\section{Evaluation Settings and Implementation}
\label{sec:supp-methods}
\label{app:methods}

We use two kinds of serving experiments. Fixed-length replay gives every
system the same amount of output work. Natural-completion experiments let
responses end normally and check the resulting programs. We keep their
results separate because natural completion can change both output length
and task outcome.

We use American Invitational Mathematics Examination (AIME) math tasks
and LiveCodeBench (LCB)~\cite{jain2024livecodebench} programming tasks.
A numeric suffix gives the
request count of a study's replay, not the number of repeated runs.

\subsection{Current H200 Comparisons}
\paragraph{Hardware and common settings.}
The repeated studies use NVIDIA H200 GPUs with 141\,GB each and BF16
model weights and KV. Qwen3-32B uses four two-GPU replicas (DP4/TP2)
unless a study states otherwise. Llama-3.1-8B uses eight single-GPU
replicas (DP8/TP1). Each node runs one serving deployment at a time.
The natural LCB study uses one TP2 replica and reserves the whole node.

All serving systems use eager execution, GPU memory fraction 0.90,
a 40{,}960-token context limit, and 16-token KV pages. Prefix caching and
chunked prefill are enabled. The vLLM systems use a 2{,}048-token batch
budget and at most 1{,}024 sequences. SGLang uses 2{,}048-token prefill and
chunk limits and at most 1{,}024 running requests. These backend controls
have different meanings; matching their values does not make the
schedulers identical. Each formal run starts fresh engines with empty
caches. Rates refer to the entire deployment, with round-robin routing.

\paragraph{Systems.}
Recompute uses the maintained vLLM 0.9.2 build with its default scheduler,
recomputation-based preemption, and no \name predictor or CPU pool.
FIFO64 adds a client-side FIFO admission queue to this unchanged engine.
It allows at most 64 dispatched but unfinished requests per replica,
or 256 across DP4. Requests still waiting for dispatch do not occupy
these slots; their waiting remains part of arrival-based latency.

\name uses an 800\,GiB aggregate CPU KV payload budget, the exact-extent
allocator, bulk KV transfer with fused K/V kernels and double buffering,
and age-based protected recovery. Its default recovery age is 5\,s and
its protected turn is 64 fresh tokens. The SRJF victim policy uses frozen,
workload-specific predictor weights with prediction interval $K=64$.
P0 keeps the same storage, transfer and recovery stack but disables
prediction and selects physical-largest victims. Exceptions, including
P0 on Cachewise, are stated with each study.

The four-system studies also include LMCache 0.3.5 with upstream
vLLM 0.9.2, and SGLang v0.5.20 with HiCache. LMCache uses 256-token chunks,
partial-prompt saving and an 800\,GiB CPU payload budget, with decode
saving disabled and no disk or remote backend. Its resumed-request path
uses the checked prompt-resume repair. HiCache uses its native page-first
layout, kernel transfers and write-through caching, with 797.219\,GiB of
CPU payload and no disk or remote backend. It may restore a cached prefix
and rebuild newer state; neither baseline is described as preserving
every request's latest complete decode snapshot.

\paragraph{Memory accounting.}
Equal payload budgets do not mean equal host-memory use. For Qwen3-32B
DP4/TP2, \name exposes 100\,GiB per worker, divided across 64 layer tensors.
The pinned allocator rounds each 1.5625\,GiB allocation to 2\,GiB, giving
a source-derived 1{,}024\,GiB backing requirement for the 800\,GiB payload.
The extra 224\,GiB is outside the KV allocator's usable space.
GPU staging buffers are separate from this host backing.

We also record a sampled memory guard: the larger of node-used memory
and the observed cgroup charge at each sample. Its peak includes startup,
serving and teardown, and can include file-cache and other charges.
It is not process RSS or an isolated measure of the KV pool.
Systems retain their native GPU KV capacities. In the IMO-AnswerBench study,
these are 721{,}680 KV tokens per TP2 replica for recompute/\name,
720{,}080 for LMCache and 768{,}480 for HiCache. The comparison therefore
includes software and resource differences between complete systems.

\subsection{Workload Populations and Repetitions}
\paragraph{Fixed-length reasoning replay.}
The 1,000-request AIME replay (AIME1000) repeats 355 distinct development
tasks: 178 from the original
predictor validation fold and 177 from its test fold, without training-task
overlap. The 880-request LiveCodeBench replay (LCB880) contains 880 tasks,
including 300 used in predictor
training. These are development-exposed serving populations.
Each system receives the same saved prompt tokens and output budgets,
ignoring EOS until the recorded length is reached. This fixes the amount
of output work; it does not evaluate answer quality.

The main recompute/FIFO64/\name comparison uses three matched arrival
seeds (30--32) per workload at 4 requests/s. Each block uses the same
inputs, arrivals and routing across systems, with rotated system order.
The continuing-arrival study uses three successive 900-second phases at
0.25, 0.5 and 1 request/s, then drains all outstanding requests.
The $K$ sensitivity study runs fresh $K=1,64,128$ configurations on
LCB880 for each of the same three seeds. It does not reuse a main-study
run as its $K=64$ reference.

\paragraph{Additional workloads.}
IMO-AnswerBench~\cite{superhumanRepository} contributes all 399
usable questions and 7{,}279{,}811 prescribed output tokens per run.
Four systems run three matched blocks at 4 requests/s, with seeds 50--52.
The prompt and common output-length freeze are shared across systems.
The predictor is trained on 100 of these questions; the overlap analysis
is in Appendix~\ref{app:superhuman-exposure}.
Cachewise~\cite{cachewiseCodingTraces} instead preserves shared prefixes and
parent--child dependencies in 1{,}418 calls. Appendix~\ref{app:cachewise}
describes why its four root arrivals per second differ from a fixed
request rate. In both four-system studies, blocks 0 and 2 run on one
H200 node and block 1 on the other; comparisons remain within each block.

The arXiv and BurstGPT tuning confirmations each use 2{,}000 saved
requests. arXiv prompts use the \texttt{ccdv} arXiv summarization dataset.
Qwen thinking is disabled during prompt preparation. BurstGPT supplies
lengths rather than conversation text, so its prompts are deterministic
synthetic tokens. Llama/AIME uses 1{,}000 saved requests and 17{,}463{,}660
output tokens. Its 413 generation-capped profiles contribute 96.44\% of
replayed output. The three repetitions of each tuning comparison reuse
the same frozen input, arrival and routing schedule; they are not new
task samples.

\paragraph{Natural completion.}
The natural-completion study uses 175 new LiveCodeBench v6 tasks
(LCB175). Recompute, P0 and
SRJF each run three seeds at 1 request/s on one H200 TP2 replica, with
45{,}105 GPU blocks per worker. P0 and SRJF use 200\,GiB of CPU KV.
Responses end at natural EOS or the 32{,}768-new-token limit.
Offline program tests use the same sandbox, a 300-second limit per task,
and one scoring task at a time. Scoring time is outside serving time.

\subsection{Metrics}
TTFT measures HTTP dispatch to the first token-bearing chunk. We also
report arrival-to-first-output time, which includes any admission queue.
For requests with at least two output tokens, TPOT is the time between
the first and last token-bearing chunks divided by the output count
minus one. Single-token responses have undefined TPOT and are excluded
from that metric. ITL pools gaps between consecutive token-bearing
chunks. It measures received chunks, not exact device-side token timing.

A request's longest output gap is one maximum per request. Its p99 differs
from both pooled ITL p99 and p99 TPOT. Completion latency includes the
entire response; its starting point is stated with each result.
Finite-replay throughput counts all output tokens from the first
scheduled arrival through final completion, including the drain.
It is not a measurement of sustainable arrival capacity.

We compute quantiles within each run. Repeated summaries average the
run-level values; percentage changes compare systems within each
matched block before averaging. The main repeated AIME/LCB and
IMO-AnswerBench figures use sample standard deviations, as do the appendix
$K$ and Cachewise figures. The retained appendix natural-completion and
tuning-confirmation figures use population standard deviations.
These show run-to-run spread, not confidence intervals. Requests within
a shared run are not independent repetitions.
Missing counters remain missing. Preempted or discarded KV-token
instances do not measure GPU work spent recomputing them.

\subsection{Earlier Load Sweeps}
The earlier load curves use the same H200 model topologies but mostly
one run per point, with arrival seed 10. Qwen AIME/LCB use 0.5, 1, 2, 4
and 8 requests/s; Llama uses 1, 2, 4, 8, 16 and 32. arXiv uses 1--16 and
BurstGPT 2--32 requests/s, each with 2{,}000 requests. Some systems run
on different node allocations. They remain distinct from the repeated
studies above. Historical LMCache restoration limits and the two new
rate-8 observations are described in Appendix~\ref{app:baseline-corrections}.
Earlier A100/H100 experiments are separated in
Appendix~\ref{supp:historical}.

\subsection{Motivation Replay}
\paragraph{Measurement setup.}
The motivation replay contains 200 LiveCodeBench requests (LCB200).
It serves Qwen3-32B in BF16 on H200 DP4/TP2, with finite Poisson
arrivals at 16 requests/s. \name uses SRJF/K64 offloading with recovery disabled,
11{,}251 GPU KV blocks per worker (16 tokens per block), and 800\,GiB of
aggregate CPU KV storage. The reduced GPU capacity exposes memory
pressure; the end-to-end comparison uses full capacity. The grouped summary
includes all 200 requests from each of three arrival seeds: 443 request
executions with no preemption, 39 with one, and 118 with at least two. Seed~10 uses the original allocation; seeds~11 and~12 use its
replacement.

For the timeline, we select the median-preemption request among those
preempted at least twice in seed~10, breaking ties by request ID. It
completes 31{,}157 output tokens with seven preemptions; the figure shows
cycles three through five to make the short decoding turns visible.
Panel (a) of the main paper's KV-thrashing figure follows the same request's replica from its
first admission to its last completion. It overlays GPU KV usage, sampled
about every 5.2 seconds, with a cumulative count of all 31 timestamped
preemptions on that replica. Usage is the fraction of the reserved GPU KV
pool in use; total device memory is largely preallocated. The sampled
curve shows how pressure changes during the replay, not the exact
occupancy at each allocation failure.

Streaming pauses are gaps between token-bearing chunks, while the orange
intervals run from eviction to the next fresh token. They include both
waiting and recovery; they are not measurements of pure RAM waiting.
These measurements associate preemption with interruptions; they do not
separate transfer time from waiting for GPU space or assign a fixed cost
to each eviction.

\subsection{Allocator Controls}
We run Qwen3-32B BF16 on one H200 TP4 replica with an 800\,GiB aggregate
CPU pool. Each design serves AIME1000 and the first 800 tasks of the
retained LCB880 replay (LCB800), at finite Poisson 4 requests/s with
three schedule seeds (10--12).

All arms run eager vLLM 0.9.2 with 104{,}935 GPU blocks per worker.
They use learned SRJF, K32 prediction, asynchronous transfers, and
age-5/Q64 recovery. Designs run serially on GPUs 0--3 with the same NUMA0 affinity.
The padding statistic describes reserved space across allocations; it
is not instantaneous fragmentation or process memory usage.

\subsection{KV Pool and Transfer Implementation}
\label{app:kv-implementation}

Each GPU worker preallocates pinned CPU buffers for its KV shard. An
extent identifies the same contiguous block range in each layer's K and
V buffers; layers remain in separate arrays. The byte budget and maximum
number of offloaded requests are separate limits. Free space in one
data-parallel replica cannot serve another. Pool backing, allocator
metadata and block-index tables all consume host memory. Metadata and
index-table costs are included in sampled host-memory use rather than
measured separately.

The scheduler supplies GPU block IDs and CPU offsets to the transfer
engine. Staging buffers are shared across requests and sized by the
maximum context length. For Qwen3-32B BF16 at TP2 with a 40,960-token
limit, one K/V staging pair uses 80\,MiB per GPU, and two pairs use
160\,MiB. Block-index buffers are separate and reused across layers.
Compatible one- or two-block transfers can bypass staging.

Within each layer, offloads are submitted before reloads. Attention
waits for both transfer streams so outgoing KV is safe and incoming KV
is ready. Empty steps and error cleanup also wait before reusing storage.
Reloaded CPU extents remain reserved until the model step completes.

\subsection{Transfer Measurements}
\paragraph{Methodology.}
We measure host--device transfers on H200 GPUs using BF16 KV,
16-token pages, eight KV heads and head dimension 128. The matrix covers
1--3{,}304 pages per layer, 1/32/64 layers, and three layouts: contiguous
GPU/CPU regions, scattered GPU pages with one CPU region, and the same
scattered pages with up to eight CPU regions. Each cell has 20 warmups and
100 measured operations in each of three fresh-process repetitions.
We report arithmetic means and population standard deviations across the
three repetition means; seeds also vary mappings and measurement order.
We time both GPU-to-CPU offload (D2H) and CPU-to-GPU reload (H2D).
Prepared timing excludes adapter mapping/index preparation, whereas
dispatch-inclusive timing includes it. Both include payload packing,
movement, scattering and completion; HiCache's internal C++ descriptor
construction remains timed in both views.

\name uses asynchronous bulk KV transfer, fused K/V gather/scatter kernels and two
staging pairs, allowing gathering or scattering of one chunk to overlap DMA
for another. In this geometry the two pairs occupy 320\,MiB.
The HiCache cohort compares fresh \name controls with SGLang v0.5.20
layer-first and native page-first paths under CUDA 12.8. The earlier cohort
compares \name with SparDA-derived and Mooncake-derived local adapters under
CUDA 12.6/12.8; the batch API is available only in the latter environment.
Each comparison keeps all methods and repetitions for a geometry on the
same host and GPU. All participating layers are ready before an operation;
there is no model computation in these microbenchmarks.

\paragraph{K/V kernels and double-buffering diagnostic.}
The four-way diagnostic moves 2\,GiB across 64 layers, between 1,024
scattered GPU blocks and eight CPU regions. Each configuration measures
ten operations. It runs on H200 GPU0 without model computation, using
Qwen TP2 rank-local geometry: BF16, four KV heads per rank, head
dimension 128, 16-token blocks, and token-major (NHD) storage. A staging
pair has one K and one V buffer; the second pair adds 80\,MiB of scratch
space. Timing runs through transfer completion, including any CPU enqueue
gaps and synchronization. The new paths share planner and metadata
changes, which are not varied separately.

\paragraph{Serving ablation.}
We compare the legacy one-pair path with fused K/V kernels and two staging
pairs on Qwen3-32B, using a 300-request AIME replay (AIME300) and a
200-request LiveCodeBench replay (LCB200), on H200 DP2/TP1 with
finite Poisson arrivals at 4 requests/s. Each arm has two runs, paired by
arrival seeds 10 and 11.
Both arms use exact extents, learned SRJF, K64, age-5/Q64 recovery,
15{,}023 GPU blocks per worker, and 800\,GiB aggregate CPU KV payload
with 1{,}024\,GiB backing. The replicas use GPU0/NUMA0 and GPU4/NUMA1.
These settings differ from the transfer diagnostic above. The earlier
TP2 four-way serving screen also had mixed latency results and only one
run per arm; it is separate from these repetitions.

\subsection{Prediction and Victim Controls}
We compare \name's frozen, workload-matched decode-time predictor with four
baselines: DSMN, a single-layer hidden-state predictor;
LTR~\cite{fu_efficient_2024}, an OPT-350M prompt-time regressor;
S3, a DistilBERT prompt-time length-bucket classifier; and
PiA, which applies LoRA adapters to the target LLM to predict output length at
prefill time.  We use Qwen3-32B generations from AIME and LiveCodeBench.
Training, validation, and test request IDs are disjoint.  Checkpoint
selection uses only validation data.  The
scatter plot uses the exact AIME300 and LCB200 development populations used by
our system studies; these populations combine validation and test requests and
have therefore been exposed to development analysis.  No continual or online
adaptation is used.

We separate the cost of running the predictor from the effect of using its estimates.
On Qwen3-32B BF16 with H200 DP4/TP2, eager execution, 800\,GiB aggregate CPU KV capacity and deployment-wide Poisson arrivals at 4 requests/s, we compare prediction disabled with K64 prediction enabled but unused by victim selection.
Both controls use physical-largest victim selection, the same GPU KV capacity, optimized transfers, and recovery age 5\,s/quantum 64.
Across three matched repetitions, enabling prediction changes mean output throughput by $-0.58\%$ on AIME1000 and $-0.26\%$ on LCB880.
These are serving-level comparisons: predictor execution can affect batching and wall-clock recovery timing even though the victim rule is unchanged.
They do not measure isolated head latency or the benefit of prediction-guided victims.

\subsection{Recovery Scheduling Details}
\label{app:recovery-implementation}

The stall timer uses a monotonic clock and starts at the first preemption
in a period without new output. Reloads, recomputation, retries and further
preemptions leave it running. A previously preempted request remains
eligible for protected recovery whether it is offloaded, waiting for
recomputation, or already running without new output. Requests never
preempted are ineligible. Selection prefers the oldest eligible stall,
with arrival time and request ID breaking ties.

Recovery reserves both a running-request slot and GPU KV space. When
choosing physical-largest victims, shared prefix blocks still in use do
not count as reclaimable memory. A saved CPU snapshot is reloaded;
discarded state is rebuilt while reusing and pinning any available cached
prefix. An already-running request retains its resident KV. If the
selected request cannot fit after reclaiming eligible memory, the
implementation reports failure. Completion or cancellation releases
protection early.

Ordinary reloads do not require the age threshold or grant protection.
After scheduling the running batch, the scheduler attempts them before
admitting new work when GPU space and scheduling budget permit, provided
the step has no preemptions. The stall timer still clears only when new
output is generated. Protected recovery handles requests that remain
stalled despite these ordinary opportunities to return.

\subsection{Recovery Controls}
\paragraph{Reduced-capacity Qwen/LCB study.}
The disabled, age-only and protected-Q64 comparison serves 50 LCB
requests on one H200 TP2 replica with Qwen3-32B and SRJF/K64.
It uses 11{,}251 GPU KV blocks per worker
and 100\,GiB CPU KV per worker. Arrivals retain replica 0's offsets from
the finite Poisson 16-request/s DP4 schedule, seed 10. It measures one
replica's routed requests, not the throughput of a full DP4 deployment.

\paragraph{Protection quantum.}
The Q64/Q1024 comparison serves Llama-3.1-8B/AIME1000 at DP8/TP1
on eight H200 GPUs, with an 800\,GiB CPU pool and finite Poisson
arrivals at 8 requests/s. Each arm and the same-node recompute control
have two runs.
The predictor, K64 cadence, transfer path, GPU capacity and 5\,s age
threshold stay fixed. Runs use the same seed-10 replay, with arm order
reversed between repetition blocks. Summaries report arithmetic means
and population SD; p99 summaries average run-level p99 values. Tuning
and confirmation use the same population, whose 413 generation-capped
profiles supply 96.44\% of output tokens. This result supports balancing
repeated recovery against other requests' waiting; it does not establish
one quantum for all workloads or a waiting-time bound under continued
arrivals.
\section{Historical Online-Training Extension}
\label{supp:online-training}
\label{app:online-training}

The main serving design uses a pretrained predictor with fixed weights and updates its length estimates as decoding reveals new hidden states.
This earlier H200 study asks whether updating the weights after completed requests helps under workload shift. Its one-replica TP2 setup and development populations are separate from the current repeated DP4 serving comparisons.
We report prediction accuracy, isolated update cost and live serving separately.

\subsection{Accuracy Under Workload Shift}
We first test whether completed requests can teach the predictor about
a changing workload. The replay gradually shifts from math to coding,
or from coding to math. Each direction uses three randomized orders of
the same 500 requests: 300 from AIME and 200 from LCB.
Table~\ref{supp:adaptation-accuracy} compares a frozen head with one that
learns from completed requests.
The source-domain share decreases from 100\% to 0\% in 10-point steps.
Each mixed phase contains 40 requests; the pure endpoint sizes are 120/20 requests for AIME-to-LCB and 20/120 for LCB-to-AIME.
The initial source-domain head is either frozen or updated with AdamW at learning rate $3\times10^{-5}$, five passes after every ten completions, batch size 128 and a FIFO window of 64 completed requests.
Each request is scored before its own completion label is available.
This stage replays saved features with serialized completions; it does not run live LLM serving.

\begin{table*}[t]
\centering
\caption{Causal feature-replay accuracy at decode token 1,000. Entries are means $\pm$ population SD over three orders of the same development population. MAE is measured in output tokens; changes compare arm means.}
\label{supp:adaptation-accuracy}
\setlength{\tabcolsep}{5pt}
\begin{tabular}{llrrr}
\toprule
Shift & Scored scope & Frozen MAE & Updated MAE & Change \\
\midrule
AIME $\to$ LCB & 494 requests/order & $4753.73\pm0.00$ & $3847.98\pm66.39$ & $-19.05\%$ \\
LCB $\to$ AIME & 494 requests/order & $4887.84\pm0.00$ & $3981.35\pm12.60$ & $-18.55\%$ \\
AIME $\to$ LCB & Final target phase & $7477.74\pm1251.80$ & $5618.72\pm1655.96$ & $-24.86\%$ \\
LCB $\to$ AIME & Final target phase & $5378.83\pm475.76$ & $3567.57\pm54.90$ & $-33.67\%$ \\
\bottomrule
\end{tabular}
\end{table*}

Only 494 requests reach token 1,000 and are eligible for that score.
The final LCB phase contains only 19--20 eligible requests per order; the final AIME phase contains 120.
Although mean error improves in both directions, the final AIME-to-LCB phase regresses by 8.74\% for seed 12.
These results support average improvement, not stable accuracy at every point in the shift.
The populations include historically used validation tasks; four LCB labels are capped served lengths, and AIME retains completion-selection bias.
The study does not establish held-out generalization, later-token accuracy or natural answer quality.

\subsection{Isolated Training Cost}
Updating the head repeatedly reuses a window of completed requests.
The packed path transfers that window to the GPU once per update,
reuses it across training passes, and reads scalar loss back at the end.
We compare it with the original path in six paired cases: both shift
directions and seeds 10--12 on one H200. Each case performs 50 updates
over 500 requests, keeping the optimizer, five passes per update, batch
size and replay window fixed. Timing includes packing and CPU-to-GPU
transfer.

Across the six pairs, total training time falls from $6.804\pm0.142$ to $3.983\pm0.043$ seconds, a 41.46\% reduction.
Mean update time falls from 136.08 to 79.67\,ms.
Every scored prediction and final weight hash matches between the two paths.
These isolated timings exclude live feature capture, completion transport, tensor-parallel weight installation and interference with serving.
They are not the total serving overhead.

\subsection{Live Serving: Three-Seed Comparison}
Better predictions and cheaper updates are useful only if they help
serving. We therefore compare frozen weights, original updates and
packed updates while the LLM serves requests. Both shift directions
use three schedule seeds (10--12) per mode, giving 18 observations.
Each run serves the same 500 requests and
4{,}914{,}049 forced output tokens on Qwen3-32B BF16, one H200 TP2
replica, eager vLLM 0.9.2, and finite Poisson arrivals at 4 requests/s.

All modes retain learned SRJF, K64 prediction, a 5\,s recovery threshold,
a 64-token protection quantum, a native GPU memory fraction of 0.90
and an 800\,GiB aggregate CPU KV payload (1{,}024\,GiB backing).
Frozen runs allocate 45{,}107 GPU KV blocks per worker; training runs
allocate 45{,}037, reflecting training memory within the same GPU budget.
Only the predictor head trains, on the TP leader GPU; Qwen weights stay
fixed. Completed features are buffered in CPU memory. Every ten
completions trigger five training passes over the last 64 requests,
using batch size 128 and learning rate $3\times10^{-5}$.
Both TP ranks install the updated weights at an engine-step boundary.

\begin{table*}[t]
\centering
\caption{Live serving with frozen weights, original updates, or packed
updates. Values are mean $\pm$ population SD across three runs per
direction/mode. Latencies are milliseconds; p99 TPOT averages the
per-run p99 values. Repetitions reuse one development population.}
\label{supp:adaptation-serving}
\setlength{\tabcolsep}{4pt}
\begin{tabular}{llrrrr}
\toprule
Shift & Mode & Output tok/s & Mean TTFT & Mean TPOT & p99 TPOT \\
\midrule
AIME $\to$ LCB & Frozen & $1712.12\pm40.97$ & $120.65\pm1.21$ & $88.80\pm0.37$ & $197.06\pm7.02$ \\
AIME $\to$ LCB & Original & $1605.02\pm63.42$ & $124.55\pm1.18$ & $86.85\pm0.24$ & $164.26\pm3.97$ \\
AIME $\to$ LCB & Packed & $1638.00\pm19.44$ & $123.39\pm1.15$ & $86.61\pm0.62$ & $172.86\pm14.31$ \\
\midrule
LCB $\to$ AIME & Frozen & $1679.15\pm97.17$ & $115.03\pm1.67$ & $94.15\pm0.78$ & $234.94\pm10.84$ \\
LCB $\to$ AIME & Original & $1690.33\pm9.06$ & $116.97\pm1.21$ & $92.34\pm0.05$ & $192.23\pm16.29$ \\
LCB $\to$ AIME & Packed & $1650.75\pm47.99$ & $116.31\pm1.34$ & $94.26\pm1.64$ & $214.94\pm13.04$ \\
\bottomrule
\end{tabular}
\end{table*}

The serving results are mixed. For AIME-to-LCB, packed updates reduce
throughput by 4.33\% while
lowering mean TPOT by 2.47\% and p99 TPOT by 12.28\% relative to frozen
weights (Table~\ref{supp:adaptation-serving}).
For LCB-to-AIME, throughput falls by 1.69\%, mean TPOT rises by 0.12\%,
and p99 TPOT falls by 8.51\%.
The reverse-direction throughput change varies across seeds:
$-6.41\%$, $+11.35\%$ and $-8.48\%$.
The mean of per-run p99 client ITL changes by $+1.15\%$ for AIME-to-LCB and
$-0.87\%$ for LCB-to-AIME. Thus, lower p99 TPOT does not imply a
uniform improvement across serving metrics.

\paragraph{Recorded update cost.}
Packed training takes 252.57\,ms per update on average. The original
path takes 303.32\,ms, so packing reduces this time by 16.73\%.
These values average six runs per update mode.
Recorded training plus weight installation accounts
for 0.424\% of packed serving time. This fraction excludes feature
capture, archival and synchronization outside those timers.
The full-system comparisons include those activities and scheduling
changes, so the timer fraction is not the total serving overhead.

The live path captures up to 16 early K64 or exceptional positions per
request through token 1{,}000, versus seven saved anchors in the
isolated replay. The longer live update time therefore cannot be
attributed solely to serving interference.

\paragraph{Delayed training labels.}
Training starts after a request finishes and its output length is known.
In the packed runs, the first update arrives on average
168.83\,s after the first arrival for AIME-to-LCB and 99.28\,s for
LCB-to-AIME. Only 6.90\% and 38.62\%, respectively, of recorded early
predictions through token 1{,}000 use updated weights.
Long requests thus delay the labels needed to improve predictions for
other requests. This is one plausible reason that the accuracy benefit
has limited reach during the replay; the measurements do not isolate
its contribution to net serving performance.

Prediction positions vary with scheduling, so pooled live prediction
error is not the fixed-anchor accuracy metric from the feature replay.
These are three orders of one development-exposed population, with
forced lengths and finite arrivals. Better prediction accuracy and
cheaper updates do not establish a consistent serving benefit here.
Online training therefore remains a supplementary extension; the
main design uses fixed-weight inference.
\section{Historical Serving Experiments}
\label{supp:historical}

These earlier experiments provide context for the current H200 evaluation.
They use different models, hardware and implementations, so they do not isolate the current predictor, CPU allocator, transfers or protected recovery.
The older A100 tables do not report repeated-run uncertainty. They illustrate earlier design tradeoffs, not the effect size of the current H200 system.

The earlier Qwen3-14B results include H100 load curves and an
output-length sweep. Their retained per-panel settings are incomplete. We leave them
out of the current evidence and use the documented H200 experiments
instead. The selected older studies below retain their own hardware,
workloads and transfer implementations.

\subsection{Recomputation and Synchronous Offload on A100}
\label{sec:eval-offload}
The historical A100-40GB testbed lists an AMD EPYC 7643 CPU and 500\,GB of system memory; the transfer microbenchmark separately uses an A100-80GB GPU.
Table~\ref{tab:recomp-vs-swap} compares QWQ-32B on 100 AIME requests at 2 requests/s and Llama-3.1-8B on 300 synthetic requests with 128 input and 1,024 output tokens.
For QWQ-32B, synchronous offload lowers p99 TPOT from 110.80 to 103.13\,ms, with throughput of 712.3 versus 702.6 tokens/s.
For the shorter Llama workload, recomputation instead reaches 3,885.8 versus 3,161.1 tokens/s.
These observations show a workload-dependent tradeoff; they do not establish its individual time costs.

\begin{table}[t]
\centering
\caption{Recompute vs.\ synchronous swap on A100. QWQ-32B uses AIME (100 requests, rate=2); Llama-3.1-8B uses random (300 requests, input=128, output=1024).}
\label{tab:recomp-vs-swap}
\setlength{\tabcolsep}{4pt}
\begin{tabular}{llrrrr}
\toprule
\textbf{Model} & \textbf{Mode} & \multicolumn{3}{c}{\textbf{TPOT (ms)}} & \textbf{Thpt} \\
 & & avg & p50 & p99 & (tok/s) \\
\midrule
QWQ-32B & Recompute & 67.28 & 61.73 & 110.80 & 702.6 \\
        & Swap      & 64.78 & 58.14 & \textbf{103.13} & \textbf{712.3} \\
\midrule
Llama-8B & Recompute & 55.93 & 57.42 & 142.91 & \textbf{3885.8} \\
         & Swap      & 69.61 & 74.50 & 144.37 & 3161.1 \\
\bottomrule
\end{tabular}
\end{table}

\subsection{Historical Transfer Microbenchmark}
Table~\ref{tab:swap-micro} reports block-transfer throughput on Llama-3.1-8B and an A100-80GB GPU; each block contains 16 tokens.
Moving more blocks per operation raises observed throughput over much of the tested range, but the trend is not strictly monotonic in either direction.
These are measurements of an older transfer path, separate from the current bulk-transfer path with fused K/V kernels and double buffering.

\begin{table}[t]
\centering
\caption{KV block swap throughput (blocks/s) on Llama-3.1-8B, A100-80GB. Each block is 16 tokens.}
\label{tab:swap-micro}
\begin{tabular}{rrr}
\toprule
\textbf{Batch (blocks)} & \textbf{GPU$\rightarrow$CPU} & \textbf{CPU$\rightarrow$GPU} \\
\midrule
1   & 1{,}068 & 1{,}103 \\
4   & 2{,}625 & 2{,}590 \\
16  & 3{,}681 & 3{,}623 \\
64  & 4{,}112 & 3{,}490 \\
256 & 4{,}237 & 3{,}496 \\
512 & 4{,}252 & 3{,}496 \\
\bottomrule
\end{tabular}
\end{table}

\subsection{Historical Victim Policies}
\label{supp:historical-victims}
Table~\ref{tab:victim-selection} uses Qwen2.5-14B, A100-40GB, and 200 synthetic requests with 128 input and 2,048 output tokens, submitted in offline mode with recomputation-based preemption.
The policies trigger similar preemption counts (446--479) but differ in p99 TPOT.
Largest-first has the lowest p99 TPOT among these four policies, while LIFO has the highest throughput.
Aggregate counts alone do not establish repeated preemption of particular requests or explain the latency differences.

\begin{table}[t]
\centering
\caption{Victim selection policy comparison (Qwen2.5-14B, A100-40GB, 200 requests). TPOT reported as avg/p50/p99 in ms.}
\label{tab:victim-selection}
\setlength{\tabcolsep}{4pt}
\begin{tabular}{lrrrrr}
\toprule
\textbf{Policy} & \textbf{\# Prepts.} & \multicolumn{3}{c}{\textbf{TPOT (ms)}} & \textbf{Thpt} \\
 & & avg & p50 & p99 & (tok/s) \\
\midrule
LIFO (last)   & 448 & 167.8 &  46.2 & 2{,}618.0 & 269.4 \\
Largest-first & 456 & 105.8 &  45.5 &   990.4 & 240.1 \\
Earliest      & 446 & 126.9 &  45.9 & 2{,}024.7 & 242.5 \\
Random        & 479 & 209.7 &  44.7 & 1{,}199.5 & 242.1 \\
\bottomrule
\end{tabular}
\end{table}

A separate historical oracle study used Qwen3-32B on 355 AIME development requests selected from completed runs. They contain 3,278,404 forced output tokens. The setting uses one H200/TP1, finite Poisson arrivals at 4 requests/s, a 400\,GiB CPU KV pool, eager execution, synchronous layerwise transfers, a 5\,s recovery age and a 64-token protection quantum.
Across three paired repetitions (seeds 0, 1, and 2), it compared admission-time oracle SRJF with predictor-on physical-largest selection; both arms retained the same predictor work.
The oracle received true lengths earlier than hidden-state predictions, so this historical comparison does not bound the headroom of the current learned policy.

\subsection{Synchronous Offload Under Sustained Pressure}
\label{sec:eval-ablation}
Table~\ref{tab:sustained-pressure} uses QWQ-32B on A100 with 40 synthetic requests, each containing 100 input and 2,000 output tokens, at 0.5 requests/s.
Synchronous offload lowers p99 TPOT from 680.7 to 249.5\,ms, while throughput changes from 46.6 to 45.9 tokens/s.
The offload run has fewer preemption events but more preempted token instances.
Those counters do not measure executed recomputation or isolate how much delay came from copying, scheduling or waiting.
These controls therefore cannot establish the current protected-recovery benefit.

\begin{table*}[t]
\centering
\caption{Sustained memory pressure (QWQ-32B, A100, 40 random requests, rate=0.5). TPOT in ms.}
\label{tab:sustained-pressure}
\begin{tabular}{lrrrrrr}
\toprule
\textbf{Mode} & \multicolumn{3}{c}{\textbf{TPOT (ms)}} & \textbf{Thpt} & \textbf{Pre-} & \textbf{Tokens} \\
 & avg & p50 & p99 & (tok/s) & \textbf{empt.} & \textbf{preempt.} \\
\midrule
Swap      & 128.0 & 116.9 & 249.5 & 45.9 & 191 & 126{,}509 \\
Recompute & 170.5 & 149.9 & 680.7 & 46.6 & 202 & 101{,}311 \\
\bottomrule
\end{tabular}%
\end{table*}

\subsection{Two Historical LMCache Corrections}
\label{app:baseline-corrections}

The new four-system studies use the checked prompt-resume path. We
selected two earlier rate-8 Qwen AIME/LCB observations for a source and
trigger audit. Both exposed the affected resume pattern. We repeat
these two points with the repair; this bounded audit does not classify
every earlier point. The reruns preserve their workloads and settings:
H200 DP4/TP2, BF16, finite Poisson arrivals at 8 requests/s, seed 10,
800\,GiB CPU KV and decode saving disabled.

\begin{table*}[t]
\centering
\caption{Historical LMCache observations and their repaired follow-ups.
Each cell is one run; old-to-new changes do not isolate the repair's
causal effect. Original observations remain available.}
\label{tab:app-lmcache-corrections}
\begin{tabular}{llrrrr}
\toprule
Workload & Observation & Output tok/s & p99 TPOT (ms) & p99 TTFT (ms) & Preemptions \\
\midrule
AIME1000 & Original & 6,844.34 & 169.02 & 282.79 & 1,732 \\
 & Repaired follow-up & 7,378.17 & 159.10 & 483.08 & 1,801 \\
LCB880 & Original & 5,159.31 & 173.51 & 277.42 & 1,103 \\
 & Repaired follow-up & 5,247.53 & 172.57 & 614.82 & 1,122 \\
\bottomrule
\end{tabular}
\end{table*}

Throughput and p99 TPOT improve, while p99 TTFT and preemption counts
increase. The new measurements are not repeated estimates and do not
retroactively validate the other historical points. The original
84-point sweep contains 30 observations with preemptions that require
source/trigger review before reuse as restoration evidence. A positive
preemption count alone does not prove a failed or corrupt restore.
The new IMO-AnswerBench and Cachewise comparisons provide the repeated
four-system evidence used for the current claims.
\section{Earlier Predictor Experiments}
\label{supp:earlier-predictors}

\subsection{Progressive Refinement on GSM8K}
\begin{figure}[htbp]
  \centering
  \includegraphics[width=\linewidth]{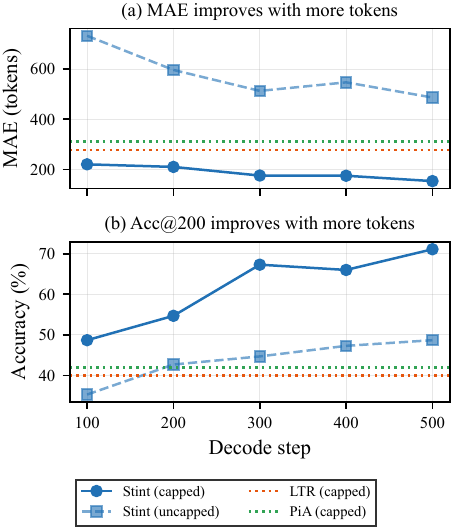}
  \caption{Earlier GSM8K trace: prediction errors and accuracy within 200 tokens as decoding advances. Horizontal lines show the prompt-time baselines. Individual steps need not improve monotonically.}
  \Description{Two line plots show mean absolute error decreasing and accuracy within 200 tokens increasing as decoding advances from 100 to 500 tokens. Horizontal lines mark fixed ahead-of-time baselines.}
  \label{fig:pred-progressive}
\end{figure}

In an earlier controlled GSM8K trace, the predictor's MAE decreases from 221 at decode step 100 to 154 at step 500 (Figure~\ref{fig:pred-progressive}).
The prompt-time baselines remain fixed after prefill.
This illustration uses a separate trace from the current common-population AIME300/LCB200 comparison and is not evidence that estimates improve monotonically for every request or workload.

\subsection{Progress Classification}
An earlier four-class head predicts progress quartiles: 0--25\%, 25--50\%, 50--75\%, and 75--100\%.
The retained exploratory results report 85.7\% overall accuracy on capped data and 79.3\% on uncapped data, with per-class accuracy between 80\% and 92\% on capped data.
These results describe a separate predictor variant; the current serving implementation uses a regression head.
Classification accuracy alone does not measure victim-ranking quality or serving benefit, and this variant is not used as evidence for the main system's performance.
The original manuscript description does not specify repeated-run uncertainty for these exploratory values.
  \clearpage

\end{document}